\documentclass[aps,twocolumn,amsmath,amssymb,showpacs,prb]{revtex4}
\usepackage{epsf}
\begin{document}
\title{Effect of the magnetic resonance on the electronic spectra of
high $T_c$ superconductors}
\author{M. Eschrig,$^{1,2}$ M. R. Norman$^{1}$}
\affiliation{ (1) Materials Science Division, Argonne National Laboratory,
Argonne, Illinois 60439}
\affiliation{ (2) Institut f{\"u}r Theoretische Festk{\"o}rperphysik,
Universit{\"a}t Karlsruhe, 76128 Karlsruhe, Germany}
\date{\today}   
\begin{abstract}
We explain recent experimental results on the superconducting state spectral 
function as obtained by angle resolved photoemission, as well as by tunneling,
in high $T_c$ cuprates.
In our model, electrons are coupled to the resonant 
spin fluctuation mode observed in inelastic neutron scattering
experiments, as well as to a gapped continuum. We show that, although the 
weight of the resonance is small, its effect on the electron self energy is 
large, and can explain various dispersion anomalies seen in the data.
In agreement with experiment, we find that these effects are a strong
function of doping.  We contrast our results to those expected
for electrons coupled to phonons.
\end{abstract}
\maketitle

\section{Introduction}

Understanding superconductivity in the cuprates is one of the 
great challenges of physics.  Determining the nature of
single particle excitations is of fundamental importance for achieving this
goal.  Two types of experiments have been extensively used to study such 
excitations:  angle resolved photoemission spectroscopy (ARPES) and 
tunneling.

In this paper, which deals with the superconducting state only,
we address the questions, what the spectral properties of fermionic 
excitations are, and how their low-energy dispersion is renormalized.
We do not directly address the question of the origin
of superconductivity in the cuprates.  Rather, we assume that
an effective pairing interaction exists, and study the additional
effects which coupling to certain collective excitations present in
cuprates have in renormalizing single particle properties.
The corresponding collective excitations responsible for
such renormalizations are most directly seen in other types of
experiments.  One of them, inelastic neutron scattering,
gives the most useful information about both phonons and
magnetic excitations in the energy range of interest ($<100 $meV). 

Motivated by earlier
work,\cite{Kampf90,Dahm96,Shen97,Norman97,Norman98,Abanov99,TKLee}
we have presented in Ref.~\onlinecite{Eschrig00} a model which 
describes the ARPES and tunneling spectra.
Here, we describe details of our calculations, and extend them by including
the effect of the spin fluctuation continuum.  In addition, we
address the issue of the doping dependence of the ARPES spectra. 
Finally, for comparison, we discuss the effect on the electrons of
coupling to a particular phonon which was recently suggested to account for the
renormalization of the ARPES dispersion in the nodal regions of the zone.

Our outline is the following: starting in Section II from the information
which experiments give about single particle properties of
low lying excitations in cuprates, we look
for a suitable collective excitation which best fits the data.
Then, we develop in Section III a model in which the collective mode is 
identified as the magnetic resonance observed in inelastic neutron scattering
experiments.  The results of calculations using this model are presented
in great detail.
Finally, in Section IV, we address the question what electron-phonon
coupling contributes to renormalization effects on the
dispersion.  Section V offers a brief summary.

\section{Experimental evidence}

\subsection{Angle resolved photoemission}

\begin{figure}
\centerline{
\epsfxsize=0.25\textwidth{\epsfbox{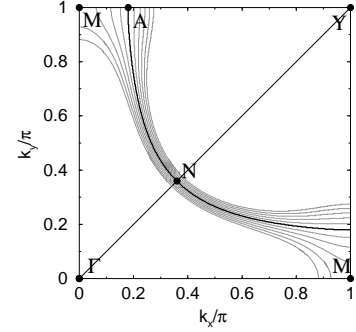}}
}
\caption{
\label{brill}
Notation used for special points in the Brillouin zone.
The Fermi surface is shown as a black curve.  Equal energy contours are shown 
in gray for energies between $\pm 50 $meV. The dispersion used here
was obtained by a 6-parameter tight binding fit to angle resolved photoemission 
dispersions in optimally doped 
Bi$_2$Sr$_2$CaCu$_2$O$_{8-\delta }$.\cite{Eschrig00}  The dispersion has a
saddle point at the $M$ point. The $N$ point corresponds to the
node of the $d$-wave order parameter in the superconducting state.
}
\end{figure}

It has been known for some time that near the
$(\pi,0)$ ($M$) point of the zone, the spectral function in the superconducting
state of Bi$_2$Sr$_2$CaCu$_2$O$_{8+\delta}$
shows an anomalous lineshape, the so called `peak-dip-hump'
structure.\cite{Dessau91,Randeria95,Ding96,Norman97}
This structure was also found recently in
YBa$_2$Cu$_3$O$_{7-\delta}$,\cite{Lu01} and in
Bi$_2$Sr$_2$Ca$_2$Cu$_3$O$_{10+\delta}$.\cite{Feng01,Sato01}
For the notation of special points in the Brillouin zone which
we use throughout this paper, see Fig.~\ref{brill}.

Extensive studies on Bi$_2$Sr$_2$CaCu$_2$O$_{8+\delta}$ 
as a function of temperature revealed
that this characteristic shape of the spectral function is closely related to
the superconducting state.
In the normal state, the ARPES spectral function is broadened
strongly in energy, the broadening increasing with underdoping.\cite{Ding96}
The width of the spectral peak quickly decreases with decreasing
temperature below $T_c$,\cite{Norman01} and sharp
quasiparticle peaks were identified
well below $T_c$ along the entire Fermi surface.\cite{Kaminski99}
When lowering the temperature below $T_c$, the coherent quasiparticle
peak grows at the position of the leading edge gap, and the incoherent
spectral weight is redistributed to higher energy, giving rise to a dip and hump
structure.\cite{Dessau91,Randeria95,Norman97} 
This peak-dip-hump structure is most
strongly developed near the $M$-point of the Brillouin zone.
The well defined quasiparticle peaks at low energies 
contrasts to the high energy 
spectra, which show a broad linewidth which grows linearly
in energy.\cite{Valla99,Yusof01} 
This implies that a scattering channel present in the normal
state becomes gapped in the superconducting state.\cite{Kuroda90}
The high energy excitations then stay broadened, since they involve
scattering events above the threshold energy.
While this explains the existence of sharp quasiparticle peaks, a gap in the 
bosonic spectrum which mediates electron interactions leads only to a weak 
diplike feature.\cite{Littlewood92}
This suggests that the dip feature is instead due to the 
interaction of electrons with a sharp (in energy)
bosonic mode. The sharpness implies a strong self energy
effect at an energy equal to the mode energy plus the quasiparticle peak
energy, giving rise to a spectral dip.\cite{Norman98} The fact that
the effects are strongest at the $M$ points implies a mode momentum
close to the $(\pi,\pi)$ wavevector.\cite{Shen97}

More clues are obtained by studying the dispersion of the
related self energy effects.
Recent advances in the momentum resolution of ARPES have led to a
detailed mapping of the spectral lineshape in the high $T_c$ superconductor
Bi${_2}$Sr${_2}$CaCu${_2}$O${_{8+\delta}}$ throughout the Brillouin
zone.\cite{Bogdanov00,Kaminski00}
The data indicated a seemingly unrelated effect
near the d-wave node of the superconducting gap, where the dispersion
shows a characteristic `kink' feature: for binding energies less than the
kink energy,
the spectra exhibit sharp peaks with a weaker dispersion;
beyond this, broad peaks with a stronger 
dispersion.\cite{Kaminski99,Bogdanov00,Kaminski00}
This kink is present at a particular energy all around the Fermi 
surface,\cite{Bogdanov00} and away from the node, the dispersion as determined
from constant energy spectra (momentum distribution curves, MDCs) shows an 
S-like shape in the vicinity of the kink.\cite{Norman01a}
The similarity between the excitation energy where the kink
is observed and the dip energy at $M$, however, suggests that these effects
are related.\cite{Eschrig00} 
Additionally, the observation that the spectral width for binding energies
greater than the kink energy is much broader than that 
for smaller energies\cite{Kaminski99,Bogdanov00,Kaminski00}
is very similar to the difference in the linewidth between the peak and
the hump at the $M$ points.
Further experimental studies supported the idea of a unique energy scale 
involved.\cite{Kaminski00}
They found that away from the node, the
kink in the dispersion as determined from constant momentum spectra (energy
distribution curves, EDCs)
develops into a `break'; the two resulting
branches are separated by an energy gap, and overlap in momentum space.
Towards $M$, the break evolves into a pronounced spectral `dip' separating
the almost dispersionless quasiparticle branch from the weakly dispersing
high energy branch (the `hump').
The kink, break, and dip features all occur at roughly
the same energy, independent of position in the zone,\cite{Kaminski00}
the kink being at a slightly 
smaller energy than the break feature.\cite{Johnson01}

The high energy dispersion is
renormalized up to at least 200 meV and does not extrapolate to the
Fermi surface crossing.\cite{Bogdanov00,Lanzara01} This lets us conclude that 
the 
continuum part of the bosonic spectrum coupling to the fermionic excitations
extends to high energies.

Finally, there is important information contained in the doping dependence
of the self energy effects. 
In underdoped compounds, there is a  pseudogap 
between $T_c$ and $T^{\ast}$;\cite{Marshall96,Ding96}
the pseudogap is maximal near the $M$-point of the Brillouin zone
and is zero at arcs centered at the $N$-points which
increase with temperature.\cite{Norman98n} 
In the pseudogap state above $T_c$, there are low energy renormalizations
in the dispersion, and some trace of the kink feature persists.
But in the recent work by Johnson {\it et al.}\cite{Johnson01}, it
was clearly shown that an additional renormalization of the dispersion
sets in just at $T_c$. This indicates that the bosonic spectrum redistributes
its spectral weight when entering the superconducting state.
The additional low energy renormalization of the dispersion
below the kink energy follows an order parameter like behavior as a
function of temperature.\cite{Johnson01} 
Arguing that the renormalization near the
nodal regions is influenced by the coupling to the same bosonic mode 
which causes the strong self energy effects at the $M$ point of the Brillouin
zone,
the above implies that some mode intensity may be present in the pseudogap
state already, but there is an abrupt increase in the mode intensity when going
from the pseudogap state into the superconducting state, and this
increase shows an order parameter like behavior as a function of
temperature below $T_c$.

The energy of the mode, as inferred from
the energy separation $\Omega_0$ between the peak and the dip,
was shown to decrease with underdoping.\cite{Campuzano99}
Similarly, the kink energy is maximal at optimal doping and decreases both with
underdoping and overdoping,\cite{Johnson01} indicating some relationship
between the kink at the nodal $N$ point 
and the peak-dip-hump structure at the $M$ point.
With underdoping, the sharp quasiparticle peak moves
to higher binding energy, indicating that the gap increases.\cite{Campuzano99} 
At the same time, the spectral weight $z$ of the peak 
drops\cite{Campuzano99,Feng00} leaving the quantity 
$z\Delta_{M}/k_BT_c$ roughly constant.\cite{Ding00} 
Also, the hump moves to higher binding energy and loses
weight with underdoping.\cite{Campuzano99}
This doping evolution of the quasiparticle peak points to an increasing
mode intensity at the $(\pi,\pi)$ wavevector with underdoping.
Again, there is a similarity to the nodal direction:
the low energy renormalization of the dispersion 
below the kink energy increases with underdoping,\cite{Johnson01}
consistent with a common origin of both effects.

\subsection{C-axis tunneling spectroscopy}
Unusual spectral dip features in tunneling data of 
Bi$_2$Sr$_2$CaCu$_2$O$_{8-\delta}$ 
are found in point contact junctions,\cite{Huang89}
in scanning tunneling spectroscopy (STM),\cite{Renner95,DeWilde98}
in break junctions,\cite{Mandrus91,DeWilde98}
and in intrinsic c-axis tunnel junctions.\cite{Yurgens99}
Consistently, these data show a peak feature, 
usually assigned to the maximal $d$-wave superconducting gap,
and a hump feature at higher bias, separated from the peak by a
pronounced dip feature. A characteristic of this dip feature in
SIN junctions is that it occurs asymmetrically around the chemical
potential, usually stronger on the occupied side of the 
spectrum.\cite{Huang89,Renner95,DeWilde98} 
This asymmetry was  
succesfully explained within the theoretical model
presented below.\cite{Eschrig00}
The dip feature has been observed in tunneling spectra of
the single Cu-O$_2$ layer compound Tl$_2$Ba$_2$CuO$_6$ as 
well.\cite{Zasadzinski00}

In order to extract information about the bosonic mode which 
would produce a dip feature in the tunneling conductance, a systematic
study as a function of doping was performed in break junction
tunneling spectroscopy by Zasadzinski {\it et al.}\cite{Zasadzinski01}
There, the doping dependence in Bi$_2$Sr$_2$CaCu$_2$O$_{8-\delta}$ of
the peak-dip-hump structure was determined over a wide range of doping. 
It was found that the dip-peak energy separation,
$\Omega $, follows $T_c$ as $\Omega = 4.9 k_BT_c$.
As expected for an excitonic mode, $\Omega $ approaches
but never exceeds $2\Delta $ in the overdoped region, and $\Omega /\Delta$
monotonically decreases as doping decreases and the superconducting gap
increases. The dip feature is found to be strongest near optimal doping.
Similar shifts of the dip position with overdoping
were reported prevously by STM.\cite{Renner96}
Together with the ARPES results, these studies give a detailed picture
about the doping dependence of the mode energy involved in 
electron interactions in the superconducting state.

\section{Coupling to the magnetic resonance mode}

There have been several theoretical treatments which assigned the anomalous
ARPES lineshape near the $M$ point of the zone
to the coupling between spin fluctuations and
electrons.\cite{Kampf90,Dahm96,Shen97,Norman97,Norman98,Abanov99,Eschrig00}
The `collective mode model' proposed by Norman and Ding\cite{Norman98}
was suggested to
account for the unusual APRES lineshapes by coupling electrons
to a dispersionless collective mode. 
The main motivation for a more detailed study of this model in 
Ref.~\onlinecite{Eschrig00} was to additionally account for the 
dispersion anomalies (`kink'), and the isotropy and robustness of this
characteristic energy scale.\cite{Kaminski00}

The minimal set of
characteristics for the collective mode we are interested in follows from
the experimental results from ARPES and tunneling. The mode is characterized by
its energy and its intensity at the $(\pi,\pi)$ wavevector (the wavevector
being suggested by the momentum dependence of the strength of the ARPES
anomalies).
Its properties from ARPES and SIS tunneling are as follows.  The energy
should be weakly dependent on momentum,
roughly 40 meV in optimally doped
cuprates, follow $T_c$ with doping, and
be constant with increasing temperature up to $T_c$.
The intensity should be maximal at the $(\pi,\pi)$ wavevector, where it
should increase with underdoping and follow an order parameter like
behavior as a function of temperature below $T_{c}$.
The mode should be absent in the normal state; a remnant can be present
in the pseudogap state, but an abrupt increase in intensity should occur
at $T_c$ with lowering temperature.

A sharp resonance with characteristics fitting the
ones extracted from ARPES and tunneling measurements
was observed in inelastic neutron scattering experiments
on bilayer cuprates in the
superconducting state, with an energy near 40 meV in optimally doped
compounds.\cite{Rossat91,Mook93,Fong95,Bourges96,Fong99}
A similar resonance feature at the $(\pi,\pi)$ wavevector 
is observed in underdoped YBa$_2$Cu$_3$O$_{6+x}$, 
but at a reduced energy.\cite{Dai96,Fong97,Bourges97,Dai99}
The resonance was also found in 
Bi${_2}$Sr${_2}$CaCu${_2}$O${_{8+\delta}}$,
both in the optimally doped\cite{Fong99,He01} and overdoped\cite{He01}
regime.  Recently, the resonance was discovered in single layer
Tl$_2$Ba$_2$CuO$_6$ compound as well.\cite{He01a}

To show how well the above criteria fit, we summarize its characteristics:
the resonance is narrow in energy and magnetic in origin.\cite{Mook93}
Its energy width is smaller than the instrumental resolution
(typically less than 10 meV) for optimally and moderately underdoped
materials.  Strongly underdoped materials show a small broadening of the order
of 10 meV.\cite{Bourges98,Fong00} 
The resonance lies below a gapped continuum, 
the latter having a signal typically
a factor of 30 less than the maximum at $\vec{Q}$ at the mode
energy.\cite{Bourges98}
The mode energy decreases with underdoping, and has its
maximal value of about 40 meV at optimal 
doping.\cite{Dai96,Fong97,Bourges97,Dai99}
In both underdoped and overdoped regimes, the resonance energy,
$\Omega_{res}$, is proportional to $T_c$, with $\Omega_{res}\approx
5-5.5 T_c$.\cite{Bourges98,Dai99,Fong99,Fong00,He01}

An additional aspect, specific to bilayer materials, is that
it only occurs in the `odd' channel, which connects the
bonding combination of the bilayer bands to the antibonding 
one.\cite{Rossat93} 
The continuum is gapped in both the even and odd scattering channels
(the even channel is gapped by $\approx $60 meV, 
even in the normal state).\cite{Reznik96}
We will address this issue further below.

The resonance is strongly peaked at the $(\pi,\pi)$ wavevector.
The momentum width of the spin fluctuation spectrum
is minimal at the resonance energy,\cite{Bourges95,Bourges96}
where it is (in contrast to the off-resonant momentum width) only weakly doping 
dependent, with a full width of about 
0.22 \AA$^{-1}$.\cite{Bourges95,Balatsky99,Dai01}
This corresponds to a correlation length $\xi_{sfl}$ of about two lattice 
spacings.

A sharp resonance is not observed above $T_c$.\cite{Fong95,Bourges00}
On approaching $T_c$ from below, the resonance energy does not
shift towards lower energy,\cite{Fong95,Bourges96,Dai96}
but its intensity decreases towards $T_c$, following an order parameter like
behavior.\cite{Rossat91,Mook93,Bourges96,Dai96,Bourges00}
With underdoping, the
intensity at $\vec Q=(\pi,\pi)$ increases from about 1.6 $\mu_B^2$
for YBa$_2$Cu$_3$O$_7$ to about 2.6 $\mu_B^2$ per unit cell volume for 
YBa$_2$Cu$_3$O$_{6.5}$.\cite{Bourges98,Fong00}
There is clearly an abrupt change in resonance intensity at $T_c$, even
in underdoped compounds.

Note that in underdoped materials,
an incommensurate response develops below the resonance
energy,\cite{Dai98} which however
never extends to zero energy, but instead 
the spectrum is limited at low energies
by the so called spin gap $E_{sg}$.\cite{Dai01}
This part of the spectrum behaves differently from
the resonance part as a function of doping.\cite{Balatsky99}
We will neglect this (weaker) incommensurate part of the spectrum in this paper.

The total spectral weight of the resonance is small and amounts to
about 0.06$\mu_B^2$ per formula unit at low temperatures.\cite{Dai99,Fong00}
We will show below that the smallness of the weight of the resonance is
not an obstacle to achieving large self energy effects.

\subsection{Theoretical model}

We are interested in the renormalizations of the
fermionic dispersion due to coupling of electrons to
a sharp spin fluctuation mode at low energies, equal to about
40 meV or less. We will assume that superconducting order is already
established without coupling to this resonant feature in the
spin flucutation spectrum, and thus describe the superconducting
state by an independent order parameter $\Delta_k$. This order parameter
will be chosen to have $d$-wave symmetry (here and in the following the
unit of length is the lattice constant $a$),
\begin{equation}
\Delta_k=\Delta_M (\cos k_x-\cos k_y)/2
\end{equation}
which takes its maximal value $\Delta_M$ at the $M$ point in the 
Brillouin zone. The magnitude $\Delta_M$ will be chosen
so that the calculated peak in the ARPES spectrum at the $M$ point,
after including self energy effects due to coupling to the
spin fluctuations, fits the position of the spectral peak in experimental
ARPES spectra. We stress that we do not specify the origin of
the pairing interaction responsible for the order parameter $\Delta_k$,
but the continuum part of the spin fluctuations is one of the candidates.
We also underline that, as our results will show, the
spin fluctuation resonance supports pairing, but does not cause
superconductivity in and of itself.

In the model we employ, the retarded Green's functions, $G^R_{\epsilon,k}$, 
for fermionic excitations in the superconducting state is a functional of the 
normal state electronic dispersion $\xi_k$, the order parameter $\Delta_k$, and
the self energies due to coupling to spin fluctuations, 
$\Sigma^R_{\epsilon,k},\Phi^R_{\epsilon,k}$. The term `normal state' here
refers to the state at the same temperature, but with zero order parameter.
We employ a six parameter tight binding fit for this dispersion, having the
form
\begin{eqnarray}
&&\xi_k=t_0+t_1\frac{\cos k_x+\cos k_y}{2}+t_2\cos k_x\cos k_y  \nonumber \\
&&+t_3\frac{\cos 2k_x + \cos 2k_y}{2}
+t_4\frac{\cos 2k_x \cos k_y+ \cos k_x \cos 2k_y}{2} \nonumber \\
&&+t_5 \cos 2k_x \cos 2k_y
\end{eqnarray}
Any set of six independent parameters for the dispersion
determines the parameters $t_0-t_5$. 
The six parameters we use are
the positions of the $N$ (node) and $A$ (antinode) points in Fig.~\ref{brill}, 
parameterized by
$k_{\Gamma N}=|\vec{k}_N-\vec{k}_{\Gamma }|$ 
and $k_{MA}=|\vec{k}_A-\vec{k}_M|$, the
band energies at the $M$ and $Y$ points, $\xi_M$ and $\xi_Y$, the
Fermi velocity at the $N$ point, $v_N=|\vec{v}_N|$, and the
inverse effective mass along direction 
$M-\Gamma$ at the $M$ point, $m_{M}^{-1}$. 
Table \ref{tab1} summarizes
our choices. 
For reference, the corresponding $t_i$ are (eV): $t_0= 0.0989$,
$t_1=-0.5908$, $t_2=0.0962$, $t_3=-0.1306$, $t_4=-0.0507$, and
$t_5=0.0939$.

The parameter $\xi_Y$ is not known from experiment. We set it to a reasonable
value to preserve a dispersion shape similar to that obtained from band theory.
The inverse mass at the $M$ point is known to be negative and
small in the $M-\Gamma$ direction, and it was suggested that it could be
zero, giving rise to an extended van Hove singularity.\cite{Gofron94}
Here we 
chose a finite, moderately small value. As we show, the inverse effective
mass will decrease when coupling to the spin fluctuation mode is taken
into account, and it is this renormalized inverse mass which is
experimentally observed. 
Similarly, the value of the Fermi velocity at the node is chosen
somewhat larger than the experimental value, since again, one
observes the fully renormalized velocity; in our calculation, self energy
effects renormalize this value to the moderately smaller 
value observed in experiment.
When the doping level is varied, the band filling varies
($t_0$ changes), so
that the van Hove singularity at the $M$ point, $\xi_M$, will
move relative to the chemical potential.  Also, the Fermi crossing $\vec{k}_A$ 
moves along the $M-Y$ line. All other band structure parameters are expected to 
be rather insensitive to the doping level.

\begin{table}
\begin{center}
\caption{\label{tab1} Parameters for the effective dispersion $\xi_k$.
}
\begin{tabular}{|c|c|c|c|c|c|}
\hline
\hline
$k_{\Gamma N}a$ & $k_{MA}a$ & $\xi_M$ & $\xi_Y$ & $\hbar v_N/a$ & 
$\hbar^2/m_{M}a^2$ \\
\hline
0.36$\sqrt{2}\pi$ & 0.18$\pi$ & $-34$ meV & 0.8 eV & 0.6 eV & $-0.2376$ eV
\\
\hline
\hline
\end{tabular}
\end{center}
\end{table}

The `normal state' dispersion $\xi_k$, 
and the order parameter $\Delta_k$, are phenomenological
quantities, which are already renormalized by other effects which we do 
not need to specify, but which are assumed to influence the physics only
on an energy scale large compared to the scale of interest in this paper
(50-100 meV). The self energies due to spin fluctuations will have 
a part due to the particle-hole continuum, and another part due to the
resonance. We will consider two models, a simple
form and an extended form. In
the simple form, we include the effect of the continuum part of the
spin fluctuation spectrum by a constant renormalization of the normal
state dispersion and the order parameter. This model will capture
the main physics for energies below 100 meV,  which is dominated by 
the coupling of the electrons to the resonant spin fluctuations.
The reason is the following: as we will show below,
in this energy range the imaginary part of the self energies due to the
continuum part of the spin fluctuations is zero, and the real part (divided
by $\epsilon$)
only varies weakly
both in energy and momentum. 
This allows to approximate it by a real constant
in that energy range, and thus include it into the renormalization of
$\xi_k$ and $\Delta_k$. For this case, the `normal state' reference is
defined as the state with zero order parameter, interacting with 
a spin flucutation
spectrum having no resonance part and a continuum part identical to
that in the superconducting state. The real, physical normal state will 
be different because the spin fluctuation continuum changes when going
from the normal to the superconducting state, leading to an additional
renormalization of the dispersion. 
Thus, in the simple form of the model, the 
low energy 
dispersion which enters the calculations will be
approximately proportional to the true normal state dispersion, but
the proportionality factor will not be unity.

At higher energies, the spin fluctuation continuum can be excited, and
this leads to an additional strong fermionic damping.
We will study this effect in an extended model which explicitly includes
the gapped spin fluctuation continuum.
For this extended model, the `normal state' dispersion will
have a different renormalization factor as compared to the simple
model above.
Specifically, we use for the extended model the above dispersion
scaled by the factor 1.5 and shifted back in energy, so that $\xi_M$ stays at
its original value of $-34 $meV.

\begin{figure}
\centerline{
\epsfxsize=0.24\textwidth{\epsfbox{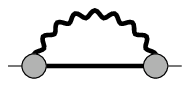}}
\epsfxsize=0.24\textwidth{\epsfbox{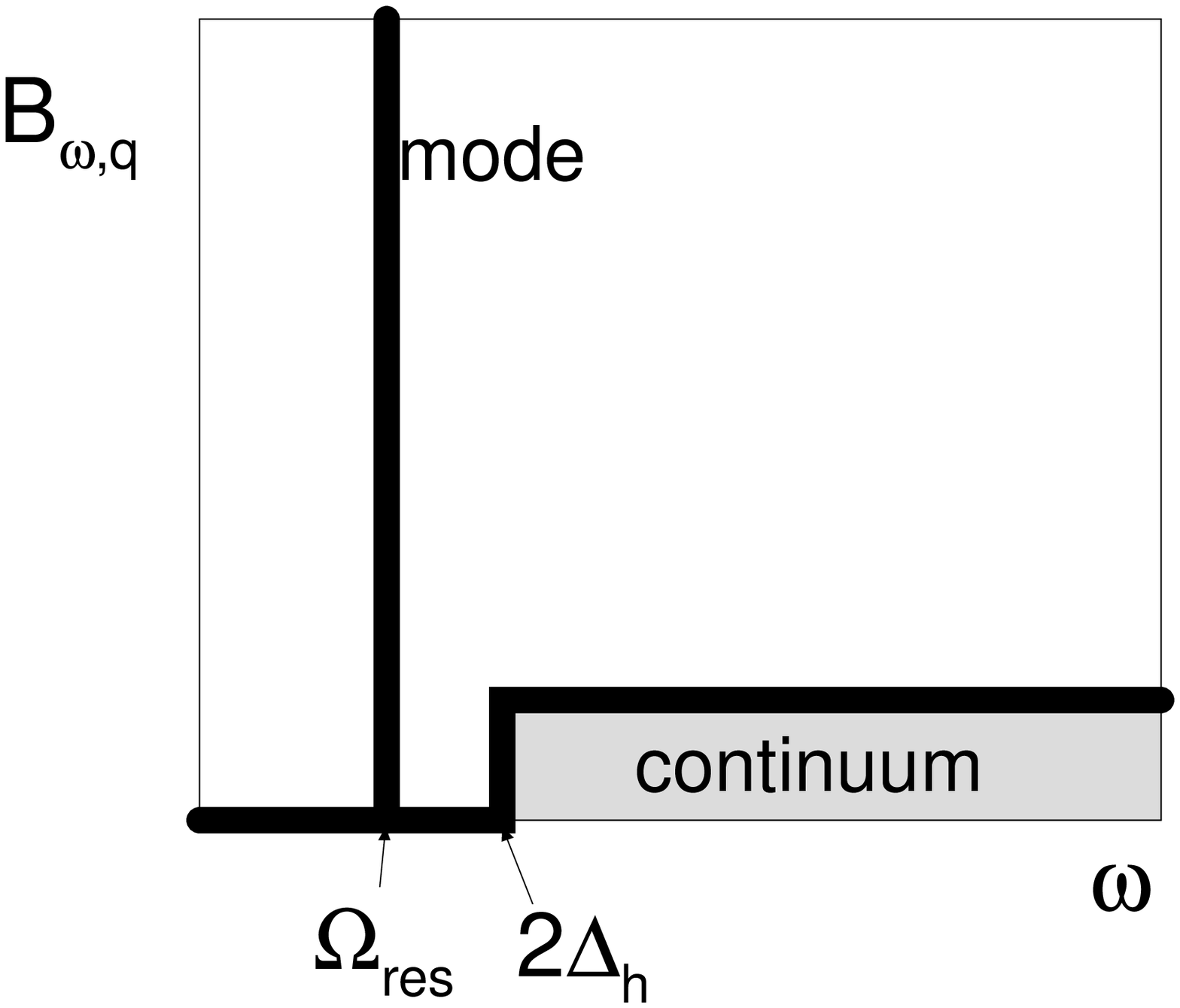}}
}
\caption{
\label{diagram}
Left: Self energy for electrons (full lines). 
The wavy line denotes a spin fluctuation.
Right: the model spin fluctuation spectrum we used for the wavy line
in the Feynman diagram.
The mode affects the low energy fermionic properties.
The continuum part only couples to electrons with higher energies, and
is neglected in the simple form of the model.\cite{Eschrig00} 
Damping of electrons at energies above 100 meV is caused by the
continuum part, and is included in the extended model, which we also discuss
in this paper.
}
\end{figure}

We find that all essential features of the self energy effects
in the superconducting state are obtained using
a minimal model with a spin fluctuation spectrum shown in
Fig.~\ref{diagram}.

The continuum formally has to be cut-off at high energies. This cut-off
only affects the real part of the self energy, and variation of the
cut-off leads to only a weakly energy dependent contribution to the 
renormalization factor 
which can be absorbed in the dispersion $\xi_k$ as described above.
We discuss the choice of this cut-off later.

The retarded Green's function in spectral representation is given
as a function of the self energies as,
\begin{eqnarray}
\label{green}
G^R_{\epsilon,k} [\Sigma^R_{\epsilon,k},\Phi^R_{\epsilon,k}] &=& 
\sum_{\nu=\pm} \frac{A^{\nu}_{\epsilon,k}}{\epsilon-E^{\nu}_{\epsilon,k}+
i\delta } \\
F^R_{\epsilon,k} [\Sigma^R_{\epsilon,k},\Phi^R_{\epsilon,k}]&=& 
\sum_{\nu=\pm} \frac{C^{\nu}_{\epsilon,k}}{\epsilon-E^{\nu}_{\epsilon,k}+
i\delta }
\end{eqnarray}
with excitation energies $E^{\nu}_{\epsilon,k}$ and
coherence factors $A^{\nu}_{\epsilon,k}, C^{\nu}_{\epsilon,k}$,
\begin{eqnarray}
E_{\epsilon,k}^{\pm} &=& \pm \sqrt{\bar \xi_{\epsilon,k}^2+|\bar 
\Delta_{\epsilon,k}|^2} + \delta \Sigma_{\epsilon,k} \\
A_{\epsilon,k}^{\pm} &=& \frac{1}{2}\pm
\frac{\bar \xi_{\epsilon,k}}{2\sqrt{\bar \xi_{\epsilon,k}^2+|\bar 
\Delta_{\epsilon,k}|^2}} \\
C_{\epsilon,k}^{\pm} &= & \quad \pm
\frac{\bar \Delta_{\epsilon,k}}{2\sqrt{\bar \xi_{\epsilon,k}^2+|\bar 
\Delta_{\epsilon,k}|^2}}
\end{eqnarray}
The renormalized dispersion and gap function are given in terms of
the diagonal 
($\Sigma^R_{\epsilon,k}$) and off-diagonal ($\Phi^R_{\epsilon,k} $)
in particle-hole space self energies, as
\begin{eqnarray}
\bar\xi_{\epsilon,k}&=&\xi_k + 
\frac{\Sigma^R_{\epsilon,k}+\Sigma^R_{-\epsilon,-k}}{2} \\
\bar \Delta_{\epsilon,k} &= & \Delta_k+
\frac{\Phi^R_{\epsilon,k} +\Phi^{R\ast}_{-\epsilon,-k}}{2} \\
\label{dSig}
\delta \Sigma_{\epsilon,k}&=&
\frac{\Sigma^R_{\epsilon,k}-\Sigma^R_{-\epsilon,-k}}{2}
\end{eqnarray}
We will couple electrons to the spin fluctuation spectrum with
a coupling constant $g$, which we assume to be independent of
energy and momentum.
The self energies for our model are then given in terms of the spectral function
of the spin fluctuations with energy $\omega $ and momentum $\vec{q}$, 
$B_{\omega,q}$, by the expressions (we chose a representation especially
well suited for numerical studies, see App.~\ref{app1})
\begin{eqnarray}
\label{self1}
\Sigma^R_{\epsilon,k}&=&\sum_{\omega,q} 
\rho^T_{\omega,\epsilon-\omega}
g^2B_{\omega,q}
G^R_{\epsilon-\omega,k-q} 
\nonumber \\
&&- T \sum_{\epsilon_n,q} G^M_{k-q}(i\epsilon_n)g^2D^M_q(\epsilon-i\epsilon_n)
\end{eqnarray}
\begin{eqnarray}
\label{self2}
\Phi^R_{\epsilon,k}&=&\sum_{\omega,q} 
\rho^T_{\omega,\epsilon-\omega}
g^2B_{\omega,q}
F^R_{\epsilon-\omega,k-q} 
\nonumber \\
&&- T \sum_{\epsilon_n,q} F^M_{k-q}(i\epsilon_n)g^2D^M_q(\epsilon-i\epsilon_n)
\end{eqnarray}
where $G^M$ and $D^M$ are the fermionic and bosonic
Matsubara Green's functions which are easily expressed in terms
of the spectral functions 
$A^\nu_{\epsilon ,k} $ and $B_{\omega ,q}$ respectively.
The Matsubara sums in the second lines of Eqs.~\ref{self1} and \ref{self2}
only contribute to the real part of the self energies.
The population factor $\rho^T_{\omega,\epsilon-\omega}$ is given in terms
of Bose ($b$) and Fermi ($f$) population functions as,
\begin{eqnarray}
\rho^T_{\omega,\epsilon-\omega}=
b_{\omega} + f_{\omega-\epsilon} = -b_{-\omega}-f_{\epsilon-\omega}
\end{eqnarray}
We solved these equations numerically using bare Green's functions
$G^R_{\epsilon, k}[0,0]$, $F^R_{\epsilon, k}[0,0]$ for calculating
the self energies $\Sigma^R_{\epsilon, k}$ and $\Phi^R_{\epsilon, k}$.
We show later that feedback effects
give no significant changes within our model.

Although we solve the equations above numerically without further
approximations, some general remarks are in order.
The function $\rho^T_{\omega,\epsilon-\omega}$ as a function
of $\omega $ is at zero temperature
nonzero only between $\omega =0$ and $\omega = \epsilon $, and
is equal to 
sign$(\epsilon )$ 
in this range.  Because
the spin fluctuation spectrum is gapped by much more than the thermal energy
in the superconducting state, we can put
for all practical reasons $b_{\Omega_{res}}=0$.
That means that we can neglect thermally excited modes, and only
allow for emission processes at the resonant mode energy.
For any gapped spin fluctuation spectrum with gap 
$\Omega$, the first terms
in Eq.~\ref{self1} and \ref{self2} are negligible in the range
$-\Omega< \epsilon < \Omega$ (apart from temperature smearing near the
value $\pm \Omega$). Thus, assuming that the spin fluctuation spectrum
is gapped below the resonance energy,
at zero temperature scattering of 
electronic excitations is disallowed in the
interval $-\Omega_{res}< \epsilon < \Omega_{res}$. This is an expression of
the fact that at least an energy $\Omega_{res}$ must be spent in order
to emit one spin fluctuation mode. This is the case for optimally
and overdoped cuprates.
For strongly underdoped cuprates, scattering is disallowed
only in the range 
$-E_g < \epsilon < E_g$, where $E_g$ is the spin gap which is smaller
than $\Omega_{res}$.
Also, as an implication, the renormalization function, determined by the
real part of the self energy, is given in the low energy range by the
second terms of Eqs.~\ref{self1} and \ref{self2} only.
In the following, we first consider the simple form of the model,
which uses only the mode part of the spin fluctuation spectrum.
After having gained some insight about the features caused by the
resonance mode, we study the extended model which includes the
continuum part as well.

\subsection{Contribution from the spin fluctuation mode}
For a sharp bosonic mode the spectral function is given by,
\begin{equation}
g^2B_{\omega,q} = 2 g^2w_q \left( \delta (\omega- \Omega_{res} ) -
\delta (\omega + \Omega_{res}) \right)
\end{equation}
where $w_q$ is the energy integrated weight of the spin fluctuation mode,
which is assumed to
be enhanced at the $\vec{Q}=(\pi,\pi)$ point. Using the correlation length 
$\xi_{sfl}$, we write it as
\begin{equation}
w_q=\frac{w_Q}
{1+4\xi_{sfl}^2 (\cos^2\frac{q_x}{2}+\cos^2\frac{q_y}{2})}
\end{equation}
We will show below that it is a good approximation to assume the
mode as perfectly sharp in energy, as corrections due to the finite energy
width of the mode are negligible.
From neutron scattering data obtained on
Bi${_2}$Sr${_2}$CaCu${_2}$O${_{8+\delta}}$,
the energy integrated weight of the resonance mode was determined 
as 1.9 $\mu_B^2$,\cite{Fong99}
leading (after dividing out the matrix element $2\mu_B^2$) to $w_Q=0.95$.
We fit ARPES data near optimal doping,\cite{Eschrig00}
giving $g^2w_Q=0.4 $eV$^2$. This implies that the coupling constant
is equal to $g=0.65$ eV. This is a reliable value as discussed
in Ref.~\onlinecite{Abanov01}.
In Table \ref{tab2}, we present our minimal parameter set entering the
model (we only include the parameter $\xi_M$ from the band structure
tight binding fit, as the results are insensitive to reasonable
variations of the other parameters), together with the values we used for
optimally doped compounds.

\begin{table}
\begin{center}
\caption{\label{tab2} Minimal parameter set used in the calculations.}
\begin{tabular}{|c|c|c|c|c|}
\hline
\hline
$\Delta_{M}$ & $\Omega_{res}$ & $\xi_M$ & $\xi_{sfl}$ 
& $g^2w_Q$ \\
\hline
35 meV & 39 meV & $-34$ meV & 2a  & 0.4 eV$^2$
\\
\hline
\hline
\end{tabular}
\end{center}
\end{table}

\subsubsection{Electron Scattering}

\begin{figure*}
\centerline{
\epsfxsize=0.95\textwidth{\epsfbox{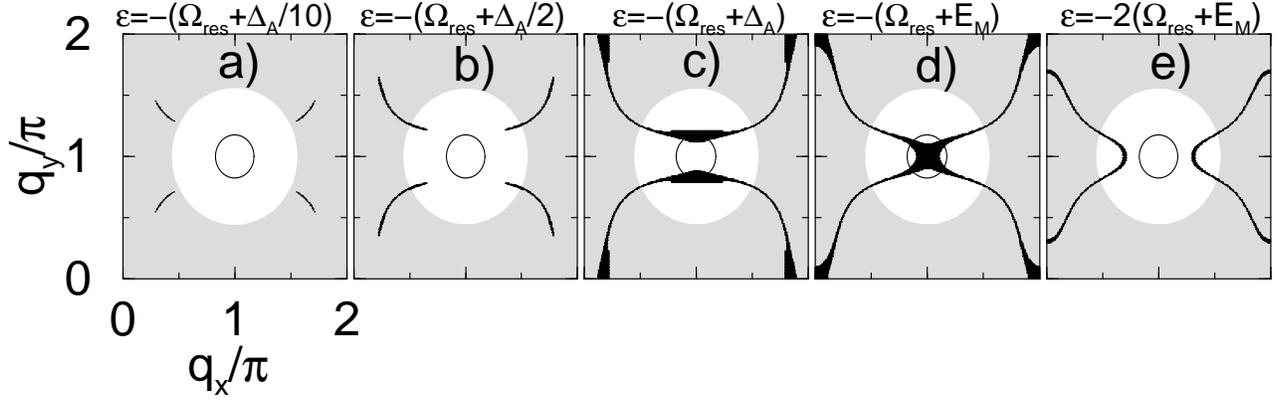}}
}
\caption{
\label{schematic1}
The black regions denote the part of the first Brillouin zone
of the spin fluctuation momentum $\vec{q}$ which participates
in scattering of electrons with momentum $\vec{k}_M$ and 
energy $\epsilon $, as indicated above each picture.
The amount of scattering events is controlled by the form factor
for the resonance mode $w_q$,
which takes its maximal value ($w_Q$ )
in the center of the Brillouin zone at $\vec{Q}=(\pi,\pi) $.
Inside the black circles $w_q>w_Q/2$, and inside the white region
$w_q>w_Q/10$.
For small energies, a), only nodal electrons are scattered.
For energies equal to $\Omega_{res}+E_M$, d), a large region
around the $M$-point of the fermionic zone
participates in scattering events. Scattering electrons with
this energy and momentum involves spin fluctuations with maximal weight,
and thus almost exhausts the entire weight of the mode part of
the spin fluctuation spectrum.
Pictures c) and d) correspond to the special energies $\Omega_{res}+\Delta_A$
and $\Omega_{res}+E_M$, leading to cusp features in the energy 
dependence of the imaginary part of the self energy.
}
\end{figure*}

We first discuss phase space restrictions for electron scattering
in the $d$-wave superconducting state, and how they relate to the issue
of whether the small relative weight of the resonance part of the spin 
fluctuation spectrum leads to comparable in magnitude effects as
in strong coupling superconductors.
Using bare Green's functions, the  
self energy at zero temperature can be written as
\begin{eqnarray}
\label{self11}
\mbox{Im} \Sigma^R_{\epsilon, k}&=&
-\sum_{q}  g^2w_q A^{-}_{k-q} 
\delta (\epsilon+\Omega_{res}+E_{k-q}) 
\nonumber \\
&&-\sum_{q}  g^2w_q A^{+}_{k-q} 
\delta (\epsilon-\Omega_{res}-E_{k-q})
\\
\label{self12}
\mbox{Re} \Sigma^R_{\epsilon, k}&=&
-\sum_{q}  \frac{g^2w_q}{\pi} \cdot
\frac{\epsilon + \left( 1+\frac{\Omega_{res}}{E_{k-q}}\right)\xi_{k-q}}{
(\Omega_{res}+E_{k-q})^2 - \epsilon^2}
\end{eqnarray}
where $E_k=\sqrt{\xi_k^2+|\Delta_k|^2}$, $A^{\pm}_k=(1\pm\xi_k/E_k)/2$.
The sum over $\vec{q}$ extends over the first Brillouin zone for the
spin fluctuation momentum. For negative energies, only the first
sum in Eq.~\ref{self11} is nonzero. The sum is a weighted average
of the expression $A^{-}_{k-q}  \delta (\epsilon+\Omega_{res}+E_{k-q})$
with weight factors $w_q$. For given fermion energies, $\epsilon $, and
momenta, $\vec{k}$, the delta function restricts the
allowed spin fluctuation momenta $\vec{q}$.
Similar zero temperature formulas hold for the off diagonal self energy,
\begin{eqnarray}
\label{self21}
\mbox{Im} \Phi^R_{\epsilon, k}&=&
-\sum_{q}  g^2w_q C_{k-q} 
\Big[ \delta (\epsilon-\Omega_{res}-E_{k-q}) 
\nonumber \\
&&\qquad \qquad - \delta (\epsilon+\Omega_{res}+E_{k-q})  \Big]
\\
\label{self22}
\mbox{Re} \Phi^R_{\epsilon, k}&=&
-\sum_{q}  \frac{g^2w_q}{\pi} \cdot
\frac{\left( 1+\frac{\Omega_{res}}{E_{k-q}}\right)\Delta_{k-q}}{(\Omega_{res}
+E_{k-q})^2-\epsilon^2} 
\end{eqnarray}
with $C_k=\Delta_k/2E_k$.

In Fig.~\ref{schematic1}, we plot for $\vec{k}=\vec{k}_M$ and for
several energies these restricted regions in $\vec{q}$-space.
The corresponding weights for these regions, given by $w_q$, are
maximal at $\vec{q}=\vec{Q}$
($q_x=q_y=\pi$), and decay away from that momentum.
For reference, we define the regions inside the black circle, where
$w_q>w_Q/2$, and the white regions, where $w_q> w_Q/10$.
The calculations were done for finite $T=40 $K, and with a broadening
parameter $\delta = 5$ meV in Eq.~\ref{green}.

For energies $-\Omega_{res} (=-39 \mbox{meV})< \epsilon < 0$, there is no phase
space available for scattering. Scattering of electrons by
the spin fluctuation mode sets in for
$\epsilon = -\Omega_{res}$ at $\vec{q}$ corresponding
to the wavevectors $\vec{q}=(\vec{k}_M-\vec{k}_N)$ mod $(\vec{G})$,
connecting the $M$ point to the nodes
($\vec{G} $ denotes a reciprocal lattice vector).
In picture a) of Fig.~\ref{schematic1},
we show for $\epsilon = -(\Omega_{res}+\frac{1}{10}\Delta_A) $ the mode
wavevectors involved in scattering events. The weight for such
events is very small, as can be seen from the fact that these wavevectors
are outside the white region. Going further away from the chemical
potential with $\epsilon $, 
the allowed mode wavevector regions increase, as shown in
picture b) for $\epsilon = -(\Omega_{res}+\frac{1}{2}\Delta_A) $.
When the special point $\epsilon = -(\Omega_{res}+\Delta_A)$ is reached
($=-71.2$ meV in our case),
the arcs of $\vec{q}$-regions involved in scattering events close at
the points $\vec{q}=(\vec{k}_M-\vec{k}_A)$ mod $(\vec{G})$, as
shown in picture c), and electrons are scattered strongly between
the $M$ point and the $A$ points.
This leads to a cusp (or peak for very small quasiparticle broadening)
in the energy dependence of the imaginary part
of the self energy at this energy. Going further in energy, another
special point is reached at $\epsilon = -(\Omega_{res}+E_M)$
(with $E_M=\sqrt{\xi_M^2+\Delta_M^2}$), at which
scattering events between the $M$ points involving spin fluctuations
with momentum $\vec{q}=\vec{Q}$ (and with $\vec{q}=\vec{0}$) are
allowed. We show the corresponding regions in $\vec{q}$-space
in picture d).
This picture is important for understanding the
large effect we obtain. First, the weight factor $w_q$ is large
in the patches of phase space for allowed scattering events around
$\vec{Q}$. Furthermore, because of the van Hove
singularity in the band dispersion, these patches have a large area, almost
filling the area inside the black circles in Fig.~\ref{schematic1}.
This has as consequence that 
a large part of the weight of the resonance is
exhausted for scattering electrons with energies equal to
$\epsilon=-(\Omega_{res}+E_M)$, which amounts to $-87.8 $ meV for our parameter
set. 
Going even further in energy, as shown in
picture e), the amount of scattering events quickly decreases.
The area which is involved in electron scattering events is
maximal for energies between 70 meV and 90 meV. For these energies,
the involved spin fluctuations are also near the $\vec{q}$-region where 
almost all their weight is concentrated. Thus, the strongest
renormalization effects will take place in the energy range
70-90 meV.

Let us compare this discussion with the case for conventional
isotropic electron-phonon coupling. In this case, the weight
factors $w_q$ are constant. The relative amount of phonon
wavevectors involved in scattering events is then equal to
the ratio between the black areas shown in Fig.~\ref{schematic1}
and the total area of the Brillouin zone. This ratio is for
the maximal case, picture d), equal to 5\%. That means that only
5\% of the total phonon weight contributes to the imaginary
part of the self energy. It is well known that electron phonon
coupling easily leads to renormalization factors of the order of 2.
In our case, the spin fluctuation weight of the mode is only about
5\% of the total spin fluctuation weight, but it is concentrated
in the region inside the black circles in Fig.~\ref{schematic1}.
Almost the total area inside the black circle contributes in the case of
picture d), showing that the same amount of only a few percent
of the bosonic spectrum is involved as well for spin fluctuations in 
high $T_c$ cuprates as for
phonons in conventional strong coupling superconductors.
Thus, the renormalization of the fermionic dispersion is
expected to be of the same order of magnitude, and our explicit
calculations confirm this.

\begin{figure}
\centerline{
\epsfxsize=0.25\textwidth{\epsfbox{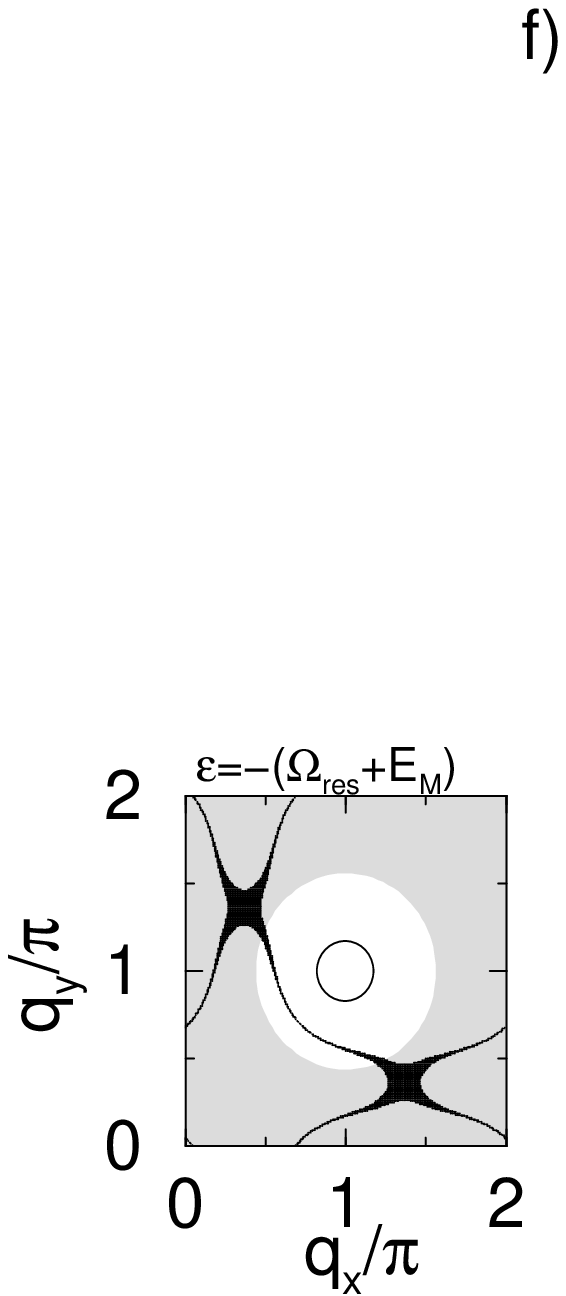}}
}
\caption{
\label{schematic}
The same as Fig.~\ref{schematic1} d) for a fermionic wavevector
at the nodal point, $\vec{k}=\vec{k}_N$. Because the allowed region
for scattering events is outside the region of enhanced spin fluctuations,
the corresponding cusp feature in the imaginary part of the 
self energy is weaker than for electrons
with momenta near the $M$-point.
}
\end{figure}

In Fig.~\ref{schematic}, we show the $\vec{q}$-space areas corresponding
to Fig.~\ref{schematic1} d), but for electrons near the nodal wavevector. 
As can be seen, the feature due to the van Hove singularity region
is now weighted by a smaller value of $w_q$.
Because of this, for nodal electrons, the corresponding peak
in the self energy is smaller than for momenta near the $M$ point.
It turns out that for the nodal electrons, the feature
at $- (\Omega_{res}+\Delta_{A})$ is more pronounced than
that at $- (\Omega_{res}+E_M)$.

\subsubsection{Renormalization factor and electron lifetime}

The self energy has a characteristic shape as a function of energy,
which is conserved qualitatively for all points in the Brillouin zone.
This is a consequence of the fact that all points are coupled via
the spin fluctuation mode, which has
a finite width in momentum, to
all special points in the Brillouin zone with their corresponding
characteristic energies. These special points are the nodal $N$ points, and
the van Hove singularities at the $M$ points and the $A$ points (the latter
is a dispersion maximum in the superconducting state).
Because the general shape of the energy dependence of the self energy
does not vary much with momentum (although the overall intensity does),
it is sufficient to discuss the important features in 
the energy dependence of the self energy at the $M$ point.

\begin{figure}
\centerline{
\epsfxsize=0.45\textwidth{\epsfbox{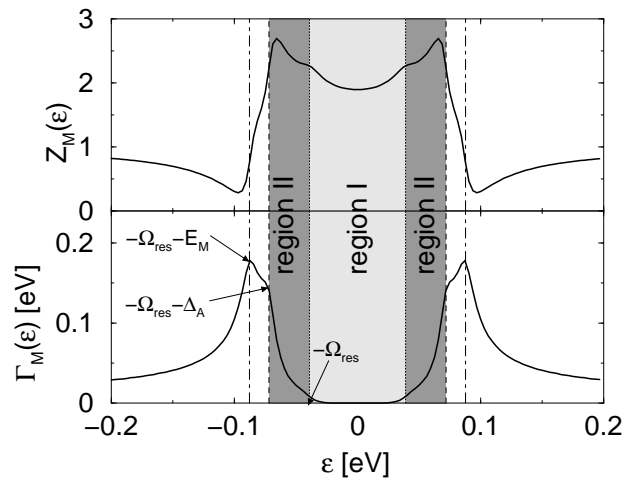}}
}
\caption{
\label{Zfac}
Renormalization factor at the $M$ point, $Z_M(\epsilon)$ (top) and
electron scattering rate at the $M$ point,
$\Gamma_M(\epsilon )$ (bottom). The thin lines denote 
some characteristic energies: $\pm \Omega_{res}$ 
(dotted), $\pm (\Omega_{res}+\Delta_{A})$ (dashed),
and $\pm (\Omega_{res}+E_M)$ (dot-dashed).
Electrons at low temperatures are scattered only if their
energy is larger than $\Omega_{res}$, so that they are able to
emit a collective mode excitation. 
The parameters used are: $T=40$K, $\Omega_{res}=$39 meV, $\Delta_{M}=$35 meV.
}
\end{figure}

We numerically evaluated the self energy, using a
broadening parameter $\delta = 5$ meV.
In Fig.~\ref{Zfac}, we show the results for the renormalization function and
electron scattering rate at the $M$ point,
\begin{equation}
Z_M(\epsilon ) = 1 - \frac{\mbox{Re} \, \delta \Sigma_M (\epsilon )}{\epsilon }
\qquad
\Gamma_M(\epsilon ) = -\mbox{Im} \, \delta \Sigma_M (\epsilon )
\end{equation}
as a function of energy.
 
There are three characteristic energies (in addition to temperature,
which smears all features by $k_B T$). Region I is bounded by
the resonance energy, $\Omega_{res}$, and has zero scattering
rate at zero temperature (this statement is true
for electrons at any point in the Brillouin zone).
At finite temperature, a region $k_B T$
around $\pm \Omega_{res}$ allows for a small amount of 
scattering, even in region I. Because states are
occupied near the $M$ point, we will only
discuss negative energies in the following.
At $\epsilon=-\Omega_{res}$, scattering for
all electrons in the Brillouin zone sets in due to coupling
to nodal electrons via emission of a spin fluctuation mode.
Absorption processes are
negligible due to the large (compared to temperature) mode energy.
In region II, a larger and larger area around the nodes participates
in scattering events, 
(as can be seen from pictures a and b in Fig.~\ref{schematic1}), 
until finally the point at the zone boundary with
maximal gap, $\pm \Delta_{A}$, is reached (picture c in Fig.~\ref{schematic1}).
This point corresponds 
in Fig.~\ref{Zfac} 
to a cusp feature in the imaginary part of the self energy at
$-(\Omega_{res} + \Delta_{A})$.
The third feature,
at $-(\Omega_{res} + E_M)$,
corresponds to the van Hove singularity at the $M$ point of the
Brillouin zone, which is close to the chemical potential in cuprates
(picture d in Fig.~\ref{schematic1}).
The proximity of this van Hove singularity leads to a stronger peaked
feature in the scattering rate near $\pm(\Omega_{res} + \Delta_{A})$
compared to the case where this van Hove singularity at the $M$ point is
absent. The renormalization factor is rather constant in region I
as a consequence of its connection to the imaginary part via Kramers-Kronig
relations. The enhancement in regions I and II compared to unity comes from two
step features at $\pm (\Omega_{res} + E_M)$ and at 
$\pm (\Omega_{res} + \Delta_{A})$. Note that the step feature due
to the van Hove singularity at the $M$ point contributes about 50\%
to the total enhancement. The small features at $\pm \Omega_{res}$
are due to the finite lifetime
of the electrons involved in scattering processes 
as discussed below. 
The onset
of scattering at the emission edge for the spin fluctuation mode 
occurs as a jump if the electrons involved have a finite 
spectral width.
At even higher energies, corresponding to Fig.~\ref{schematic1} e), 
the scattering due to the spin fluctuation mode becomes
less effective.
Note that the spectral peak of the electrons at $\pm \Delta_k$
is either in region I or in region II. Thus, quasiparticles near the
nodal regions are always sharper in energy then quasiparticles near the
maximal gap regions. In overdoped cuprates, the maximal gap is
usually smaller than the mode energy, so that for the broadening of the
quasiparticle peaks, the spin fluctuation mode is not relevant.

For the following discussion, it is useful to derive approximate analytical 
expressions. At zero temperature, using Eq.~\ref{self12},
we obtain 
\begin{equation}
Z_M(\epsilon ) =
1+\sum_{q}  \frac{g^2w_q}{\pi}
\frac{1}{(\Omega_{res}+E_{k_M-q})^2 - \epsilon^2}
\end{equation}
The main contribution comes from the regions where
$E_{k_M-q}$ is less than 100 meV. We can estimate those regions
by the requirement that $\vec{k}_M-\vec{q}$ is in the area 
around the $M$ points deliminated by $\pm0.35\pi$ in $M-Y$ direction
and by about 0.3$\pi$ along the $M-\Gamma $ direction.
Then, replacing $\Delta_{k_M-q}$ by $-\Delta_M$, and $E_{k_M-q}$ by $E_M$,
we perform the $\vec{q}$-sum over that area of the function $w_q$.
We denote $\sum_q w_q$ over this area by $I_0$. 
For our model we have
$I_0=0.035 $.
Using this approximation, we obtain
\begin{equation}
\label{ZM}
Z_M(\epsilon ) \approx
1 + \frac{g^2I_0}{\pi} 
\frac{1}{(\Omega_{res}+E_M)^2-\epsilon^2}+\lambda_{M}^{(N)}(\epsilon)
\end{equation}
Here, $\lambda_{M}^{(N)}(\epsilon )$ denotes the contributions coming from
the regions where $\vec{k}'=\vec{k}_M - \vec{q}$ is outside of the above range.
It is dominated by contributions where $\vec{k}'$ is near the nodal
regions of the Brillouin zone, thus the relevant spin fluctuation 
momentum is $\vec{q}=(\vec{k}_M-\vec{k}_N)$ mod $\vec{G}$.
The contribution $\lambda_{M}^{(N)}$ is smaller than the first term
in Eq.~\ref{ZM}, but not negligible.
Because Eq.~\ref{ZM} neglects the dispersion
between $(\Omega_{res}+\Delta_A)$ and $(\Omega_{res}+E_M)$
near the $M$ point, it should be used for energies not too close
to the region between these two values. We will make use of this
formula below for energies near $\epsilon = E_M$, where this 
formula gives a good approximation.

\begin{figure}
\centerline{
\epsfxsize=0.45\textwidth{\epsfbox{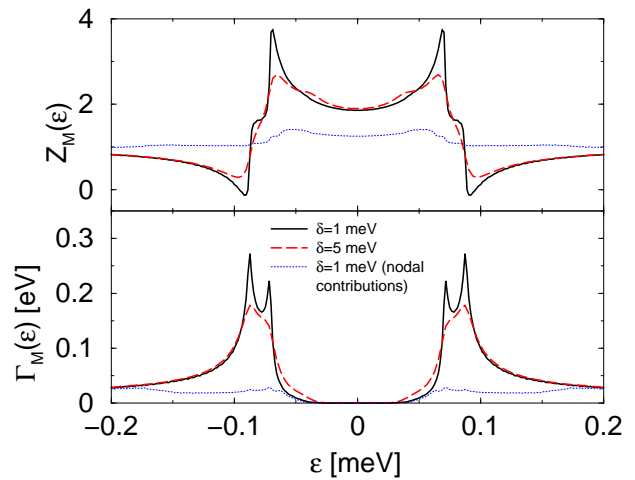}}
}
\caption{
\label{gamma}
Renormalization factor at the $M$ point, $Z_M(\epsilon)$ (top) and
electron scattering rate at the $M$ point, $\Gamma_M(\epsilon )$ (bottom). 
The picture compares results for two different
residual quasiparticle linewidths: $\delta= 1$ meV (full lines) and
$\delta =5 $ meV (dashed lines). 
As dotted lines the nodal contributions,
when restricting the quasiparticle momenta to the regions outside the
area around  the $M$ points discussed in the text, is shown
for $\delta= 1$ meV.
}
\end{figure}

\subsubsection{The quasiparticle scattering rate}

For overdoped materials the quasiparticle peak at the $M$ point is
situated below the onset of scattering due to emission of spin fluctuations.
In this case the width is determined by other processes, and we model
this residual quasiparticle width by a parameter $\delta $. In
Fig. \ref{gamma}, we show the influence on the renormalization
factor and the scattering function of the residual quasiparticle width.
We compare the results for $\delta = 5$ meV with those for $\delta = 1 $ meV.
For very small quasiparticle broadening (full lines) the cusp features
in the imaginary part of the self energy turn into peaks
(which ultimately evolve into square root singularities for perfectly sharp
quasiparticles and resonance). The second feature to mention is that
the scattering rate near the onset points, $\pm \Omega_{res}$, is 
influenced strongly by the residual quasiparticle width. 
Because this onset region governs the quasiparticle width in 
underdoped cuprates, as we show later, we study it in the following in more
detail. In the lower part of Fig. \ref{gamma} we show as a dotted line
the contribution to the electron scattering rate coming from the
final states not too close to the $M$ points (the regions which
determine $\lambda_M^{(N)}$, introduced above) as compared to the
full scattering rate (full line). It is clearly seen that the sharp
features come from the $M$ point regions, whereas the nodal regions
contribute to the onset of electron scattering
and provide a smooth constant background at higher energies.

The behavior of the imaginary part of the self energy near the
onset points, $\pm \Omega_{res}$, in Figs.~\ref{Zfac} and
~\ref{gamma} is determined by the nodal
electrons. For larger residual quasiparticle widths ($\delta = 5 $ meV,
dashed lines in Fig. ~\ref{gamma}) there are
states available at the chemical potential (coming e.g. from impurity
scattering), which increase the number of final states for scattering
events. Thus, the onset in Fig. ~\ref{gamma} for the electron scattering
rate is stronger in this case than for $\delta = 1$ meV. For zero temperature
there will be a jump at energy $\pm \Omega_{res}$ in the imaginary part
of the self energy, which causes the small cusps at the same energy
in the renormalization factor (top panel in Fig.~\ref{gamma}).
For $\delta =0 $ the onset is linear in energy.

We will estimate analytically the onset behavior near these points for
the case $\delta =0$ now.
For this we use Eqs.~\ref{dSig} and ~\ref{self11}. We replace $\vec{q}$ by
$\vec{k}'-\vec{k}$, approximate $w_q$ by 
$w_{MN}=w_{k_M-k_N}$, and linearize the dispersion around the nodes,
$\Delta_k=\vec{v}_{\Delta}(\vec{k}'-\vec{k}_N)$, $\xi_k=
\vec{v}_N(\vec{k}'-\vec{k}_N)$. Here $\vec{v}_{\Delta}=\partial_{k} \Delta_k $
and $\vec{v}_N= \partial_{k} \xi_k$ taken at the $N$ point. For our
model, we have $v_{\Delta } = \Delta_M \sin(k_{xN})/\sqrt{2}$, which is
valid near optimal doping.
(But note that for underdoped cuprates, $v_{\Delta }$ was experimentally
shown to be smaller than that value, perhaps scaling with
$k_B T_c $ instead of with $\Delta_M$.\cite{Mesot99}) 
Performing the $\vec{k}'$ sum and summing over all four nodes, we arrive at
\begin{eqnarray}
\label{ImSnode}
\mbox{Im} \delta \Sigma_M^{(N)} (\epsilon ) &=& -\frac{g^2w_{MN}}
{\pi v_Nv_{\Delta}}
(|\epsilon + \Omega_{res}| \Theta (-\epsilon-\Omega_{res}) \nonumber \\
&&\qquad +|\epsilon - \Omega_{res}| \Theta (\epsilon-\Omega_{res}) )
\end{eqnarray}
Here, the $\Theta $-function is unity
for positive argument and zero otherwise. Thus, the slope of the
scattering rate at $\epsilon=\pm \Omega_{res}$ is given by
$\mp g^2w_{MN}/\pi v_N v_{\Delta }$.  
For the parameters in Tables \ref{tab1} and \ref{tab2},
the magnitude of this slope is equal to 9.5 $w_{MN}/w_Q \approx 0.56$.
Note that Eq.~\ref{ImSnode}
gives a good approximation of the scattering rate in
the interval $\Omega_{res}<|\epsilon|< \Omega_{res}+\Delta_A/2$.
For energies further away from the onset, the change of the
quantity $v_{\Delta }$ (which goes to zero at the $A$ point) leads to a
stronger increase.
Finally, for underdoped cuprates the excitation energy at the $M$ point,
$E_M$, is larger than $\Omega_{res}$. Then, the quasiparticle linewidth
at the $M$ point is given by $-\mbox{Im} \delta \Sigma_M^{(N)} 
(-E_M ) /Z_M(-E_M)$.
Thus, for underdoped cuprates it is given by,
\begin{eqnarray}
\label{Gamma_u}
\tilde\Gamma_{M} &=& \frac{g^2w_{MN}}{\pi v_Nv_{\Delta}}
\frac{E_M - \Omega_{res}}{ Z_M }
\end{eqnarray}
with $Z_M \equiv  Z_M (-E_M)$. 
Near the nodes, on the contrary, the quasiparticles will stay relatively 
sharp even in underdoped compounds because
the peaks positions are then below the onset energy $\pm \Omega_{res}$.

\subsubsection{The coupling constant and the weight of the
spin resonance}

One potential criticism of a model which assigns the observed
anomalies in the dispersion to coupling of electrons
to the spin resonance mode is the spectral weight of the resonance, $I_0$,
which amounts to only a few percent of the local moment sum rule.\cite{Kee01}
Our calculations show that
this is not an obstacle,\cite{Abanov01} as we obtain a dimensionless coupling 
constant of order one, as observed experimentally.

Here we estimate $\lambda_{M} $, given by $Z_M(0)-1$, for the resonance mode. 
From Eq.~\ref{ZM}, it is equal to
\begin{eqnarray}
\lambda_{M}  &\approx &
\frac{g^2I_0}{\pi} \cdot 
\frac{1}{(\Omega_{res}+E_M)^2}  + \lambda_{M}^{(N)}(0)
\end{eqnarray}
Using values for optimal doping (Table \ref{tab2}),
the first term in this sum is equal to $17.44 I_0$, which amounts to
about 0.61 (in our model $I_0=0.035$). This is already a large part
of the total coupling constant, which from
Fig. \ref{Zfac} is $\lambda_{M} \approx 0.9$.
The contribution $1+\lambda_{M}^{(N)}(\epsilon ) $ is shown as dotted line in
the upper part of Fig. \ref{gamma}. $\lambda_{M}^{(N)}$ is not negligible, but 
contributes about 30\% to the total coupling constant.

We obtain an analytic formula for the low energy correction to the
renormalization factor due to scattering between nodal points and $M$ points, 
$\lambda_{M}^{(N)}(\epsilon )$,
by a Kramers-Kronig transform of $\mbox{Im} \Sigma_M^{(N)} (\epsilon )$,
in which only energies up to a cut-off $\pm(\Omega_{res}+\Delta_A)$ are taken 
into account, and replacing $\mbox{Im} \Sigma_M^{(N)} (\epsilon ) $ above
this cut-off by a constant (see the dotted lines in Fig. \ref{gamma}) equal
to its value at the cut-off.
The result for $\epsilon=0$ is,
\begin{equation}
\label{ZSnode}
\lambda_M^{(N)} (0) \approx \frac{g^2w_{MN}}{\pi v_Nv_{\Delta}}
\frac{2}{\pi}\ln\left(1+\frac{\Delta_A}{\Omega_{res}}\right)
\end{equation}
For our parameter set this amounts to $\lambda_M^{(N)} (0) \approx 0.21$.
Note that $\lambda_{M}^{(N)} $ increases with decreasing $\Omega_{res}$.

To summarize, dimensionless coupling constants (comparable to those for 
strong-coupling electron-phonon systems) are easily achieved with reasonable 
parameters by coupling electrons to the spin resonance.

\subsubsection{Particle hole asymmetric renormalizations}

From Eq.~\ref{self12}, we see that the second
term in the numerator, proportional to $\xi_{k-q}$,  affects the
band dispersion $\xi_k$. The resulting renormalization is given by,
\begin{equation}
\label{xbar}
\bar\xi_{\epsilon ,k}= \xi_k 
-\sum_{q}  \frac{g^2w_q}{\pi} \cdot
\frac{\left( 1+\frac{\Omega_{res}}{E_{k-q}}\right)\xi_{k-q}}{
(\Omega_{res}+E_{k-q})^2 - \epsilon^2}
\end{equation}
From this formula, it is clear that notable renormalizations of the
Fermi surface only take place if $\xi_{k-q}$ is 
not too far from
(but also not at)
the chemical potential. Thus, the largest renormalizations are expected 
at the $M$ point regions of the Brillouin zone.

\begin{figure}
\centerline{
\epsfxsize=0.25\textwidth{\epsfbox{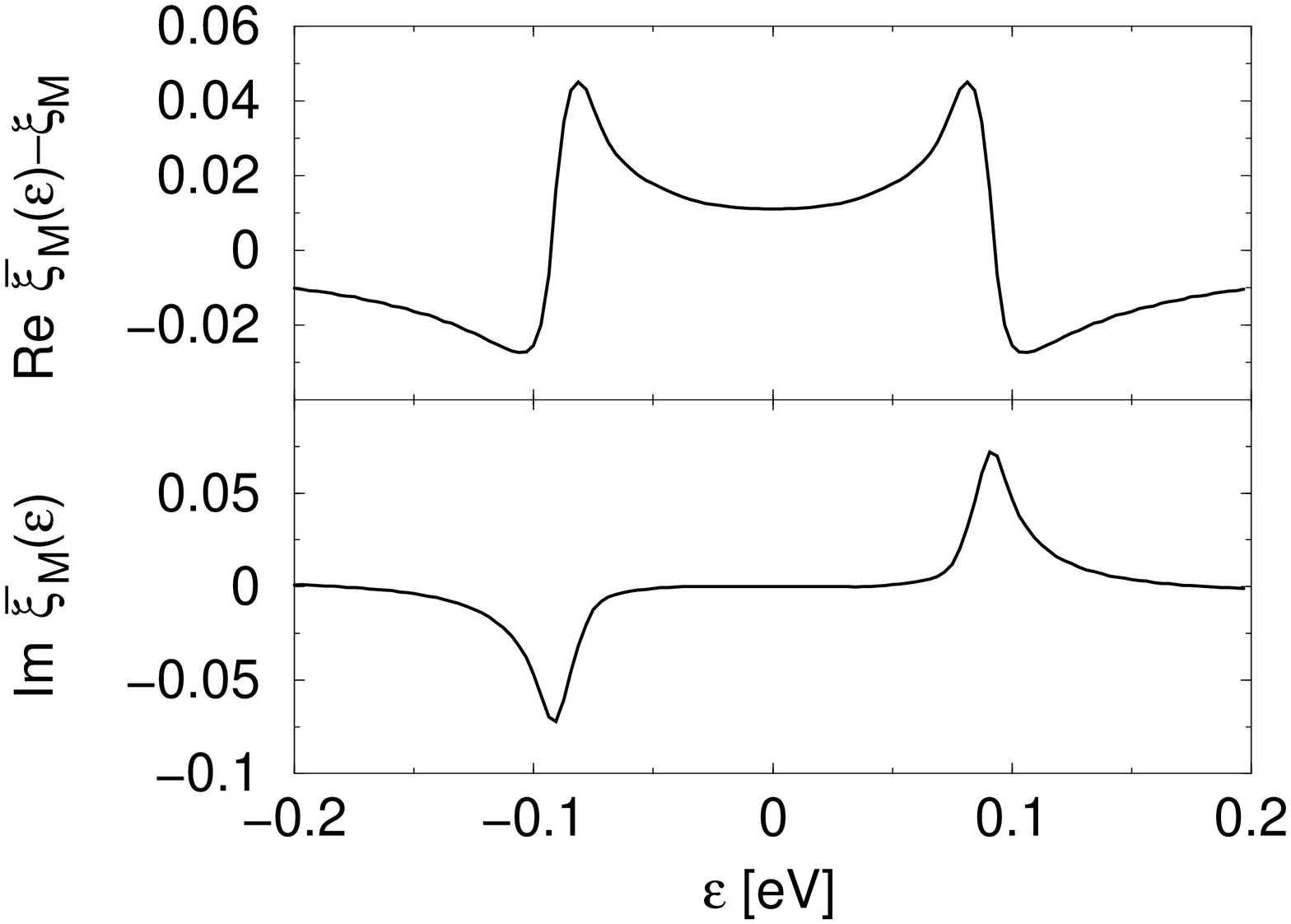}}
\epsfxsize=0.22\textwidth{\epsfbox{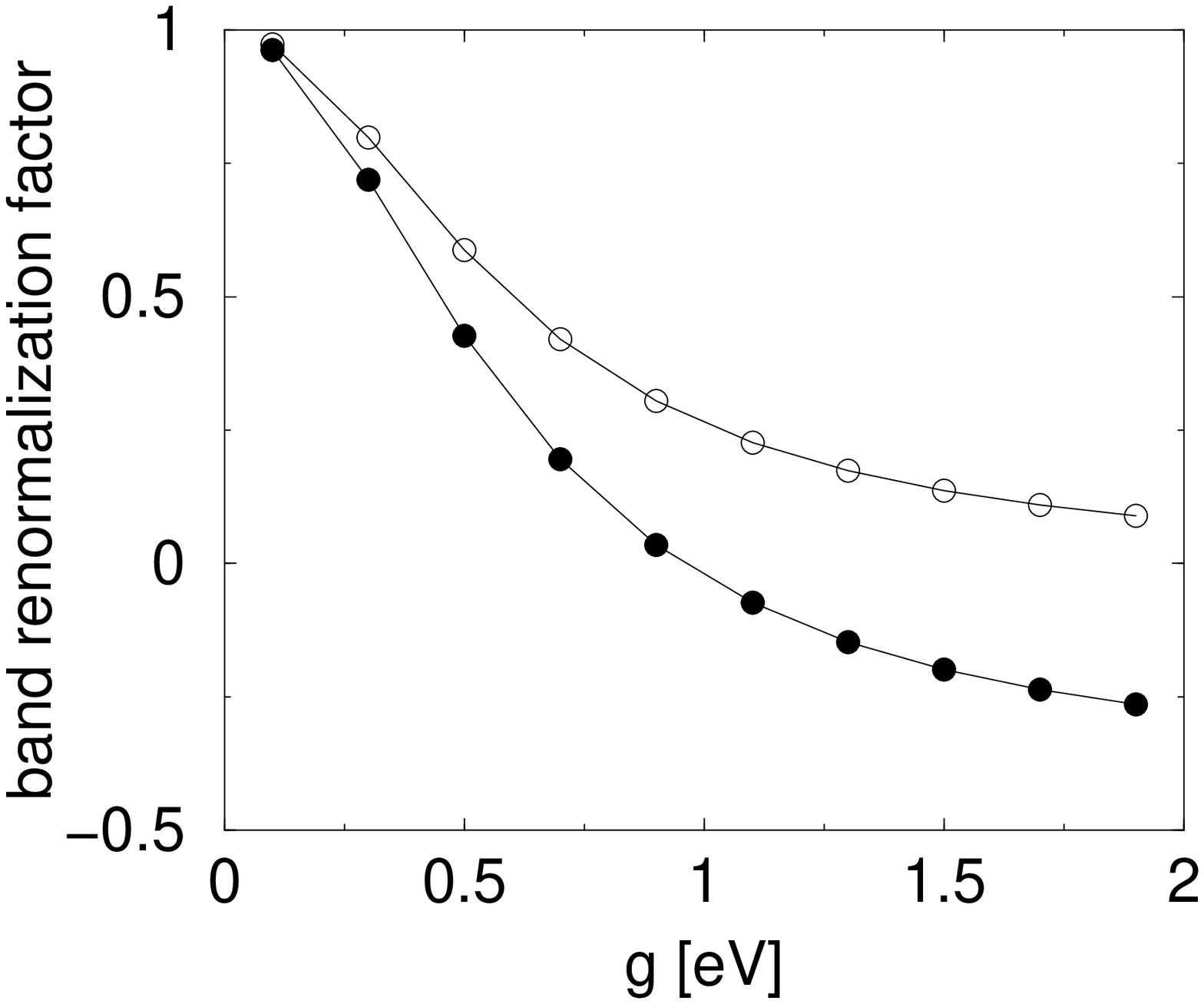}}
}
\caption{
\label{Sigx}
Left: The part of the self energy (eV) defining the band renormalization at
the $M$ point as a function of energy,
for $\Delta_{M}$=35 meV, $\Omega_{res}$=39 meV, and $g$=0.65 eV.  
Right:  The quantity $\bar \xi_M(\epsilon )/(Z_M(\epsilon ) \xi_M)$ (filled
circles) and $1/Z_M(\epsilon )$ (empty circles)
for $\epsilon = -\Delta_M$ as a function of the coupling constant $g$.  }
\end{figure}

In Fig.~\ref{Sigx} (left), we show the particle-hole asymmetric
part of the self energy as a function of $\epsilon $ for electrons
at the $M$ point of the Brillouin zone. The imaginary part shows
a peak due to the van Hove singularity at the $M$ point, but the 
cusp feature due to the $A$  points is missing, because points where
$\vec{k}-\vec{q}$ is on the Fermi surface
do not contribute to the sum in Eq.~\ref{xbar}. The real part 
indicates that the renormalization of the dispersion is confined to
energies between $-\Omega_{res}-E_M$ and $\Omega_{res}+E_M$.
Using the same approximation procedure as above, we obtain for the
renormalization at the $M$ point,
\begin{equation}
\label{xbar1}
\bar\xi_M(\epsilon ) \approx
\xi_M\left(1-\frac{g^2I_0}{\pi}\frac{1}{E_M}
\frac{\Omega_{res}+E_M}{(\Omega_{res}+E_M)^2-\epsilon^2}\right)
\end{equation}
The first important point is that the renormalization has opposite sign
to $\xi_k$, thus the band is renormalized towards the chemical
potential. In particular, there is a `pinning' effect of
the van Hove singularity at the $M$ point
to the chemical potential, as long as $\xi_M$ is of the order of $\Omega_{res}$.
Furthermore, the renormalization factor $Z_M(\epsilon )$ from Eq.~\ref{ZM} 
increases this effect, as $\bar\xi_M/Z_M$
defines the quasiparticle dispersion.

In order to quantify this, we show in the right panel
of Fig.~\ref{Sigx} the relative changes of the dispersion,
$\bar \xi_M(\epsilon )/(Z_M(\epsilon ) \xi_M)$ (filled circles), in comparison
to the inverse renormalization factor $1/Z_M(\epsilon )$ (empty circles). 
The latter would give the band renormalization in the absence of
particle hole asymmetric parts in the self energy.
As can be seen in this figure, the band is renormalized towards the
chemical potential and even crosses it for large coupling constants.
For coupling
constants near 0.6 eV, the renormalized band is close to the chemical
potential. Thus, the dispersion of the peak in ARPES is negligible
in the $M$ point regions as a result of the renormalization of the
dispersion. 
The renormalization of the band implies an increase in the chemical potential,
so as to keep the particle density constant. 
This effect would increase the distance between the chemical
potential and the van Hove singularity at the $M$ point, leading to
an equilibrium value in a self consistency loop.
We did not solve this self consistency problem, but assumed that
our parameter choice is close enough to the self consistent solution to
capture the main physics.

\subsubsection{Off diagonal self energy}

\begin{figure}
\centerline{
\epsfxsize=0.25\textwidth{\epsfbox{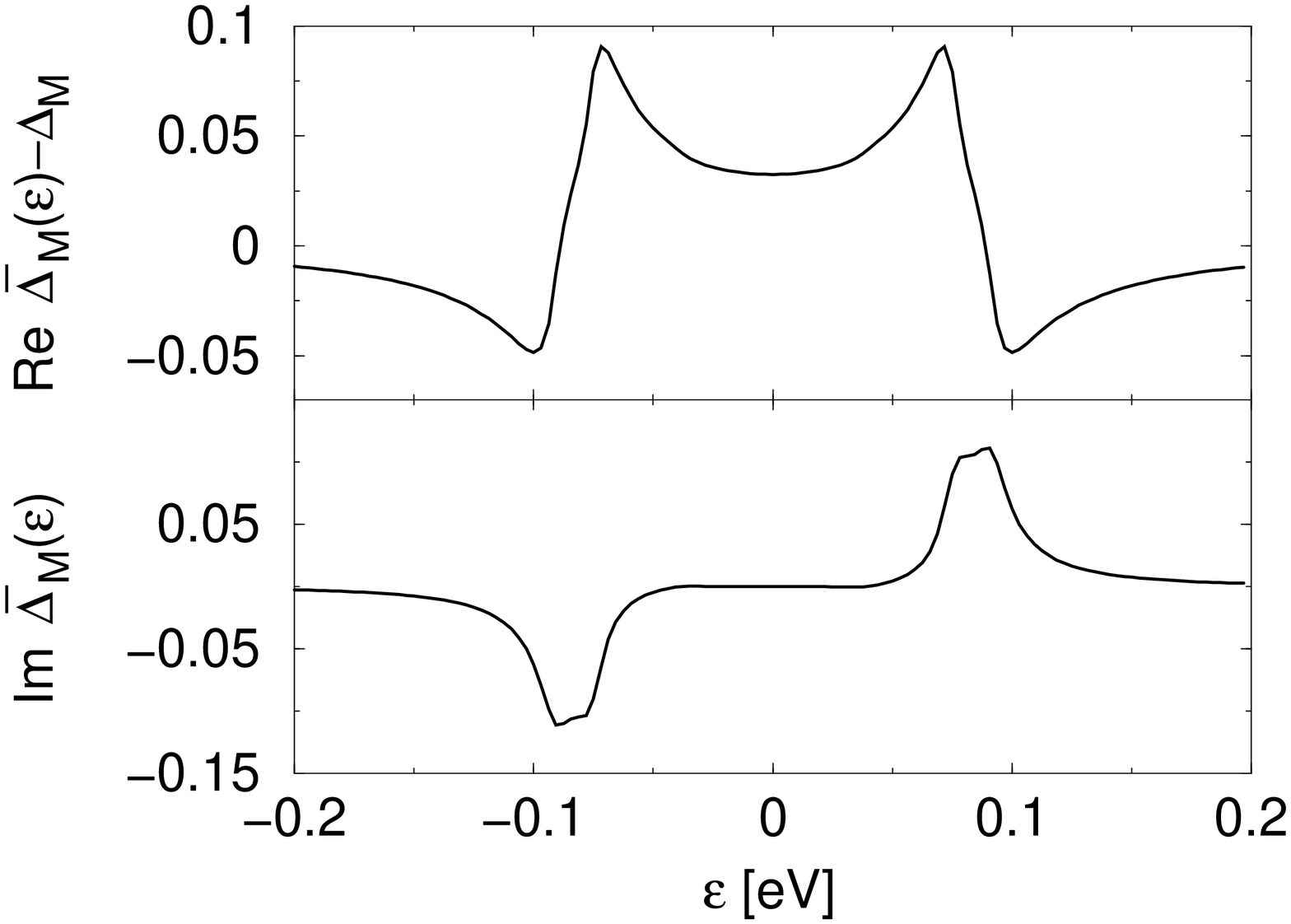}}
\epsfxsize=0.22\textwidth{\epsfbox{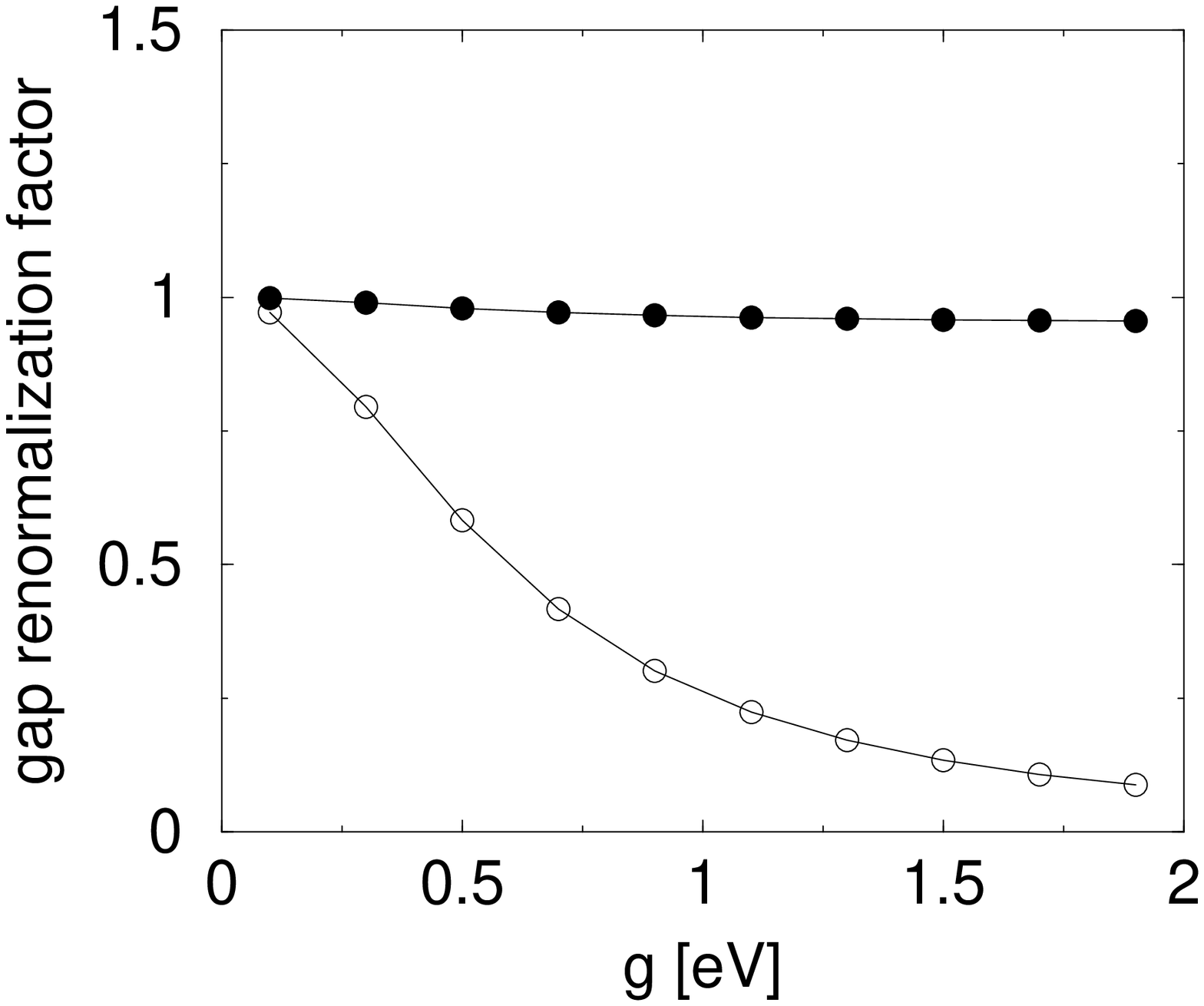}}
}
\caption{
\label{Sigd}
Left: The off diagonal self energy (eV) at the $M$ point of the
Brillouin zone as a function of $\epsilon $ is shown for coupling
constant $g$=0.65 eV.  Right: 
The quantities $\bar \Delta_M(\epsilon )/(Z_M(\epsilon )\Delta_M)$ 
(filled circles)
and $1/Z_M(\epsilon )$ (empty circles) are shown for
$\epsilon = -\Delta_M$ as a function of the coupling constant $g$.
Parameters used are $\Delta_{M}$=35 meV, $\Omega_{res}$=39 meV.
}
\end{figure}

\begin{figure}
\centerline{
\epsfxsize=0.24\textwidth{\epsfbox{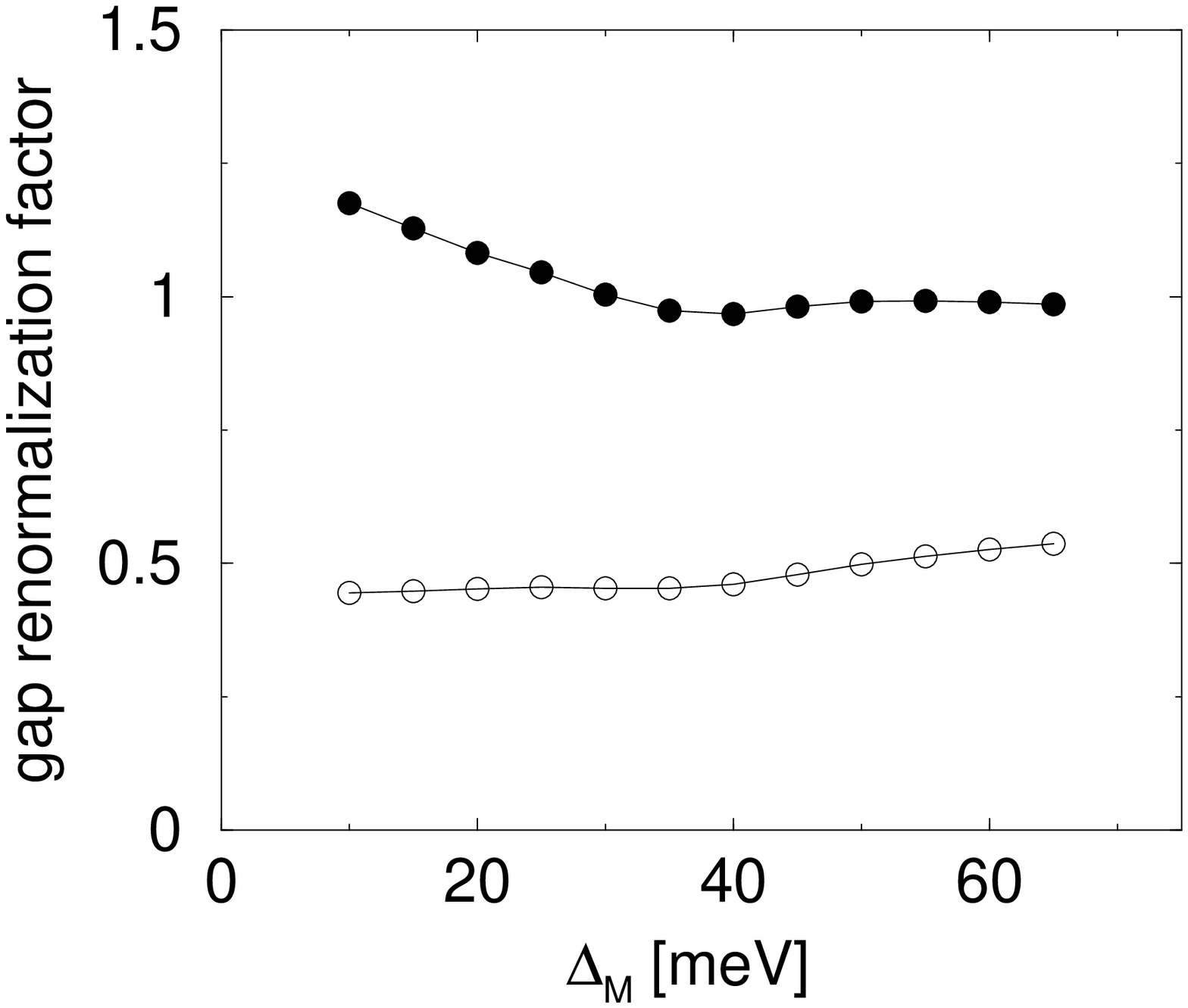}}
\epsfxsize=0.24\textwidth{\epsfbox{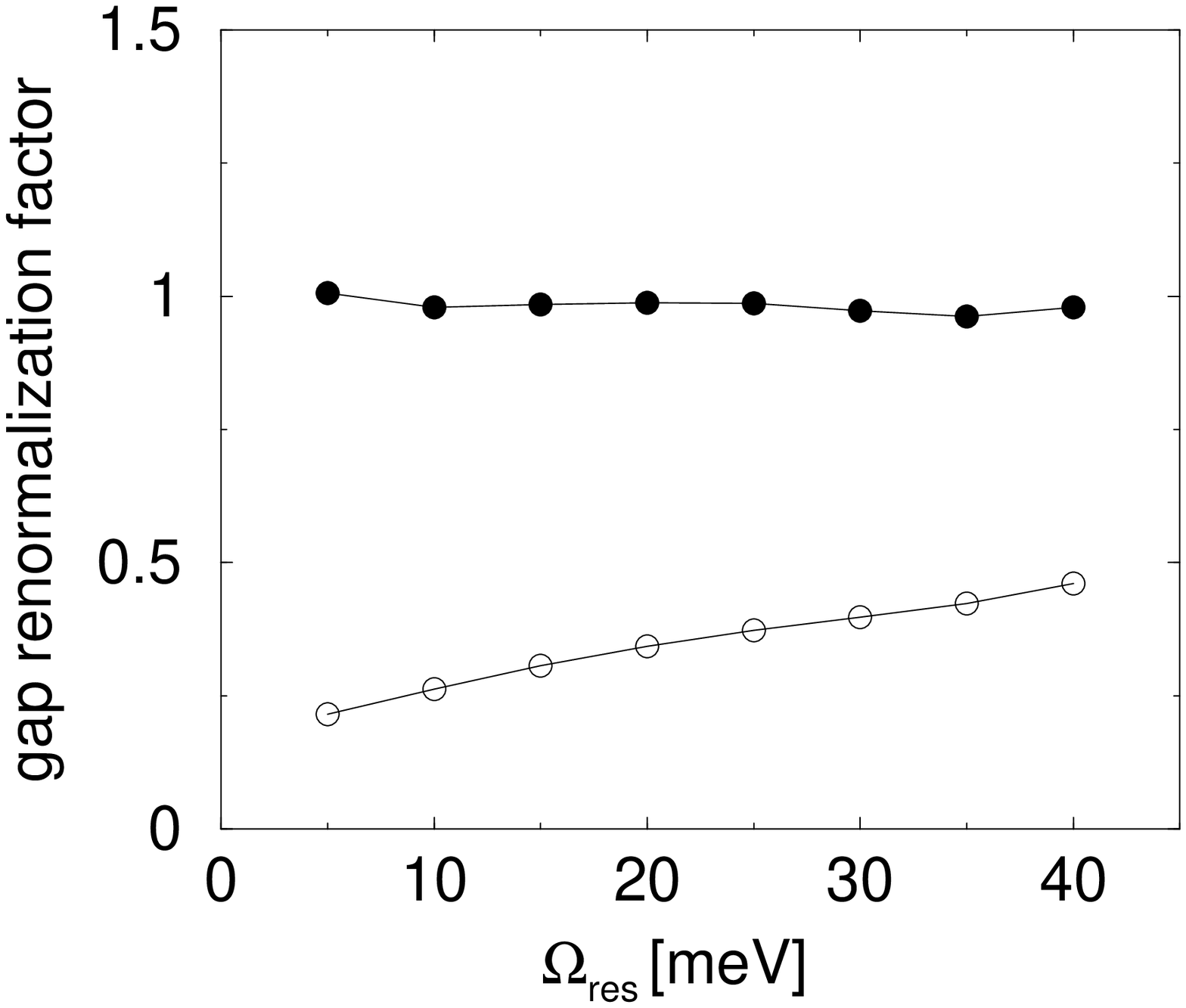}}
}
\caption{
\label{del_dop}
The quantities $\bar \Delta_M(\epsilon )/(Z_M(\epsilon )\Delta_M)$ 
(filled circles)
and $1/Z_M(\epsilon )$ (empty circles) are shown for
$\epsilon = -\Delta_M$ as a function of $\Delta_M$ (left,
for $g=0.65$ eV and $\Omega_{res}=39$ meV)
and $\Omega_{res}$ (right, for $g=0.65$ eV and $ \Delta_M=35$ meV).
}
\end{figure}

In order to understand the 
renormalization of the order parameter $\Delta_k$ due to
coupling to the resonance mode, we observe from
Eq.~\ref{self22},
\begin{eqnarray}
\label{dbar}
\bar\Delta_{\epsilon,k}= \Delta_k 
-\sum_{q}  \frac{g^2w_q}{\pi}
\frac{\left( 1+\frac{\Omega_{res}}{E_{k-q}}\right)\Delta_{k-q}}{
(\Omega_{res}+E_{k-q})^2 - \epsilon^2} 
\end{eqnarray}
This formula is very similar to that for the band renormalization, except
that the order parameter at momentum $\vec{k}-\vec{q}$ now determines
the renormalization effect.
Note that if $w_q $ were independent of $\vec{q}$,
no renormalization would take place due to the $d$-wave symmetry of
the order parameter.  
Since the spin fluctuation continuum, which we discuss later, is very
broad in momentum, the renormalization effects in the
off-diagonal components is dominated by the resonance contribution.
As the order parameter vanishes at the node, we 
concentrate on the renormalization near
the $M$ point region again. Adopting the approximations as above
(note that contributions from the nodal regions cancel because of
the $d$-wave symmetry), and using Eq.~\ref{self22}, we arrive at 
\begin{eqnarray}
\label{dbar2}
\frac{\bar\Delta_M(\epsilon)}{\Delta_M}\approx
1+\frac{g^2I_0}{\pi}\frac{1}{E_M}
\frac{\Omega_{res}+E_M}{(\Omega_{res}+E_M)^2-\epsilon^2}
\end{eqnarray}
The positive sign is due to the fact that $\Delta_{M+Q}=-\Delta_M$.
As a result of this, there will be a compensating effect
when calculating the quantity $\bar \Delta_M(\epsilon ) 
/Z_M(\epsilon )$, which determines the peak position.
In Fig.~\ref{Sigd} (left), the real and imaginary parts of the
off diagonal self energy at the $M$ point are shown. The imaginary part
is  
relevant only
for energies with absolute value $> \Omega_{res}+\Delta_A$.
For smaller energies, the main effect is to increase the magnitude
of the order parameter $\Delta_k$ in the energy range
$-\Omega_{res}-\Delta_A< \epsilon < \Omega_{res}+\Delta_A$.
Note that the self energy due to coupling to the resonance mode
has $d$-wave symmetry, like the order parameter.
Thus, the coupling to the resonance mode supports superconductivity.
In order to quantify the amount that the resonance mode contributes to
the spectral gap, we show in Figs.~\ref{Sigd} (right) and \ref{del_dop}
the quantity $\bar \Delta_M(\epsilon )/(Z_M(\epsilon )\Delta_M)$
(together with $1/Z_M(\epsilon )$ for comparison) as a function of three
different parameters: $g$, $\Delta_{M}$, and $\Omega_{res}$.

As can be seen from these figures, although the renormalization
factor $Z_M$ would reduce the order parameter considerably, the
off diagonal contribution to the self energy from coupling 
of electrons to the resonance mode restores the gap to
its original value. Thus, the resonance contribution to the
gap is as big as that from other sources, and starts to dominate
if the coupling constant exceeds about 0.5 eV.

The reason why 
$\bar \Delta_M(\Delta_M )$ is so close to $Z_M(\Delta_M )\Delta_M $
is that the additional factor $1+\Omega_{res}/E_M$
in Eq.~\ref{dbar2} compared to Eq.~\ref{ZM} is approximately canceled by
the presence of the additional $\lambda_M^{(N)}(\Delta_M)$ in Eq.~\ref{ZM}.
An analogous term in Eq.~\ref{dbar2} is missing due to
the sign change of the order parameter at the node. 
The degree to which this cancellation holds is a surprising numerical
result and allows us to avoid a self consistency loop for the determination
of $\Delta_M$ near optimal doping.
Thus, the experimental parameters which enter our calculations
are already sufficiently self consistent.

\subsubsection{Spectral functions at the $M$ point}

In this part, we discuss the spectral lineshape, which is an
experimentally accessible quantity. The main features of the spectral
lineshape are captured in the simple model neglecting the continuum
part of the bosonic spectrum. We discuss in the following the influence of
the different parameters of the theory on the spectral function, 
\begin{equation}
A (\epsilon , \vec{k}_M) = -2 \mbox{Im} G^R(\epsilon, \vec{k}_M)
\end{equation}
and
will discuss changes due to the continuum part of the spin fluctuation
spectrum later. In our numerical studies, we used a broadening parameter 
$\delta = 5 $ meV. 
This accounts for processes not covered by scattering by spin
fluctuations.

\begin{figure}
\centerline{
\epsfxsize=0.45\textwidth{\epsfbox{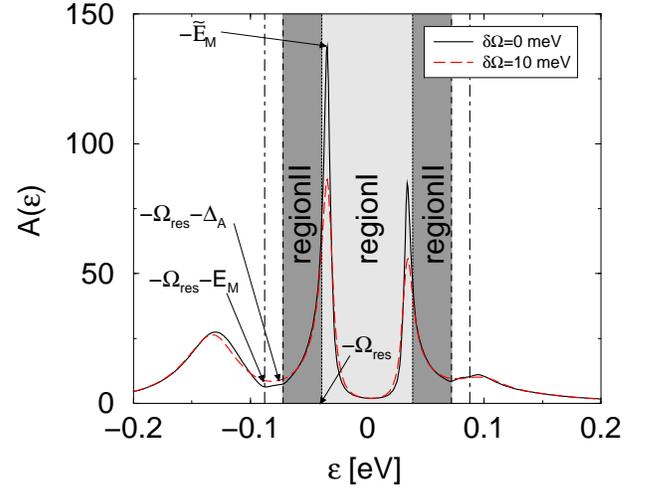}}
}
\caption{
\label{Wwidth}
Spectral functions at $M$ for a perfectly sharp resonance (full line) and for
a resonance with a finite energy width of 10 meV (dashed line).
Parameters are for optimal doping. The finite width
of the mode has very little influence on the ARPES spectra,
and can be neglected for most purposes.
}
\end{figure}

In Fig.~\ref{Wwidth}, we present the results for the spectral function
at the $M$ point of the Brillouin zone for both a perfectly sharp resonance 
and for
a finite width of the resonance of 10 meV. It is obvious that
the energy width of the resonance has very little effect on the ARPES 
spectra, except a slight reduction of the peak height.

Thus, we will concentrate all our following
discussions on a perfectly sharp resonance mode.
The main features of the spectral function is the dip feature at an energy
of about the resonance energy relative to the peak. 
The peak position at $-\tilde E_M$ is renormalized by self energy effects
discussed 
above,
and is shifted from the bare $-E_M$ to be near $-\Delta_M$.
The dip feature is actually
spread out over a range of size $E_M-\Delta_A$, and it is the onset of
this dip feature which defines the resonance energy, $\Omega_{res}$.
The dip feature is followed by a hump at higher binding energies, and the
position of the hump maximum is very sensitive to the 
coupling constant and to damping due to the spin fluctuation continuum, as
we show later. Thus, we concentrate in the following on the peak-dip structure.
Another feature worth mentioning is the asymmetry of the {\it lineshape}
at positive and negative binding energies, with a relatively weak 
dip feature on the unoccupied side compared to the occupied side.

\begin{figure}
\centerline{
\epsfxsize=0.24\textwidth{\epsfbox{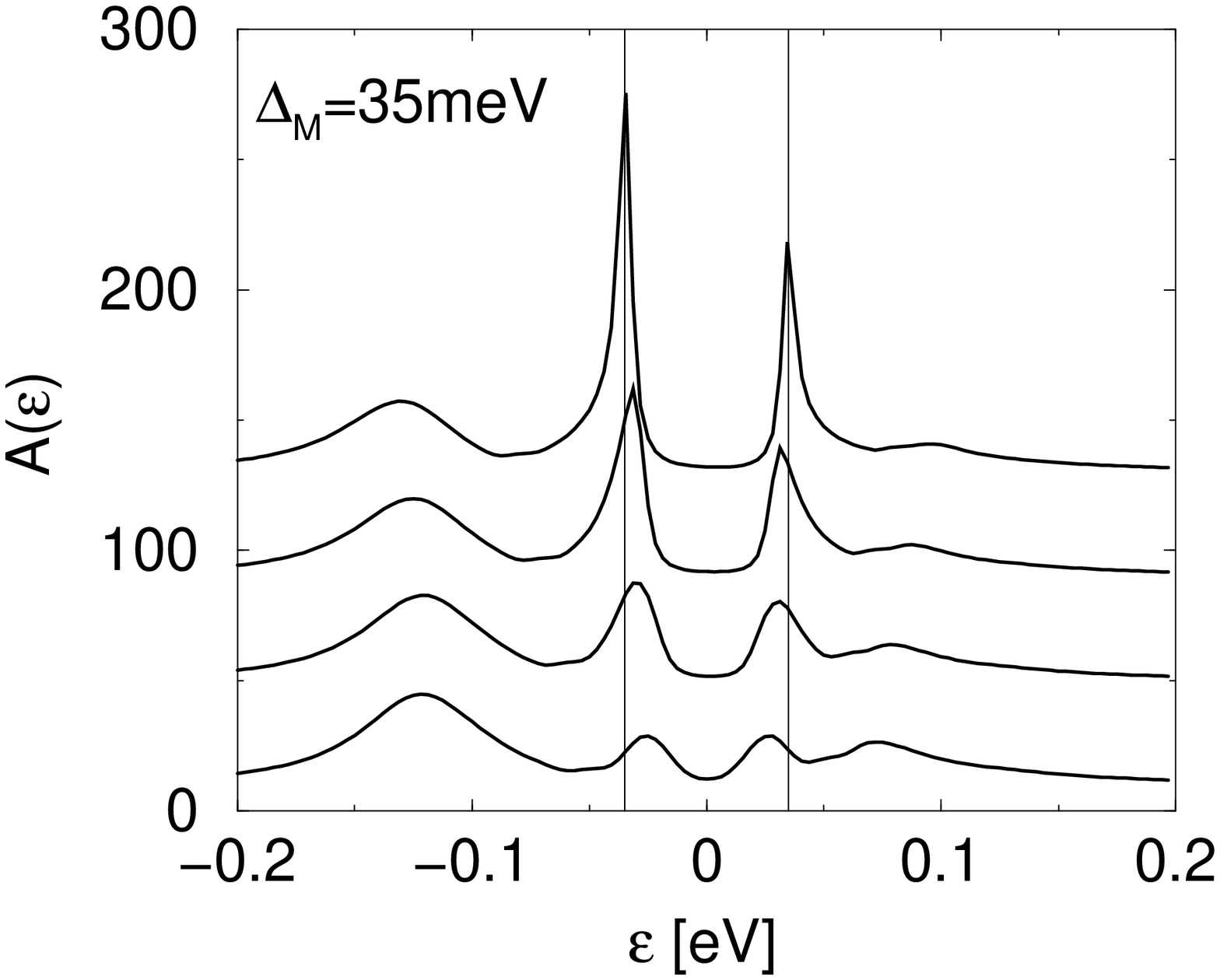}}
\epsfxsize=0.24\textwidth{\epsfbox{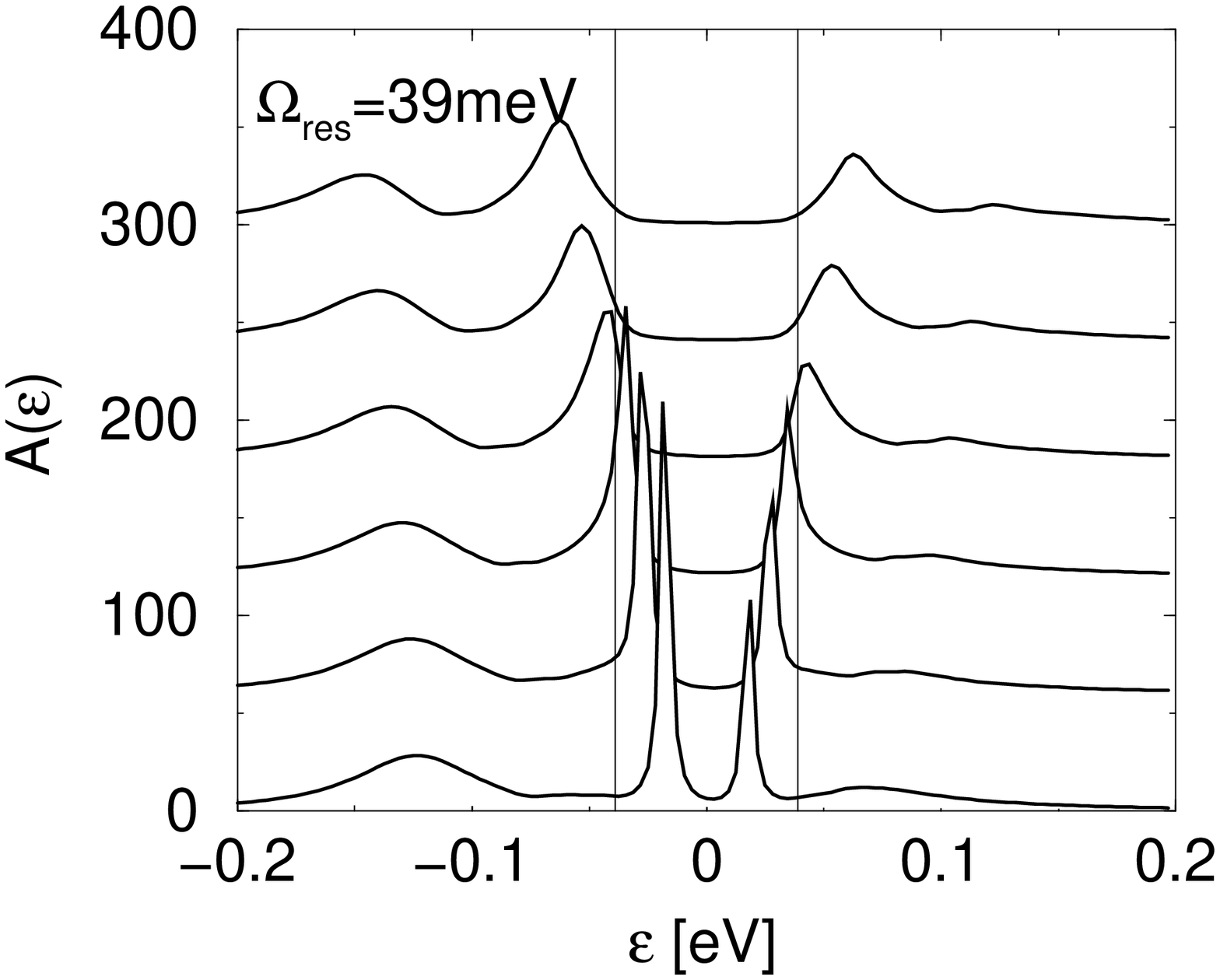}}
}
\caption{
\label{figspecom}
\label{figspecde}
Spectral functions at the $M$ point for varying 
$\Omega_{res}$ (left, for 10 meV, 20 meV, 30 meV and 40 meV from bottom
to top; the thin lines denote the value $\pm \Delta_{M}$), 
and for varying
$\Delta_{M}$ (right, for 15 meV, 25 meV, 35 meV, 45 meV, 55 meV,
and 65 meV from bottom to top;
the thin lines denote the value $\pm \Omega_{res}$).
All other parameters are kept fixed at their optimal doping values.
The spectra are offset for clarity.
}
\end{figure}

In Fig.~\ref{figspecom} (left), the effect of a varying resonance energy
$\Omega_{res} $ (keeping all other parameters at their values for optimal
doping) is shown.
The spectral function shows two effects. 
First, the peak weight is reduced with decreasing mode energy. 
Second, as soon as the quasiparticle excitation energy
exceeds $\Omega_{res} $, strong damping sets in. 
We can understand these results in the light of the discussion for the
self energy. As we mentioned above, 
the scattering rate has a gap equal to $\Omega_{res} $. Thus, as long as the
spectral peak is situated below that energy, in region I of
Figs.~\ref{Wwidth} and \ref{Zfac}, there will be no
damping, and the peak width is set by the residual broadening due to 
other processes. 
If the peak is positioned above $\Omega_{res}$ (region II
in Fig.~\ref{Wwidth}), it
feels the self energy in region II of Fig.~\ref{Zfac}, and will
be broadened. Because in region II the self energy is dominated by
scattering processes involving nodal electrons,
the width in this region is set by the imaginary part of the self energy 
divided by the renormalization factor, and is given in Eq.~\ref{Gamma_u}.
At the same time, for decreasing resonance mode energy, the incoherent
part of the spectral function grows, taking weight from the
quasiparticle peak.
\begin{figure}
\centerline{
\epsfxsize=0.24\textwidth{\epsfbox{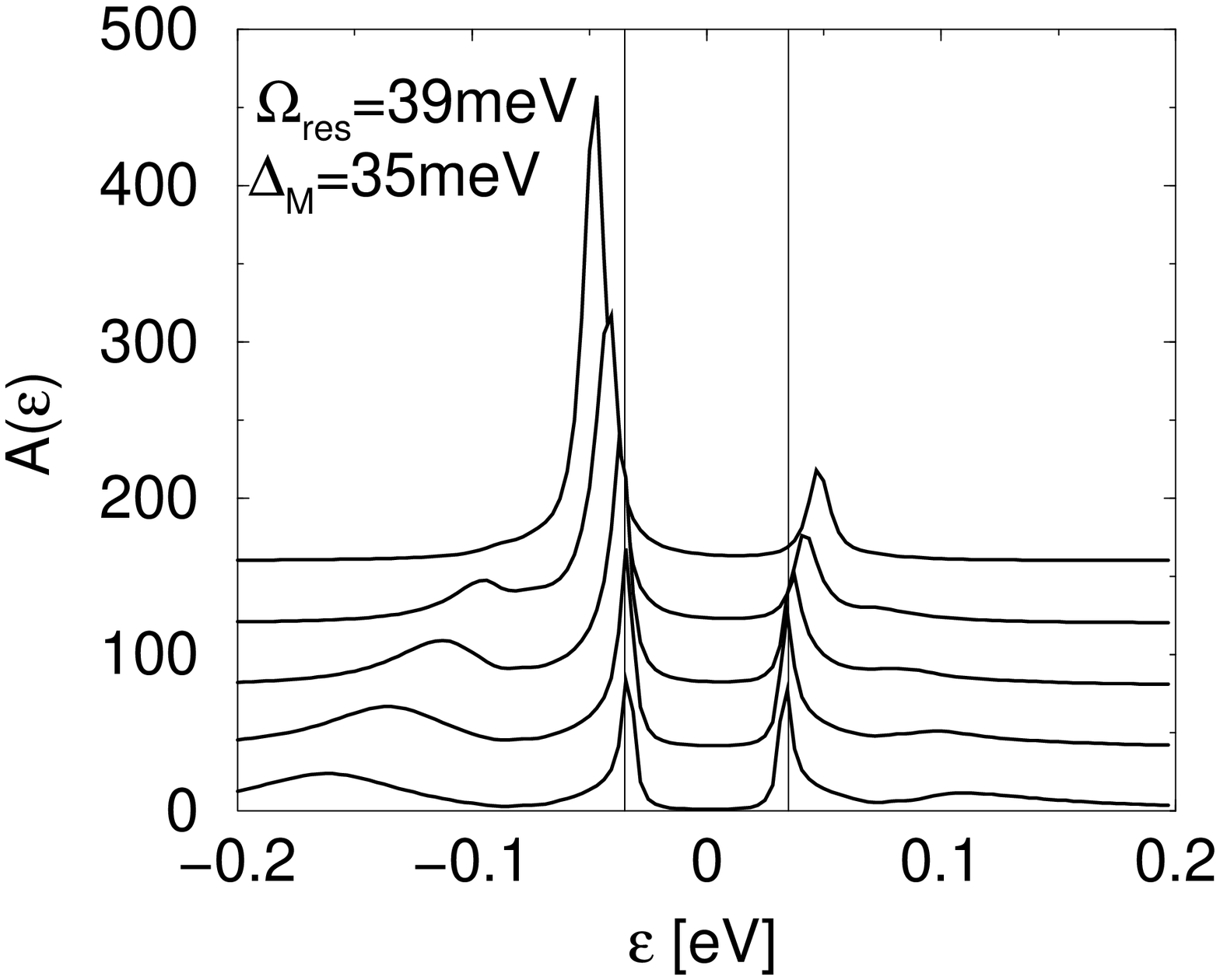}}
\epsfxsize=0.24\textwidth{\epsfbox{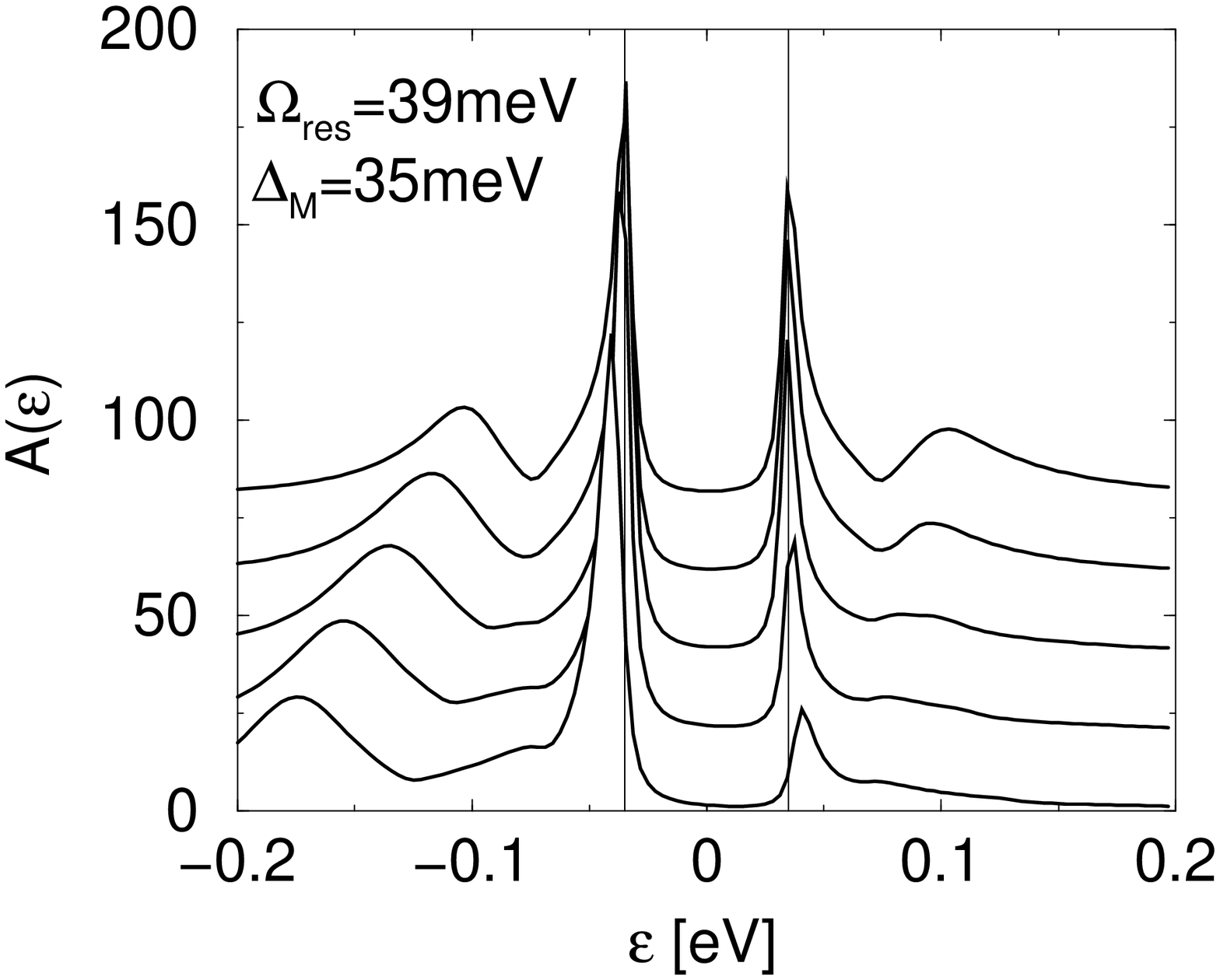}}
}
\caption{
\label{figspecxi}
\label{figspecg}
Spectral functions at the $M$ point for varying coupling constant
(left, $g=$ 0.1 eV, 0.3 eV, 0.5 eV, 0.7 eV, and 0.9 eV from top to bottom;
thin lines denote $\pm \Delta_{M}$)
and for varying distance of the van Hove singularity at the $M$ point
from the chemical potential (right, for $-\xi_M=$ 0 meV, 20 meV, 40 meV,
60 meV, and 100 meV from top to bottom; thin lines denote $\pm \Delta_{M}$).
All other parameters are kept
fixed at their optimal doping values. The spectra are offset for clarity.
}
\end{figure}

Thus, in Fig.~\ref{figspecom}, which is for $\Delta_M=$35 meV, 
the quasiparticle weight increases from
the lowest curve (for $\Omega_{res}=$ 10 meV) to the uppermost curve
(for $\Omega_{res}=$ 40 meV). Simultaneously the broadening decreases.
As the onset of quasiparticle damping and the loss of the coherent part
of the spectrum is a result of a decreasing resonance mode energy relative to 
the gap, the same effect is 
expected
by increasing the gap keeping the
resonance mode energy constant. This is shown in Fig.~\ref{figspecde} 
(right).  In this case, the onset of quasiparticle damping
is always at the same energy $\Omega_{res} =$39 meV, but for the lowest
curve, corresponding to a small gap of 15 meV, 
quasiparticle peaks are well established,
whereas for the uppermost curve, corresponding to a large gap of 65 meV,
the quasiparticle peaks are strongly broadened. 
However,
in this case, the weight
of the peak is affected only weakly, as we will discuss below.

Finally, we show in Fig.~\ref{figspecg} the influence of increasing
coupling $g$, and of an increasing distance of the van Hove singularity from
the chemical potential, $\xi_M$. In both cases, the hump energy is strongly
affected, moving to higher binding energy with increasing coupling and
increasing $\xi_M$. In the left panel, one can also see that the
weight of the peak is strongly reduced with increasing coupling constant.
This is not the case with varying $\xi_M$, as seen from the right panel in
Fig.~\ref{figspecg}, and will be discussed in more detail below.

\subsubsection{The coherent quasiparticle weight of the ARPES spectrum}

Although one can define a quasiparticle residue via the renormalization
factor $Z(\epsilon )$, in light of the experimental
studies, we will in this part study the weight of the quasiparticle peak 
in the ARPES spectrum, 
determined by numerically integrating over the peak region.
For strongly renormalized spectra,
this experimentally motivated quantity will differ from the first.
We note that due to coupling
to the mode, the peak weight is reduced and redistributed to the hump.
Because the peak weight in the experimental literature is often
referred to as the 
`coherent quasiparticle weight', 
we will use the same terminology here.

\begin{figure}
\centerline{
\epsfxsize=0.24\textwidth{\epsfbox{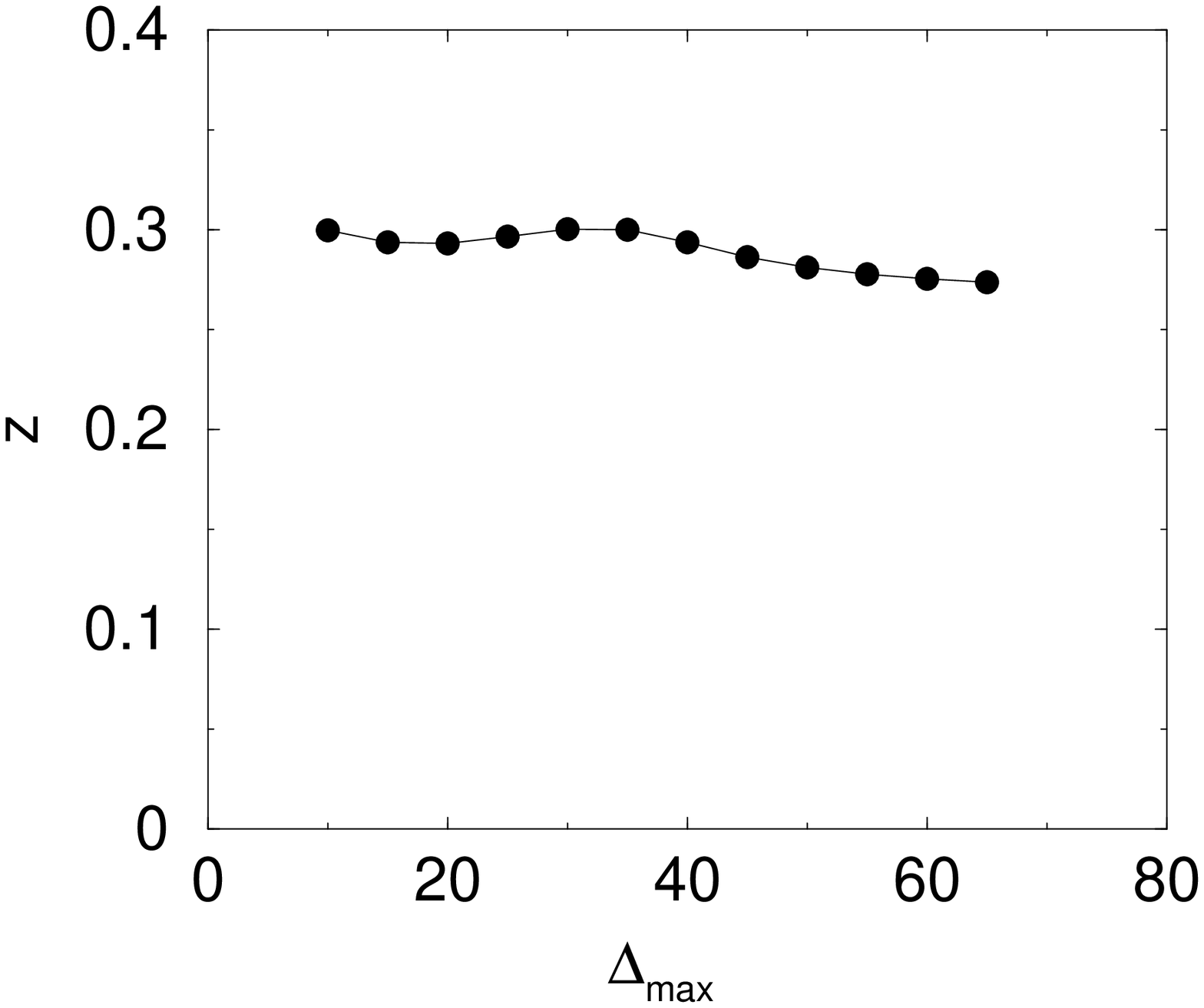}}
\epsfxsize=0.24\textwidth{\epsfbox{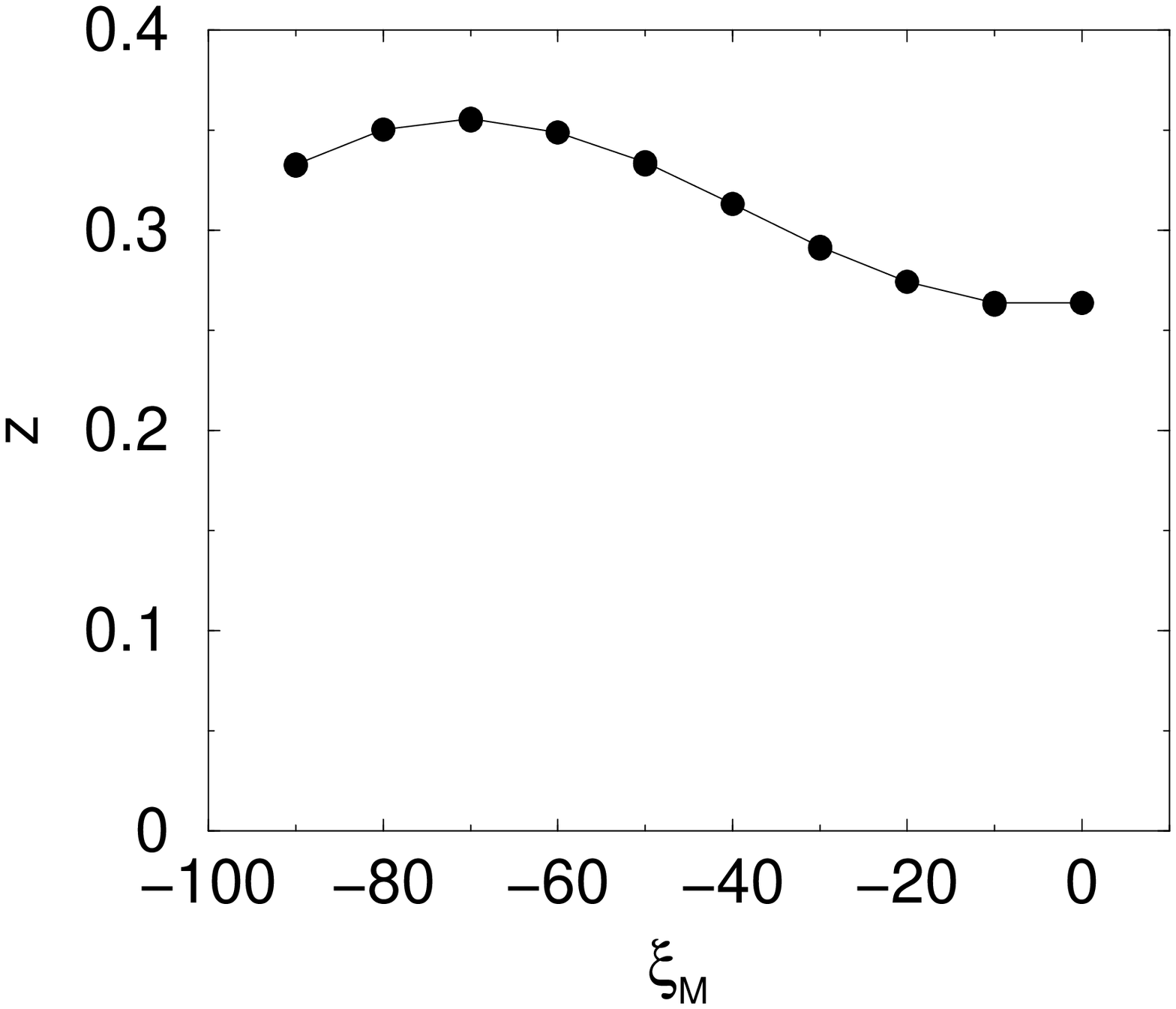}}
}
\caption{
\label{z1}
The 
coherent quasiparticle weight
as a function of $\Delta_{M}$ 
(left) and $\xi_M$ (right) for $\Omega_{res}=$39 meV.
Although the peak width changes considerably as a function of $\Delta_{M}$, 
the peak weight is only weakly dependent on $\Delta_{M}$ and $\xi_M$. 
}
\end{figure}

\begin{figure}
\centerline{
\epsfxsize=0.22\textwidth{\epsfbox{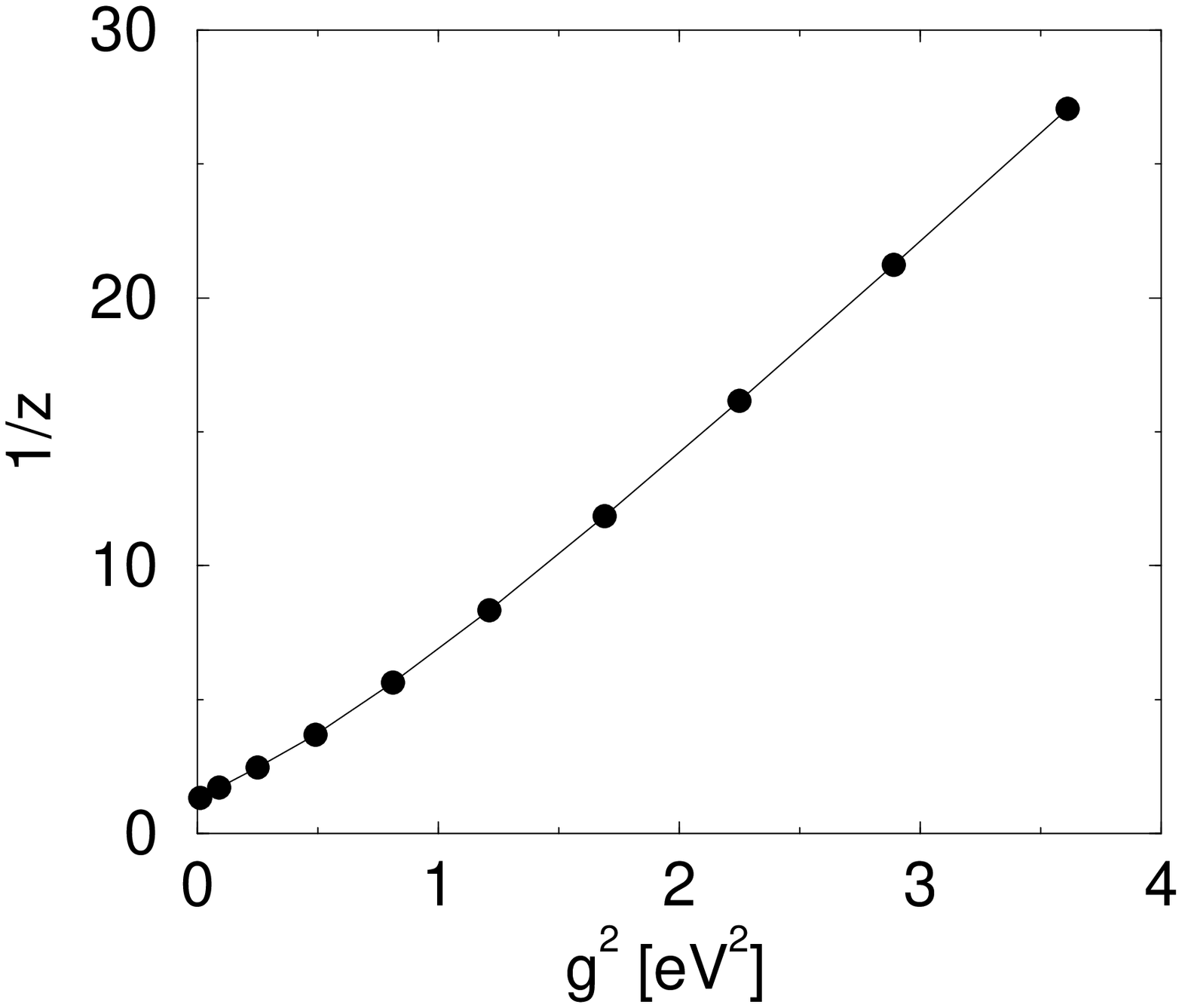}}
\hfill
\epsfxsize=0.22\textwidth{\epsfbox{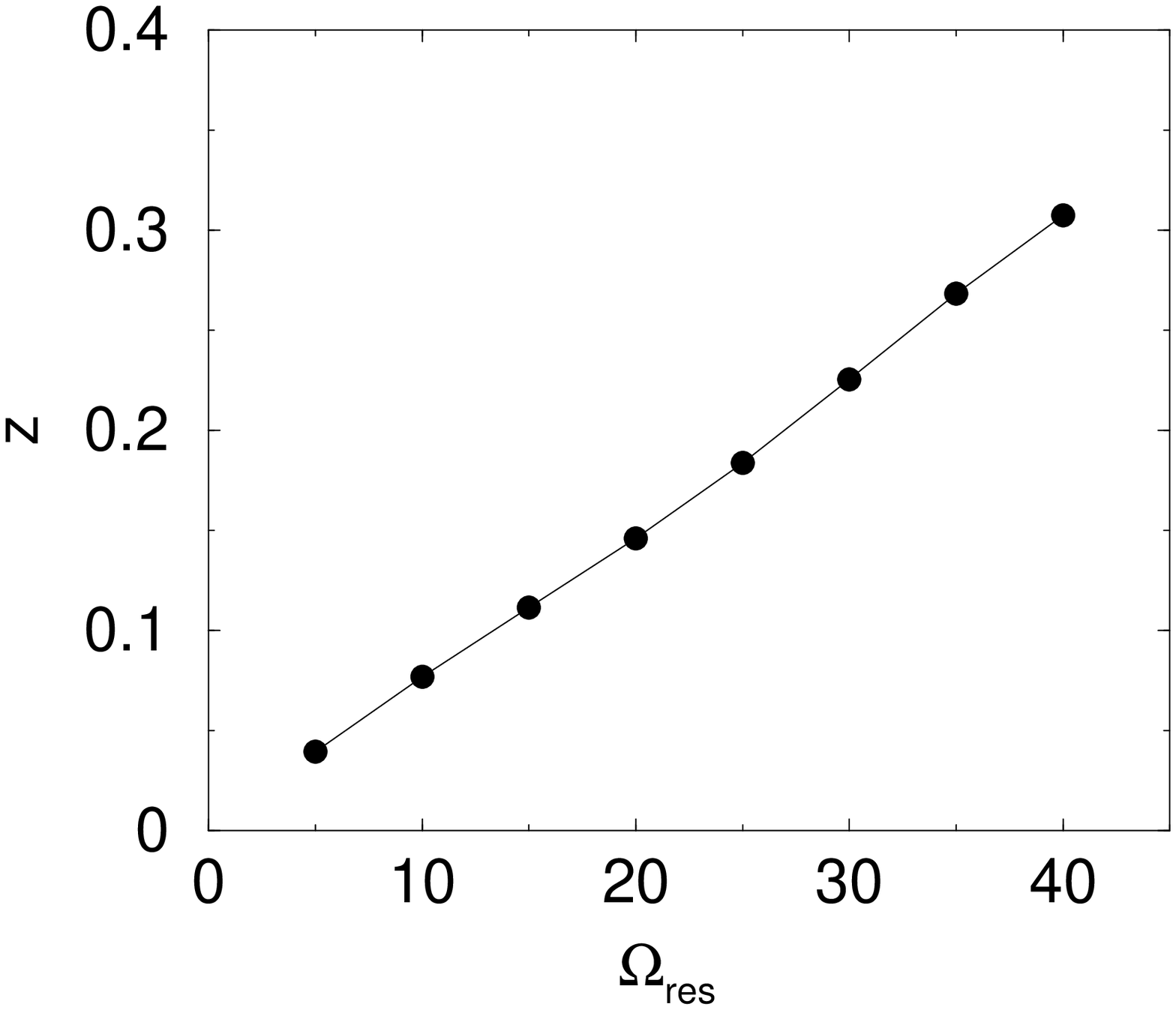}}
}
\vspace{0.3cm}
\caption{
\label{z2}
The inverse of the 
coherent quasiparticle weight 
$1/z$ is approximately a linear function of $g^2w_Q$ (left).
Here we have chosen $\Omega_{res}=$39 meV and $\Delta_{M}=$35 meV.
The right panel shows that $z$ is proportional to $\Omega_{res}$.
}
\end{figure}

We consider the spectral function at the $M$ point of the Brillouin zone.
Because the peak is separated from the hump by a dip which extends
from $-\epsilon_1=-(\Omega_{res}+\Delta_{A})$ to 
$-\epsilon_2=-(\Omega_{res}+E_M)$, we
define as the coherent quasiparticle weight 
the quantity,
\begin{equation}
z_M=
-\frac{1}{\pi } 
\int_{-\epsilon_1}^{0} \; d\epsilon \; 
\mbox{Im}G^R_M(\epsilon )
\end{equation}
Without interactions between the quasiparticles, 
$z = 0.5$ at the Fermi surface, because
the quasiparticle peaks at $\pm \Delta $ in BCS theory each have one half of 
the total weight; the value at negative energy is somewhat larger than 0.5
at the $M$ point because it is an occupied state. Coupling of the
quasiparticles to the mode reduces $z$.
In Figs.~\ref{z1} and \ref{z2}, our numerical studies are summarized.
The results are as follows: (1) $z_M$
is only weakly dependent on the gap and the band structure in the relevant
parameter range; (2) $z_M$ is proportional to the mode energy $\Omega_{res}$;
together with the experimental finding $\Omega_{res}\propto k_BT_c$, this
means $z_M\propto k_BT_c$; (3) for coupling constants of order the band width 
or larger, $z_M\propto 1/(g^2w_Q)$;
for smaller coupling constants, $1/z_M \sim A+Bg^2w_Q$ with 
$A$ and $B$ constants;
(4) $z_M$ weakly decreases with increasing antiferromagnetic correlation length 
$\xi_{sfl}$.
We can understand some of these features using the approximate expression
of Eq.~\ref{ZM}. Evaluating $Z_M(\epsilon )$ at $\epsilon=-E_M$, and
taking into account the coherence factor at the $M$ point,
$A^{-}_M \equiv A^{-}_M(-E_M)$, and the nodal renormalization factor
$Z_M^{(N)}\equiv 1+\lambda_M^{(N)}(-E_M)$,
gives
\begin{eqnarray}
\label{renfactor}
z_M &\approx & \frac{\Omega_{res}A^{-}_M}{
Z_M^{(N)}\Omega_{res}+\frac{g^2I_0}{\pi(\Omega_{res}+2E_M)} }
\end{eqnarray}
which defines the constants $A$ and $B$.

In the underdoped region, where $\Omega_{res}$ is much smaller than
$2E_M$, we can approximate further to obtain
\begin{eqnarray}
\label{renfactor1}
z_M &\approx & \frac{2\pi E_M \Omega_{res}A^{-}_M}{ g^2I_0}
\end{eqnarray}
Here, we neglected the first term in the denominator of Eq.~\ref{renfactor} 
compared to the second, which is justified when $z_M$ is small.
In the overdoped region, where $g^2I_0 $ decreases and $\Omega_{res}$
approaches $2\Delta_h$ (where $\Delta_h$ is the gap at the hot spots), this 
scaling with $\Omega_{res}$
should break down according to Eq.~\ref{renfactor}.
Note that experimentally, the relation $\Omega_{res} \approx 4.9 k_B T_c$ was 
shown,\cite{Zasadzinski01}
and also the relation
\begin{equation}
\left(\frac{z_M E_{M}}{k_BT_c} \right)_{(exp)} \approx 0.5
\end{equation}
was experimentally found.\cite{Ding00}
Thus, our expression Eq.~\ref{renfactor1} would be consistent with the
experimental finding if with doping $E_M^2$ scaled with $g^2I_0$. 
Within our theory this experimental finding can be interpreted
as an indication that the phenomenological order parameter 
$\Delta_k$ is governed by the same coupling constant $g$.

\subsection{Contribution of the spin fluctuation continuum}
At energies higher than that corresponding to the 
continuum edge of the spin fluctuation spectrum,
additional broadening due to coupling to that part of the
spectrum sets in. Because the continuum
extends to electronic energies ($\sim $ eV), the introduced
scattering rate will increase continuously with energy up to
electronic energies as well. 
We model the continuum part by
\begin{equation}
\label{Bcont}
g^2B^c_{\omega,q} = 2 g^2c_q \left( \Theta (\omega- 2\Delta_h ) -
\Theta (-\omega - 2\Delta_h ) \right)
\end{equation}
where the gap in the continuum spectrum is given by
$2\Delta_h$.
This form for the gapped continuum is similar to the gapped marginal
Fermi liquid spectrum considered earlier by other 
authors. \cite{Littlewood92,Norman98}
The momentum dependence takes into account the experimentally observed
flatter behavior around the $(\pi,\pi)$ wavevector at higher energies,
and is modeled as
\begin{equation}
c_q=c_Q \left(\frac{1+(32\xi_c^4)^{-1}}
{1+16\xi_{c}^4 (\cos^4\frac{q_x}{2}+\cos^4\frac{q_y}{2})} -(32\xi_c^4)^{-1}
\right)
\end{equation}
with a correlation length $\xi_{c}=0.5a$ compatible with experimental
findings.
We subtracted a background term, so that the response far away from
the $(\pi,\pi)$ wavevector is small, as experimentally observed
(we have chosen this background term so that $c_q$ is zero at $\vec{q}=0$).

For the chosen correlation length, the momentum average of $c_q$ gives
$0.5 c_Q$.
The constant $c_Q$ can be obtained from the experimental values
for the momentum averaged susceptibility at 65 meV, which was found
to be $6 \mu_B^2/$eV for underdoped YBa$_2$Cu$_3$O$_{7-\delta}$
in the odd channel, and about $3 \mu_B^2/$eV in the even channel.\cite{Fong00}
Dividing out the matrix element $2\mu_B^2$, this gives
$c_Q \approx 6/ $eV and $3/$eV respectively. The corresponding values
near optimal doping should be smaller.
We use in our calculations 
$c_Q=5.6/$eV
and $g=0.65$ eV.
The choice of this value is motivated by the ARPES measurements
on optimally doped Bi$_2$Sr$_2$CaCu$_2$O$_{8-\delta}$
of the high energy (linear in excitation energy) part of the
momentum linewidth, which gives
$\Gamma_N=0.75 \epsilon $.\cite{Valla99,Yusof01}
This coupling includes
both the even and odd (with respect to the bilayer indices)
contributions of the spin fluctuations, in contrast
to the coupling to the mode, which is present only in the odd
channel.
Note that our value for $c_Q$ 
is about a factor 
1.6
smaller than neutron scattering measurements
give for underdoped YBa$_2$Cu$_3$O$_{7-\delta}$. 
Because in optimally doped compounds the intensity of the
spin fluctuation continuum is
smaller than in underdoped ones, this is a reasonable value
for optimal doped Bi$_2$Sr$_2$CaCu$_2$O$_{8-\delta}$

The spin fluctuation continuum is gapped in the odd channel from zero energy to
twice the gap at the `hot spots', 2$\Delta_{h}$, which is slightly less than
twice the maximal gap. This means that additional damping only sets in
for energies $|\epsilon| > 2\Delta_{h}$. This corresponds
in optimally doped compounds to about 65 meV. In the even channel,
the optical gap ($\sim$60 meV) persists into the normal state.\cite{Reznik96} 

The continuum formally has to be cut-off at high energies. This cut-off
does not affect the imaginary part of the self energy, but its choice
leaves a real term of the form $-C \epsilon $
at energies small compared to the cut-off energy scale.
This term, equivalent to a contribution to the
renormalization factor which is constant up to the high energies, 
has to be regarded as an additional phenomenological parameter.
The constant $C$ depends on the model one uses for the high
energy tail of the spin fluctuation spectrum. Because we model the
continuum by a constant, which overweights high energies,
we have chosen a relatively low cut-off of 200 meV for our model spectrum.
Because the constant $C$ is only weakly (logarithmically) dependent on the
cut-off, the exact energy of the cut-off is not crucial.

In the simple form of our model, we absorbed the renormalization from the
continuum into the band dispersion $\xi_k$. 
Now we take into account explicitly the continuum, and
thus have to start with a band dispersion not renormalized by this
contribution.  We found that we can reproduce experiment best by
rescaling the dispersion from Table \ref{tab1} in the following way:
$\xi_k^{(new)}=1.5 \xi_k - 0.5 \xi_M$. With this choice, the van Hove
singularity at the $M$ point has the same distance from the chemical
potential as before.

\begin{figure}
\centerline{
\epsfxsize=0.25\textwidth{\epsfbox{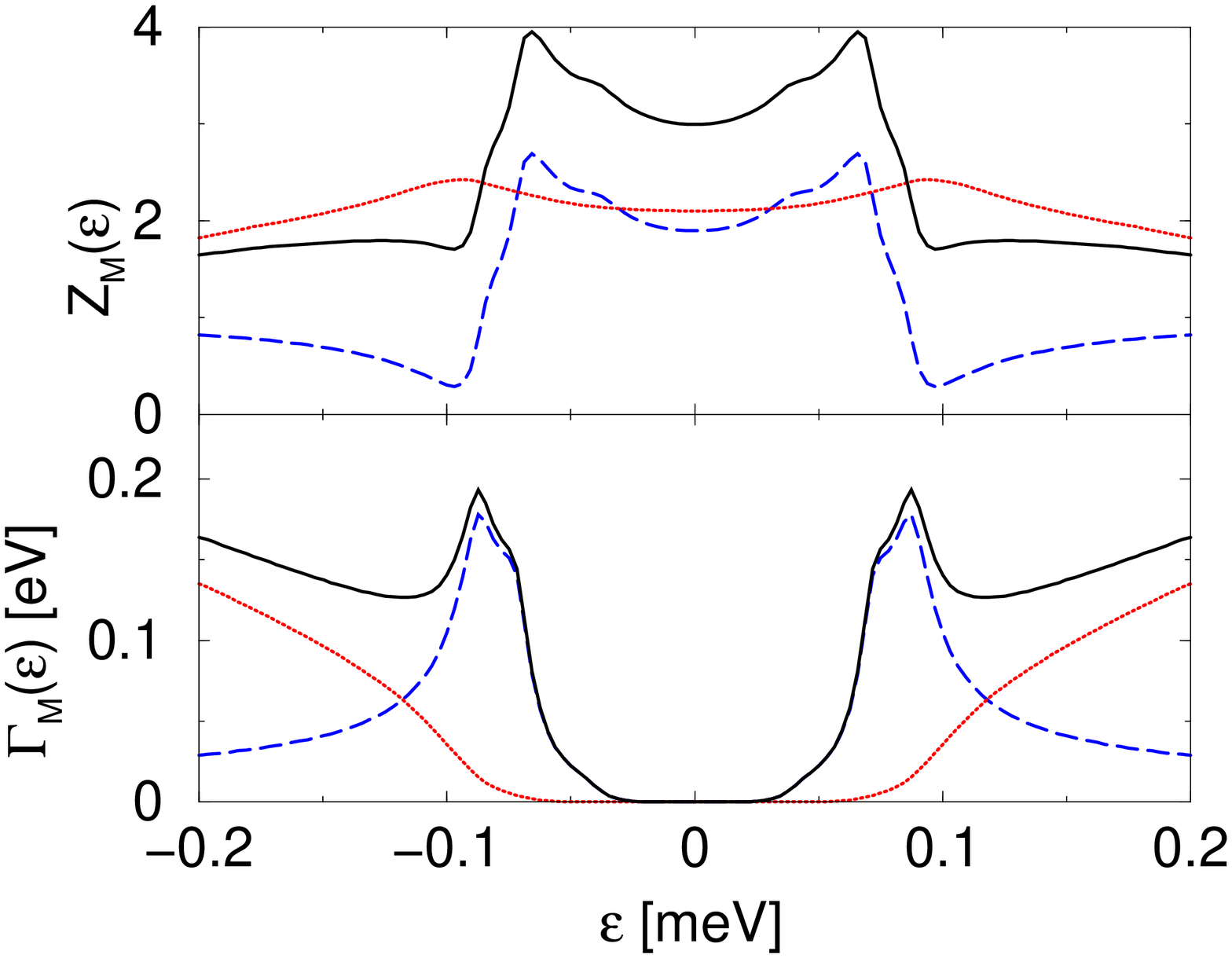}}
\epsfxsize=0.25\textwidth{\epsfbox{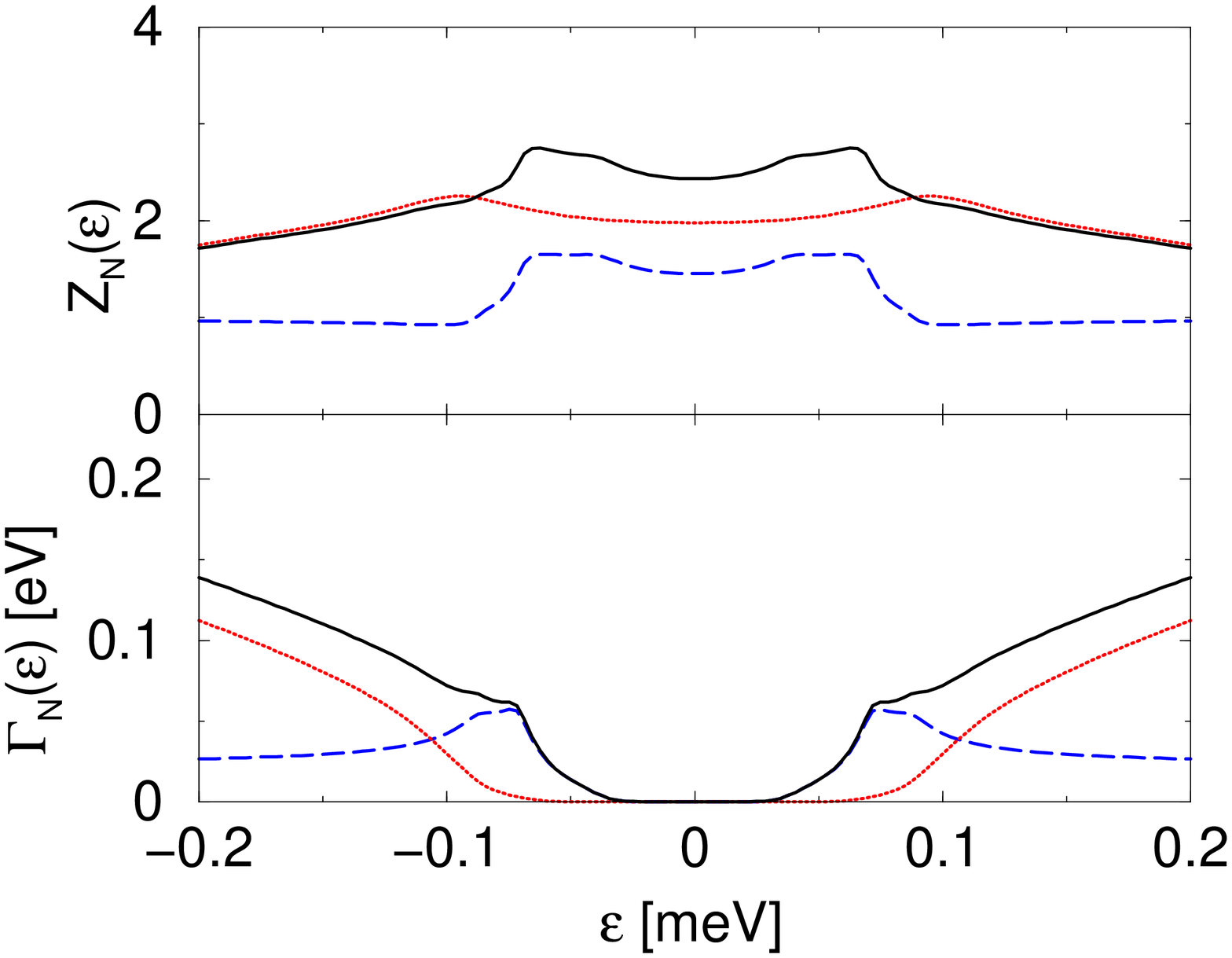}}
}
\caption{
\label{Cont_Mode}
The different contributions to the renormalization factor (top)
and the scattering rate (bottom) are shown for the $M$ point (left)
and for the $N$ (node) point (right). Dotted curves are the contribution 
from the
spin fluctuation continuum, dashed the contribution from the spin
fluctuation mode, and full both contributions.
}
\end{figure}

In Fig.~\ref{Cont_Mode}, the continuum contribution to the self-energy
is shown as a dotted line.
As can be seen from the figure, the
continuum contribution to the scattering rate sets in above the 
structures which are induced by the mode. 
It also contributes considerably
to the renormalization factor. 
As mentioned above, the renormalization does not
decay up to energies of 200 meV, consistent with experiment. 
At the nodal point, the
modification due to the continuum relative to the mode part is strongest.
The importance of the continuum contribution can be seen by
noting the strong similarity of the lower right hand panel of
Fig.~\ref{Cont_Mode} to self-energies extracted from ARPES data along
the nodal direction. \cite{Kaminski99,Kaminski00}

Finally, note that in the normal state, the even channel stays gapped.
That means that at the $N$ point, the self energy for scattering between
bonding bands and between antibonding bands (but not between bonding
and antibonding) is similar to one half 
the continuum contribution (dotted line)
in the right panel of Fig.~\ref{Cont_Mode}.
This will induce a weaker kink feature in the normal state 
at an energy equal 
to the even channel (optical) gap in the spin susceptibility, which
is around 50-60 meV. Correspondingly, the high energy renormalization
will be present in the normal state, but weaker. The difference between
the high energy renormalization in the normal and superconducting states
is mainly due to the appearance of a continuum gap 
in the {\it odd } channel. The low energy renormalization 
is mainly due to the appearance of the mode in the odd channel.

\subsection{Renormalization of EDC and MDC dispersions }

\begin{figure}
\centerline{
\epsfxsize=0.25\textwidth{\epsfbox{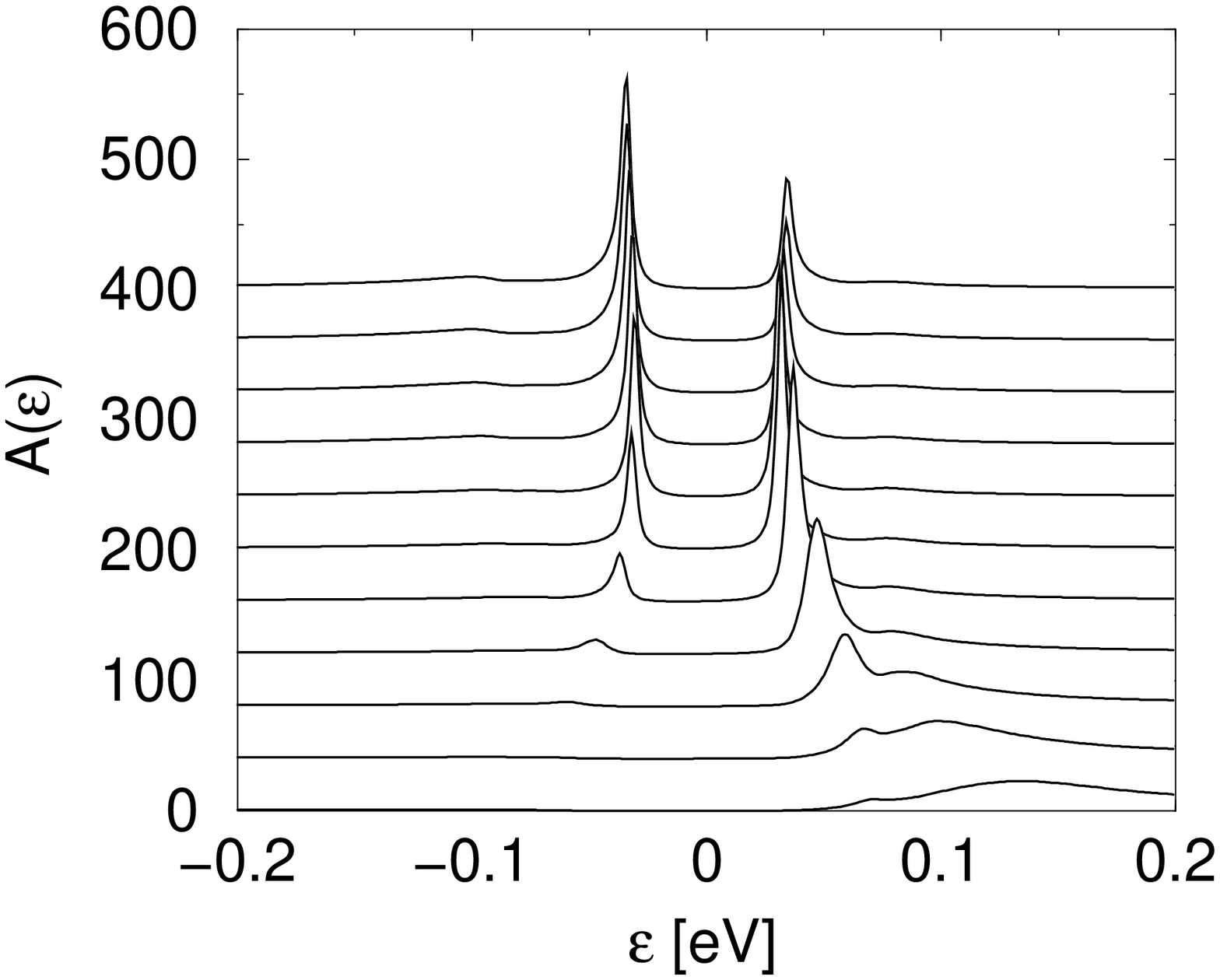}}
\epsfxsize=0.25\textwidth{\epsfbox{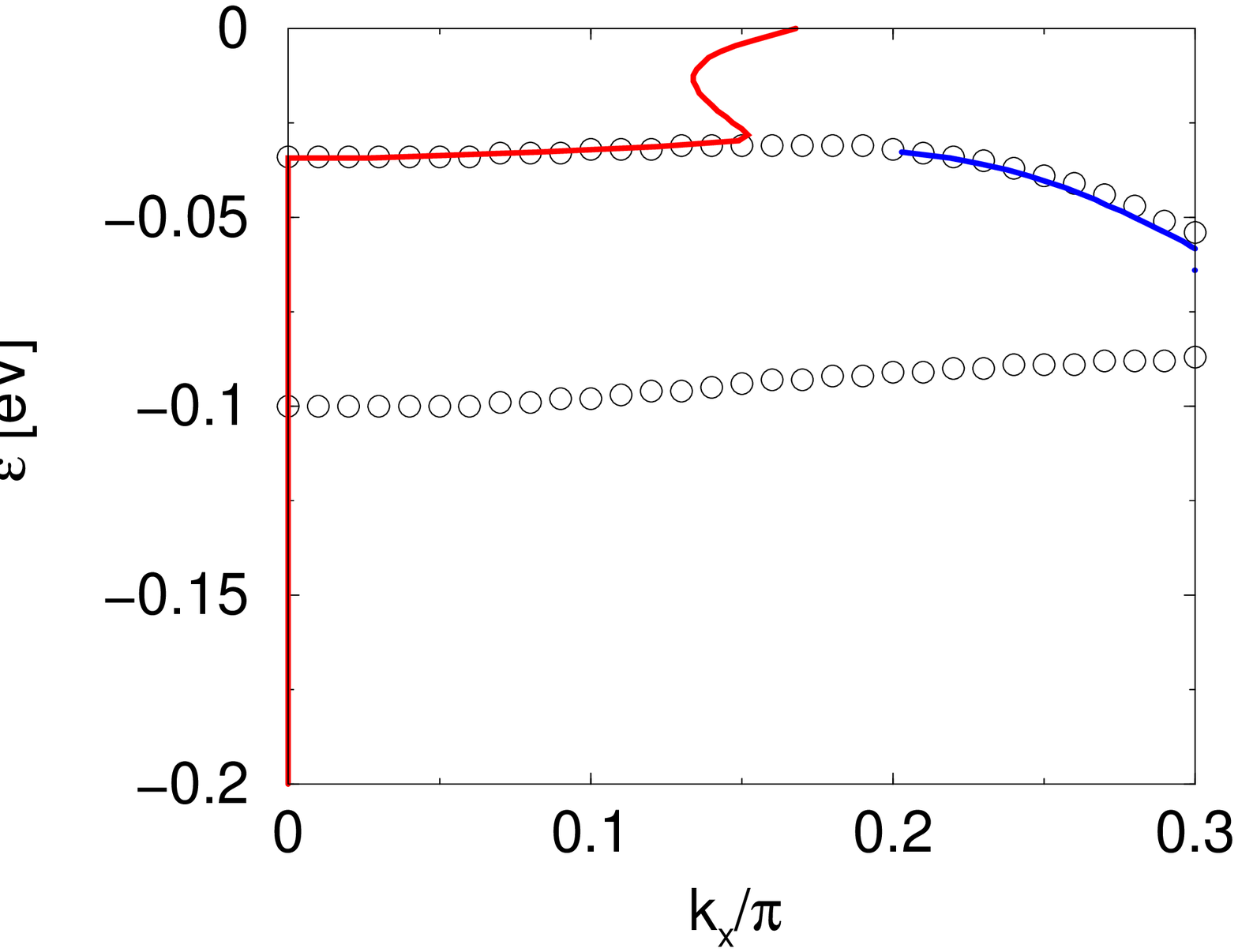}}
}
\caption{
\label{QX1}
Left: Dispersion of the spectral intensity and lineshape 
as a function of momentum along the $M-Y$ cut, ($k_y=\pi $,
$k_x=0...0.4\pi$ in steps of $0.04\pi $ from top to bottom). 
Right: EDC (circles) and MDC (curve) dispersions from maxima of
the curves shown in the left panel. In the EDC dispersion, the low
energy peak and the high energy hump with the break feature in
between is clearly visible. Because the bottom of the normal
state dispersion is at $\xi_M=-34 $meV, the MDC shows only a
broad maximum at $M$ for high energies.
}
\end{figure}

\begin{figure}
\centerline{
\epsfxsize=0.25\textwidth{\epsfbox{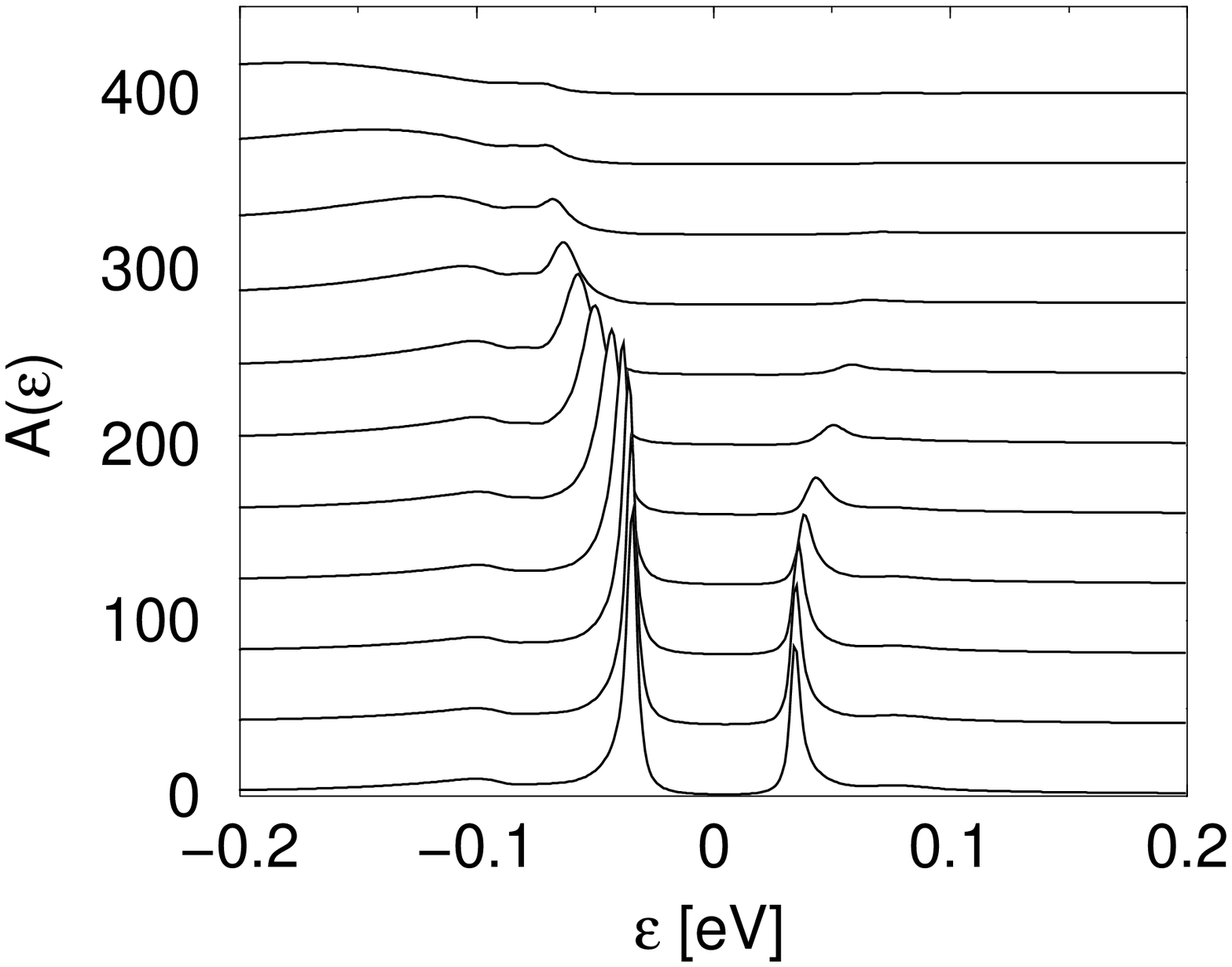}}
\epsfxsize=0.25\textwidth{\epsfbox{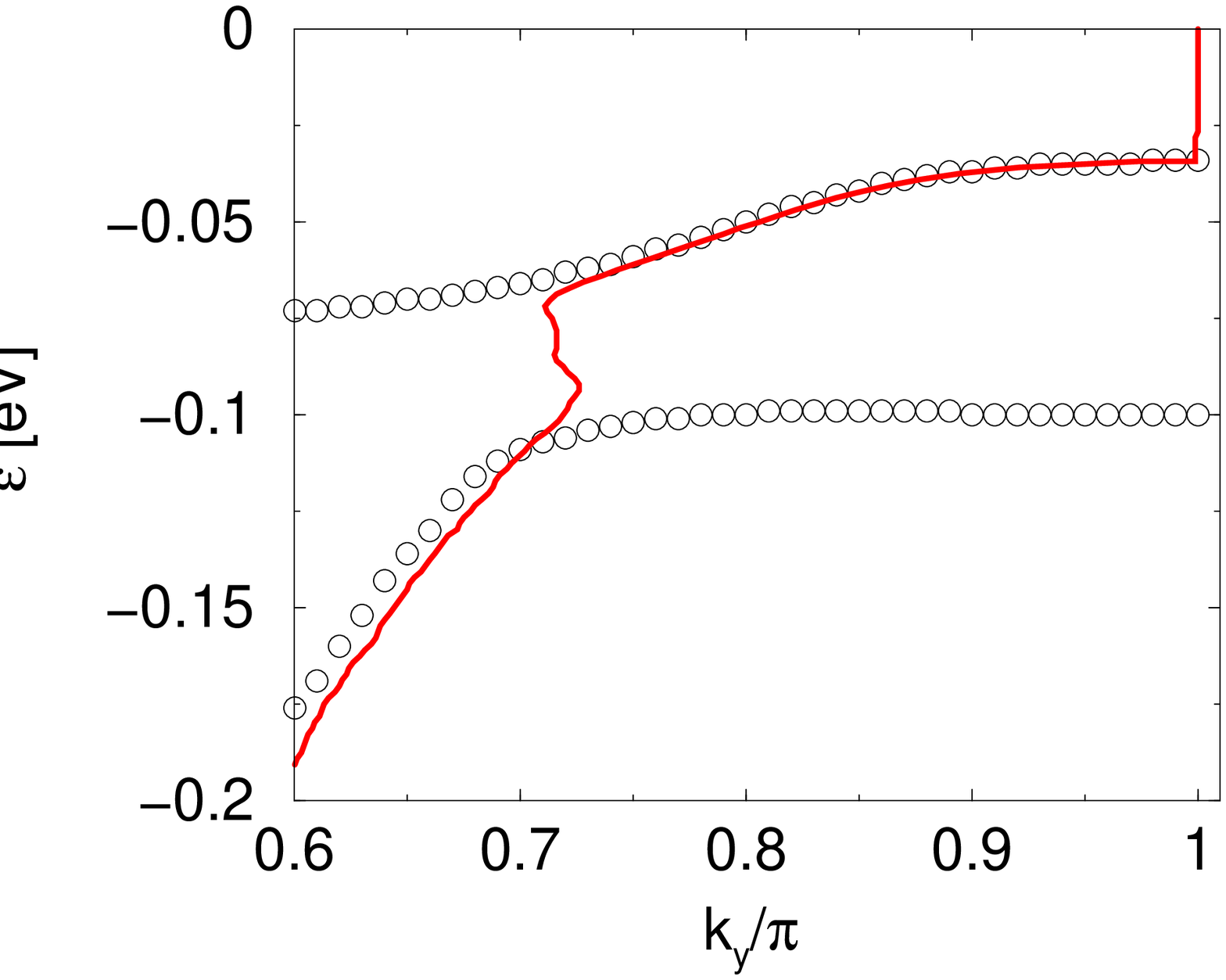}}
}
\caption{
\label{QY2}
Left: Dispersion of the spectral intensity and lineshape 
as a function of momentum along the $M-\Gamma $ cut
($k_x=0 $,
$k_y=0.6\pi..\pi$ in steps of $0.04\pi $ from top to bottom).
Right:
EDC (circles) and MDC (curve) dispersions from maxima of
the curves shown in the left panel.
}
\end{figure}

\begin{figure}
\centerline{
\epsfxsize=0.25\textwidth{\epsfbox{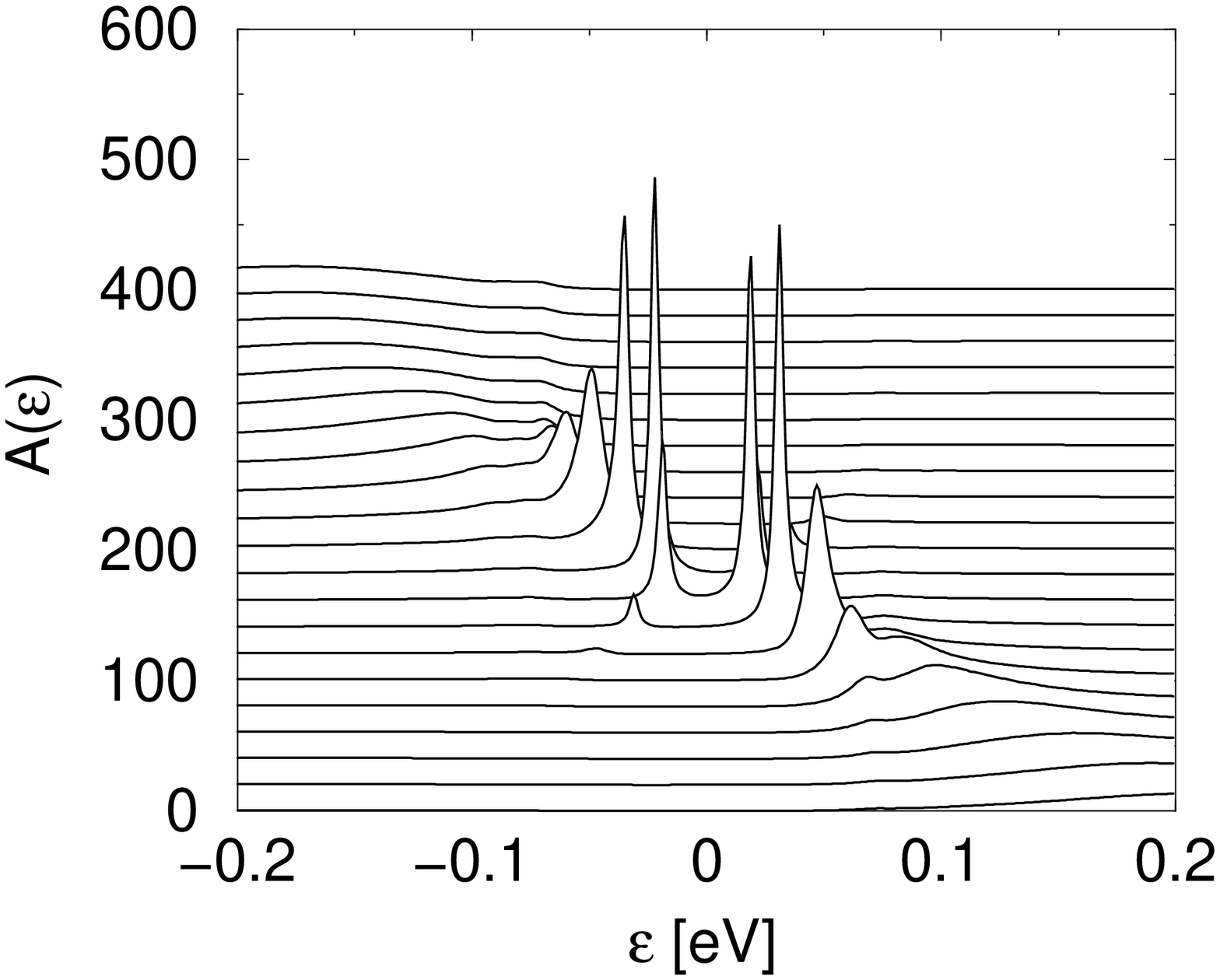}}
\epsfxsize=0.25\textwidth{\epsfbox{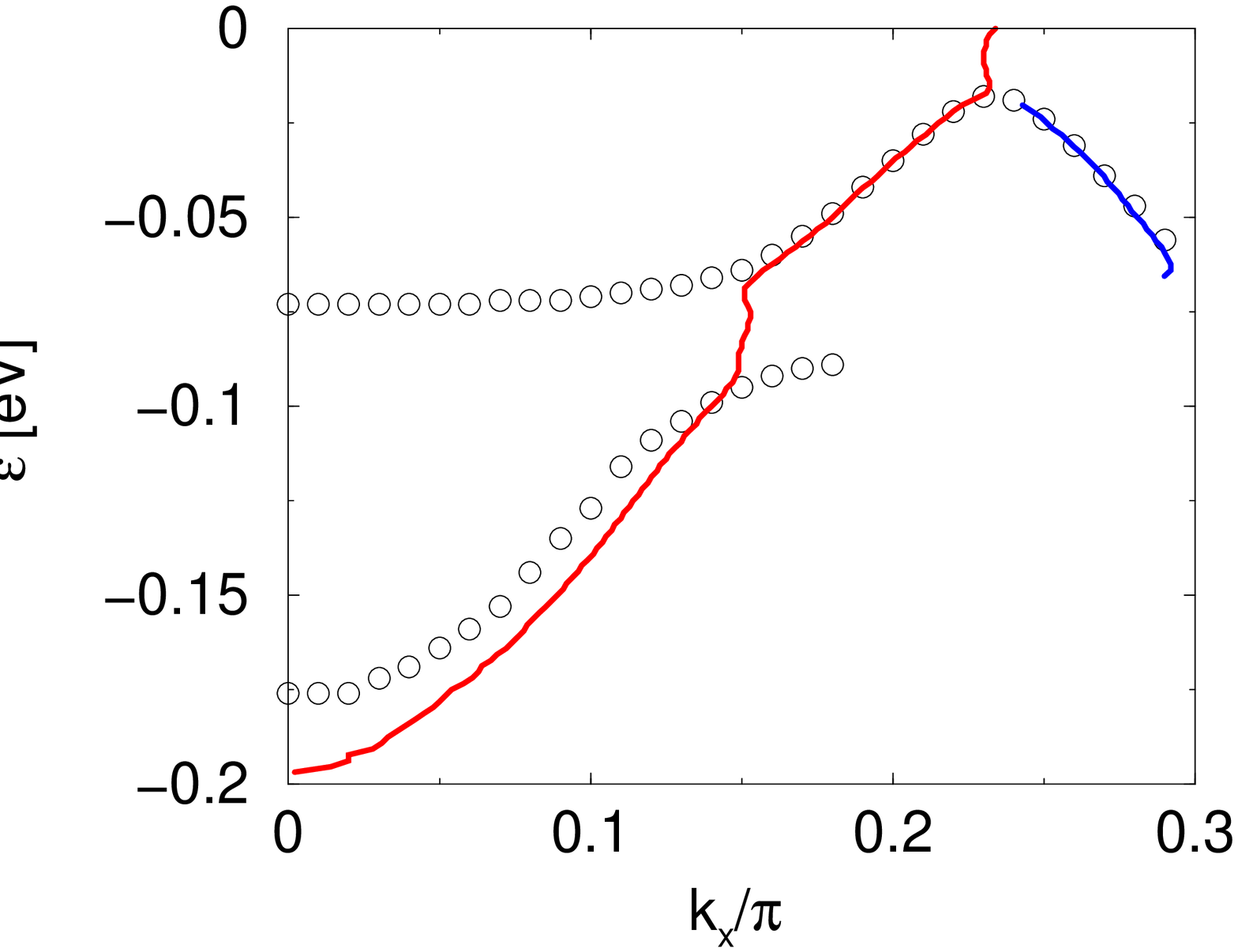}}
}
\caption{
\label{QX2}
Left: Dispersion of the spectral intensity and lineshape 
as a function of momentum $k_y=0.6\pi $,
$k_x=0...0.4\pi$ in steps of $0.02\pi $ from top to bottom. 
Right: EDC (circles) and MDC (curve) dispersions from maxima of
the curves shown in the left panel. 
}
\end{figure}

\begin{figure}
\centerline{
\epsfxsize=0.25\textwidth{\epsfbox{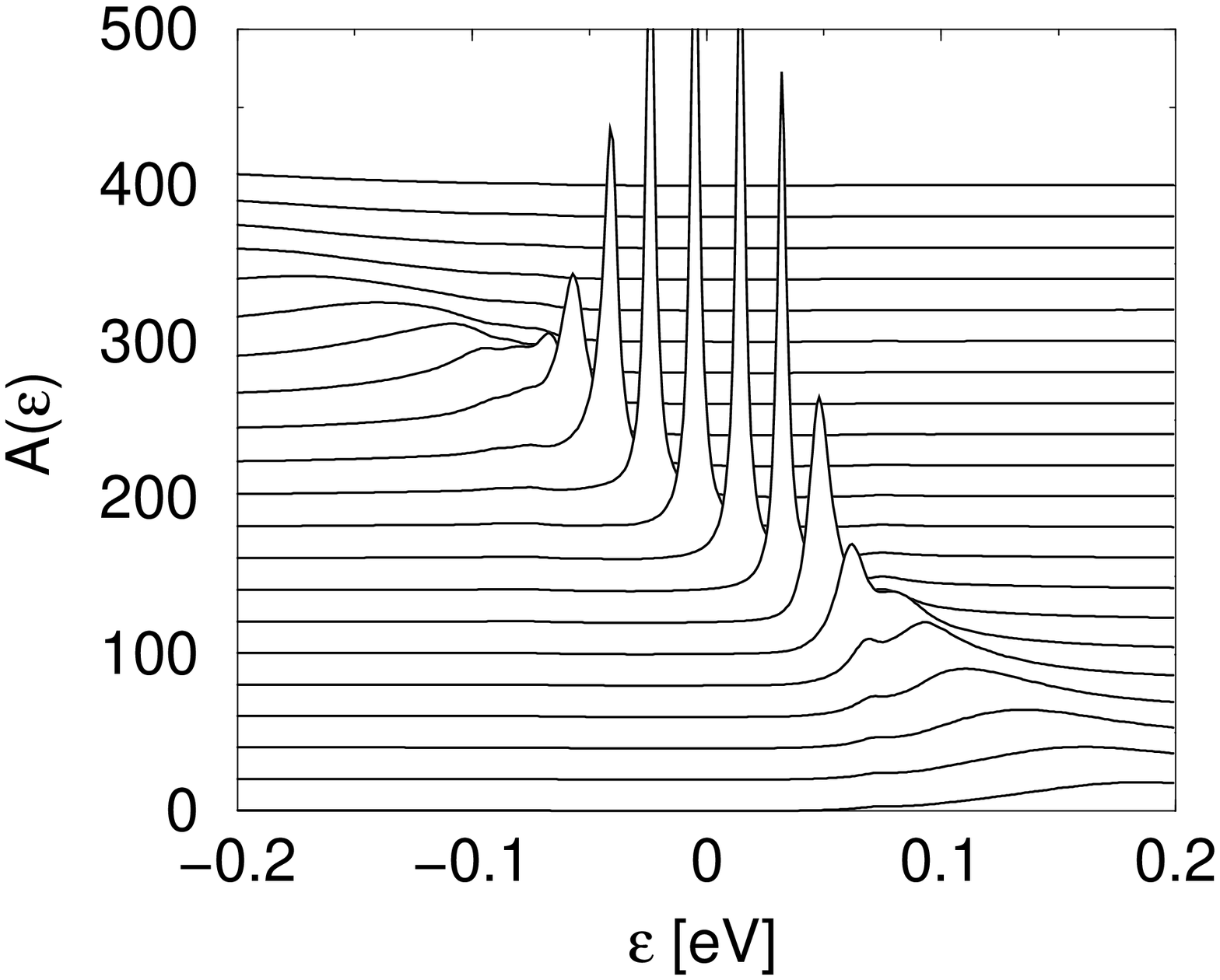}}
\epsfxsize=0.25\textwidth{\epsfbox{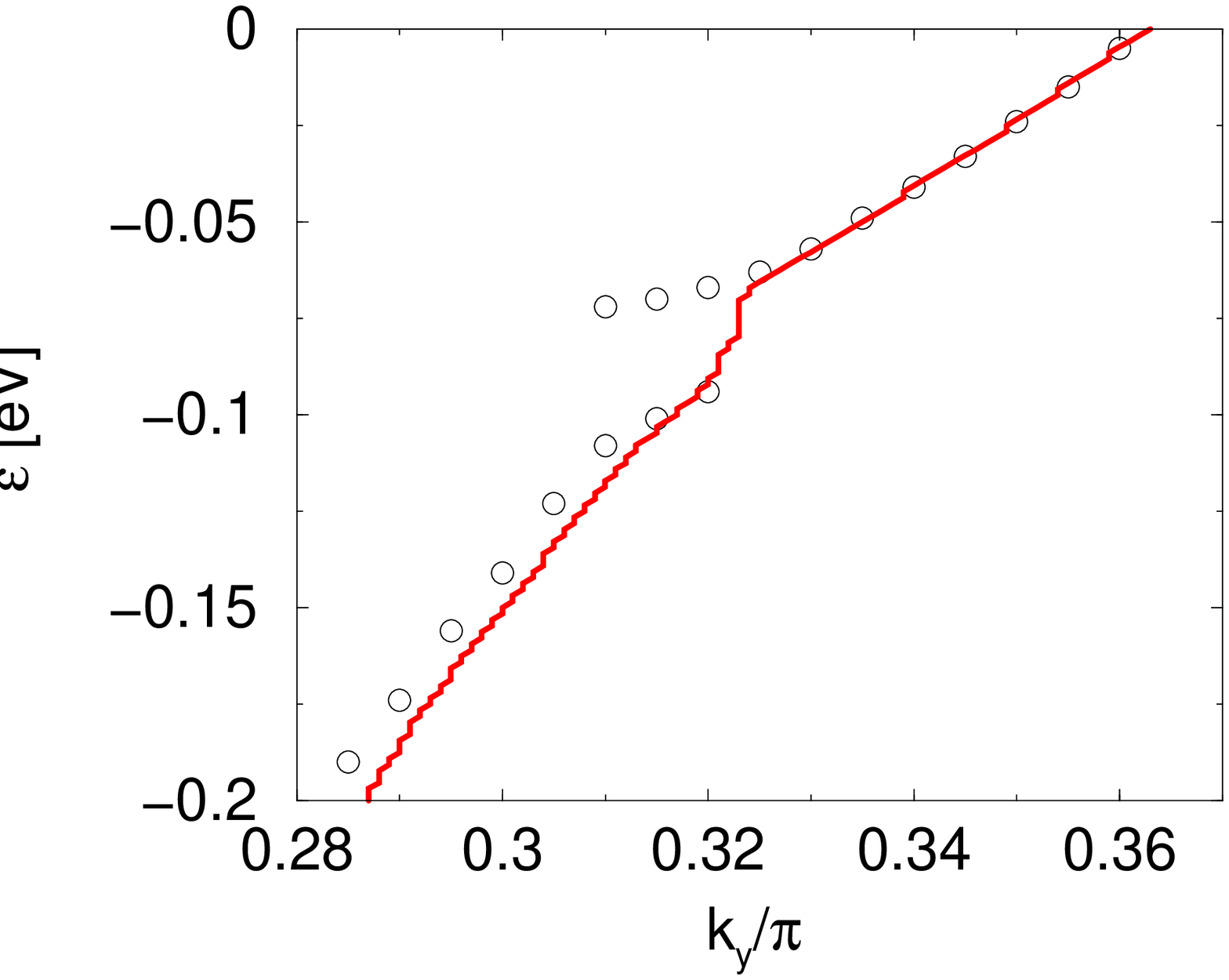}}
}
\caption{
\label{NODE2}
Left: Dispersion of the spectral intensity and lineshape
in the nodal direction ($\Gamma-Y$) as a function
of momentum $k_x=k_y=0.25\pi...0.45\pi$ in steps of $0.01\pi$ from
top to bottom.
Right: the corresponding EDC (circles) and MDC (curve) 
dispersions.
The kink is most clearly seen in the MDC dispersion. The low energy 
velocity is roughly half the high energy one. The high energy dispersion does 
not extrapolate to the Fermi surface crossing.
}
\end{figure}

In the following, we discuss the dispersion of the spectral lineshape
through the Brillouin zone and study the corresponding EDC (as determined
from the spectral maximum as a function of energy) and MDC (as determined from 
the spectral maximum as a function of momentum)
dispersions. We include both the mode and the gapped continuum of the
spin fluctuation spectrum.  In Figs.~\ref{QX1}-\ref{NODE2}, we show
dispersions of the ARPES spectra along several selected paths in the
Brillouin zone. In the left panels of the figures, 
the intensities and spectral lineshapes
can be followed, and in the right panels, the corresponding dispersions of
the peak maxima and hump maxima in the EDCs are shown as circles, and the
maxima in the corresponding MDC dispersions as curves.
A general remark concerns the linewidth of the high energy features 
compared to the low energy features. Due to the strong self energy
damping effects setting in above the dip energy (Fig.~\ref{Cont_Mode}),
the hump features are considerably broader than the peak features for
all momenta in the Brillouin zone. This holds for both EDC and MDC
dispersions. Note that even without taking into account the
lifetime effects due to the spin fluctuation continuum,
the high energy features are much broader in energy
than the low energy features.\cite{Eschrig00}
To account for the experimental MDC linewidth, however, one has to take
into account the continuum contribution.

Starting with Fig.~\ref{QX1}, we follow the dispersion along a cut going
from the $M$ point of the Brillouin zone towards the $Y$ point. The $A$
point corresponds to spectra roughly in the middle of the set.
From the left panel, we see that sharp peaks are restricted
to the momentum regions between the $M$ and $A$ points. The dip structure
is maximal at the $M$ point and much weaker at the $A$ point.
The corresponding dispersion, shown in the right panel,
reproduces the experimental findings\cite{Kaminski00} of two almost
dispersionless EDC branches, one for the peak and one for the hump.
The MDC follows the peak branch, then shows a nontrivial variation at
energies within the gap edge. This behavior is discussed in 
Ref.~\onlinecite{Norman01a}.
The Fermi crossing is only slightly shifted with respect to the
unrenormalized value of $k_x=0.18 \pi$. At higher energies, the
MDC is peaked at $M$.

Going from the $M$ point in the direction of the $\Gamma $ point,
the corresponding dispersion of the ARPES spectra is shown in
Fig.~\ref{QY2}. On the left side, one can see that the intensity
of both the peak and the hump is almost unaffected in the region between the
$M$ point and roughly $0.3\pi$ from there in direction of $\Gamma $.
In this range, the renormalized EDC dispersion of the hump is extremely flat, as
seen in the right panel, and the peak shows a moderate dispersion,
becoming almost flat between $q_y = 0.9\pi $ and $q_y=\pi$. 
When going further away from the $M$ point, the intensity of the peak
drops sharply, and a strong dispersion of the hump sets in.
There is a clear break between the peak and the hump EDC dispersion due 
to the dip.
The MDC along this cut follows the peak near the $M$ point, but 
changes over to the hump dispersion at roughly the point where the hump
starts to disperse strongly away from the chemical potential. In this
range, at energies between 70 meV and 100 meV, the MDC dispersion is
almost vertical, with a weak S-like shape.
We draw the attention to the fact that the hump shows a weakly {\it positive}
dispersion close to the $M$ point, with point of closest approach to the
chemical potential at $q_y\approx 0.85\pi$. This effect is due to
the coupling of the $(\pi,0)$ and $(0,\pi)$ points by self energy
effects, and is a result of the fact that going towards $\Gamma $ from
the $(\pi,0)$ point means going towards $Y$ from the $(0,\pi)$ point at
the $(\pi,\pi)$-displaced wavevector. As a result of this, the weakening 
of the self energy effect along the $M-\Gamma $ cut leads to a
{\it minimum} in the hump dispersion at the $M$ point.
This effect was experimentally found.\cite{Campuzano99}

In Fig.~\ref{QX2}, we show our results for a cut parallel to the $M-Y$
cut shown in Fig.~\ref{QX1}, keeping $q_y=0.6\pi $ constant.
At low energies, the spectral evolution, seen on the left part of
the figure, shows the typical BCS mixing between particle and hole
states. Concentrating on the negative energy parts,
again two branches are present, the peak branch and the hump branch,
separated by a break in the EDC dispersion. Both branches now show
considerable dispersion, but still overlap in momentum. The MDC dispersion
changes from the low energy peak branch to the high energy hump
branch at roughly the point where the intensity of the peak drops
dramatically. Note that the EDC and MDC dispersions are considerably displaced
relative to one another at high energies. Also at low energies, the MDC 
dispersion is stronger
near the break region than the EDC dispersion. This effect increases
when the residual width of the quasiparticle peak increases, and when
convolution with the experimental resolution function is taken into
account.\cite{Norman01a}

Finally, we discuss the cut along the nodal direction, shown in 
Fig.~\ref{NODE2}.
For this direction, the gap is zero as a consequence of $d$ wave symmetry,
and as a result the EDC dispersion should cross the Fermi energy.
This is seen in the left panel of the figure. Note the very strong
damping of the spectral peak as soon as it crosses the energy region
which corresponds to the break effect near the $M$ point. Actually,
the damping starts at slightly lower energies, due to the onset of
node-node scattering processes at an energy $\Omega_{res}$, as can be
seen in the left panel of Fig.~\ref{NODE2}.
The velocity renormalization for low energies and high energies differs
by a factor of roughly two, both for EDCs and MDCs, in
agreement with experiment.\cite{Kaminski00} Finally, we also reproduce
the experimental fact that the high energy dispersion does not
extrapolate to the Fermi crossing.\cite{Bogdanov00,Lanzara01} Again, note some 
shift between the EDC and MDC dispersions at high energies due to the energy 
variation of the self energy.

Clearly, the velocity break (kink) along the nodal direction and the
break between the peak and hump (dip) near the $M$ point
are occurring in the same energy range between $-\Omega_{res}-\Delta_A$
and $-\Omega_{res}-E_M$. This is an appealing result of our
theory, because it explains all features in the dispersion anomalies
in the Brillouin zone seen by ARPES with a simple model.

\subsection{Tunneling spectra}
\begin{figure}
\centerline{
\epsfxsize=0.25\textwidth{\epsfbox{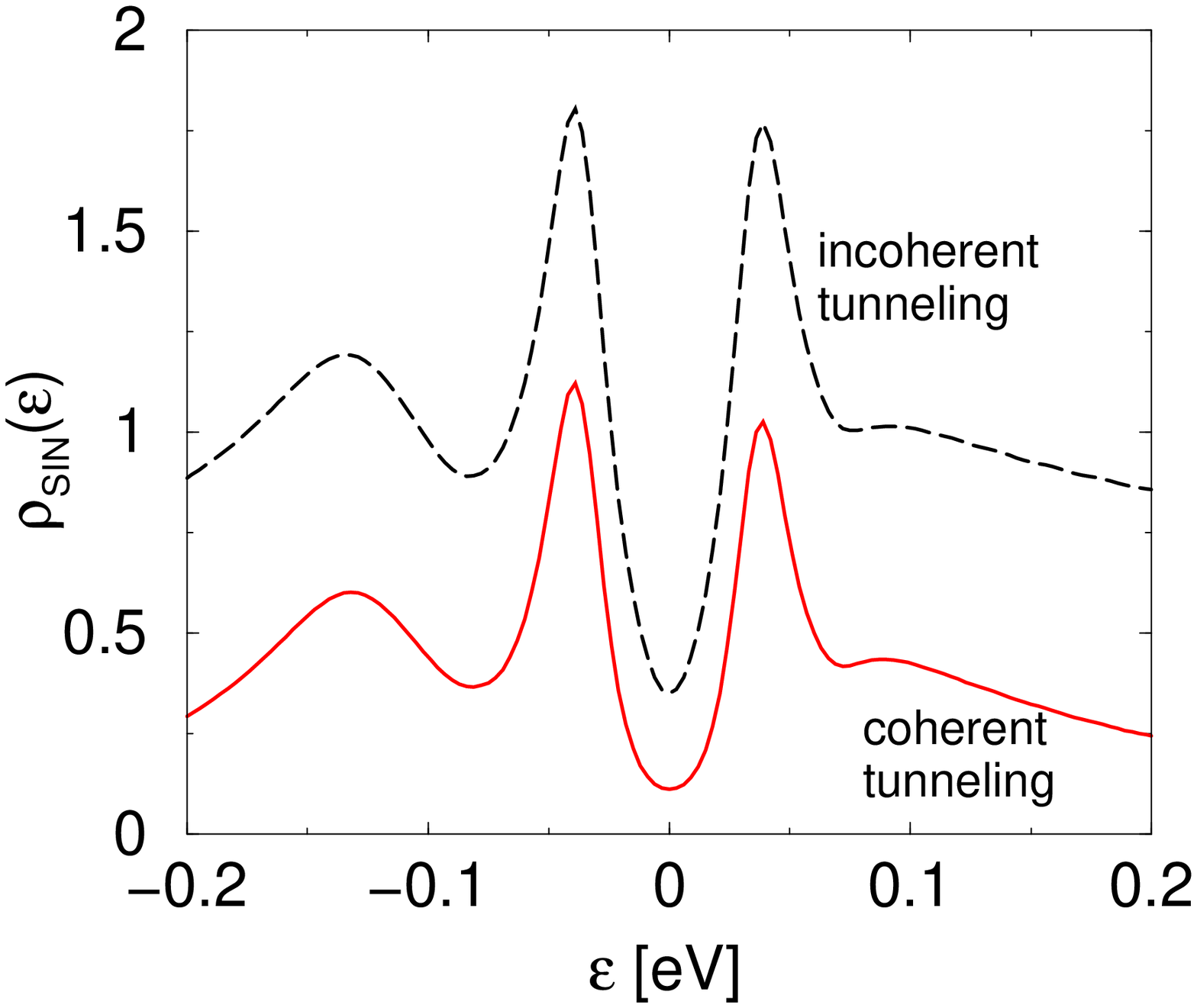}}
\epsfxsize=0.25\textwidth{\epsfbox{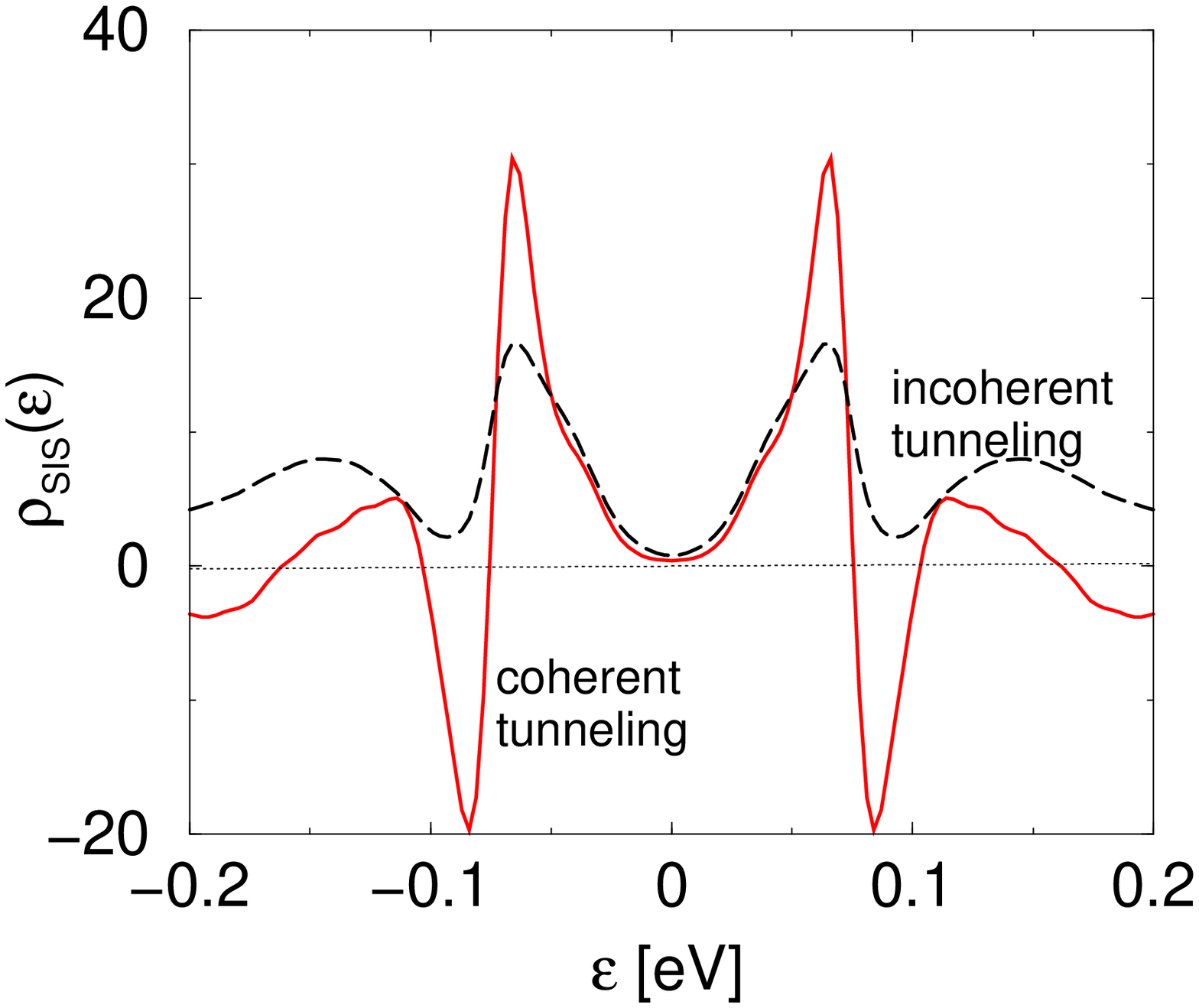}}
}
\caption{
\label{Tunneling} 
Differential tunneling conductance for SIN (left) and SIS (right) tunnel
junctions for $T=40 $ K. Units are $eM_i^2$ for SIN and $2e^2T_i^2$ for
SIS. Results for the coherent (full curves) and incoherent (dashed curves)
tunneling limits are shown. The parameters are given in Table \ref{tab2}.
}
\end{figure}

Knowing the spectral function, $A(\epsilon, \vec{k})$, throughout the zone,
we are able to calculate the tunneling spectra
given a tunneling matrix element $T_{\vec{k}\vec{p}}$.
For simplicity, we present numerical results for the simple model, neglecting
the continuum part of the spin fluctuation spectrum.
From the SIN tunneling current $I(V)$,
one obtains the differential conductance, $dI/dV$.
As usual, we neglect the energy dependence of the SIN matrix element
$|M_{\vec{k} }|^2 = 2e\sum_{\vec{p}}|T_{\vec{k}\vec{p}}|^2
A_N(\vec{p}, \epsilon )$, where $A_N$ is the spectral function of the normal
metal. The SIN tunneling current is then given by
\begin{equation}
I(V)=\sum_{\vec{k} } |M_{\vec{k}}|^2
\int_{-\infty}^\infty \frac{d\epsilon}{2\pi}
A(\epsilon,\vec{k})
\left\{f(\epsilon )-f(\epsilon+eV) \right\}
\end{equation}
We model the tunneling matrix element for two extreme cases: for
incoherent tunneling we assume a constant
$|M_{\vec{k} }|^2 =M_0^2$,
whereas for coherent tunneling we use
$|M_{\vec{k} }|^2=\frac{1}{4}M_1^2( \cos k_x - \cos k_y )^2$.
\cite{Chakravarty93}
Coherent tunneling in the c-axis direction is
strongly enhanced for the $M$ points
in the Brillouin zone compared to the regions near the zone
diagonal due to the matrix elements.\cite{Chakravarty93}
Our numerical results for SIN junctions are shown in Fig.~\ref{Tunneling} 
(left).  In both cases, we
observe a clear asymmetry, with a dip-hump structure on
the negative bias side and a very weak feature on the positive side of the
spectrum, as in experiments.\cite{Renner95,DeWilde98}
The low energy behavior of the tunneling spectrum in the coherent tunneling
limit does not show the characteristic linear in energy behavior 
for $d$-wave, because the nodal electrons have suppressed tunneling
as a result of the matrix elements. The 
peak-dip-hump features, on the other hand, are not affected by the matrix
elements, as they are dominated by the $M$ point regions which
are probed by both coherent and incoherent tunneling.

For an SIS junction, the single particle tunneling current is given in
terms of the spectral functions by
\begin{eqnarray}
I(V)=2e\sum_{\vec{k} \vec{p}} |T_{\vec{k}\vec{p}}|^2
\int_{-\infty}^\infty \frac{d\epsilon }{2\pi }
A(\epsilon, \vec{k}) A(\epsilon+eV, \vec{p}) \nonumber \\
\times \left\{ f(\epsilon ) - f(\epsilon + eV) \right\}
\end{eqnarray}
Again we show results for incoherent tunneling ($|T_{\vec{k}\vec{p}}|^2=T_0^2$)
and for coherent tunneling
with conserved parallel momentum,
$|T_{\vec{k}\vec{p}}|^2=\frac{1}{16 }T_1^2 ( \cos k_x - \cos k_y )^4
\delta_{\vec{k}_{||},\vec{p}_{||}}$.\cite{Chakravarty93}
Our results are shown in the right panel of Fig.~\ref{Tunneling}.

All structures are symmetric around the chemical potential. The
low energy part of the spectrum is strongly suppressed in the
incoherent tunneling limit already, thus there is no big difference
to the coherent tunneling limit there. At higher voltages, however,
in the coherent tunneling limit, we obtain negative differential 
conductance.  Such an effect was observed 
recently in optimally doped Bi$_2$Sr$_2$CaCu$_2$O$_{8-\delta}$
break junctions.\cite{Zasadzinski01}
We also observe negative behavior at higher bias in the coherent tunneling
limit, but note that in reality, the tunneling matrix element will have
both coherent and incoherent contributions (especially at higher voltages), 
and thus will be a weighted
average of the dashed  and full curves in Fig.~\ref{Tunneling}. In this
case, most probably only the negative behavior below 100 meV will be observable.
We note that the spin fluctuation continuum broadens the spectral functions and,
as we show below, this leads to
a positive response at higher voltages.

We give approximate expressions for the SIS differential conductances
for zero temperature. In the incoherent limit,
\begin{equation}
\label{inc}
I^{(incoh)}(V)= \frac{eT_0^2}{\pi}
\int_{-eV}^{0} d\epsilon
N(\epsilon ) N(\epsilon+eV)
\end{equation}
In the coherent tunneling limit, the tunneling matrix element very effectively
suppresses the nodal regions, thus only allowing for tunneling near the $M$
point regions. In these regions, however, the dispersion is weak, so that we
may approximate the spectral function by 
its value at the $M$ point, $A_M(\epsilon )$. Then, we
obtain in the coherent tunneling limit
\begin{equation}
\label{coh}
I^{(coh)}(V)\approx  \frac{eT_1^2}{\pi}
\int_{-eV}^{0} d\epsilon
A_M(\epsilon ) A_M(\epsilon+eV)
\end{equation}
with $T_1^2=\sum_{\vec{k}\vec{p}} |T_{\vec{k}\vec{p}}|^2$. 
Note that two different quantities are probed in the two limits. 
In the incoherent limit, it is the density of states, and in 
the coherent limit, it is the {\it spectral function} at the $M$ point
of the Brillouin zone. 

It is easy to show by differentiating
Eq.~\ref{coh} that the differential conductance can be negative,
and furthermore, can approach a negative value for large voltages.
The limiting behavior at high voltages
in the incoherent tunneling limit is proportional to $N(-)N(+)$, where
$N(\pm )$ is the density of states at large positive/negative energies.
If in the coherent tunneling limit the corresponding term
proportional to $A_M(+)A_M(-)$ is very small, then the main contribution
comes from the region where either $\epsilon \approx -\Delta_M $
or $\epsilon +eV\approx \Delta_M$, varying within a range of order $\Delta_M$
around these values. It is easy to show that this contribution is negative.
But as soon as incoherent
contributions play any role, or if $A_M$ has a considerable 
incoherent part, then their positive contributions 
will dominate at high voltages. 
Note that for SIN tunneling, the differential conductance is always positive 
definite.

\subsection{Self consistency issues}
When going towards underdoping, the spectral function deviates considerably
from the bare BCS spectrum. Self consistency issues become important then.
 
Our studies have shown that the quasiparticle peak is always reasonably well
separated in energy from the high energy incoherent part by a dip.
By coupling electrons to the spin resonance mode, weight
is shifted from the quasiparticle peak to the incoherent part
which includes the broad hump structure. Thus, when calculating the self
energy effects due to this coupling, only the quasiparticle peak part of the
spectrum with its reduced weight
will contribute to the sharp self energy features at energies
affecting the quasiparticle peak. The incoherent part of the
fermionic spectrum, which is gapped by roughly the hump energy, will
affect the low energy quasiparticle properties only in form of an
effective high energy renormalization factor, which is constant up to
energies comparable to the hump energy. This effective renormalization
adds to the one due to coupling of electrons to the spin fluctuation
continuum. Thus, we can concentrate on the renormalization equations
following from the set of equations which includes the quasiparticle
peak spectrum of reduced weight interacting with the spin fluctuation mode.
In deriving these equations, we make use of the approximate
equations for the renormalization functions derived above.

The quasiparticle part of the Green's function has in this approximation
at the $M$ point the form
\begin{eqnarray}
G^R_{\epsilon, k_M}=
\frac{1}{\tilde Z_M} \left(
\frac{\tilde A^{+}_{M}}{\epsilon-\tilde E_{M}+i\tilde\Gamma_M}
+\frac{\tilde A^{-}_{M}}{\epsilon+\tilde E_{M}+i\tilde\Gamma_M} \right) \\
F^R_{\epsilon,k_M}=
\frac{1}{\tilde Z_M} \left(
\frac{\tilde C_{M}}{\epsilon-\tilde E_{M}+i\delta_M}
-\frac{\tilde C_{M}}{\epsilon+\tilde E_{M}+i\delta_M} \right)
\end{eqnarray}
where  $\tilde E_M=\sqrt{\tilde \xi_M^2+\tilde \Delta_M^2}$
and $\tilde A^{\pm}_{M} = (1\pm \tilde \xi_M/\tilde E_M)/2$.
Here, $\tilde E_{M}$ is the measured peak position at the $M$ point, and
$\tilde\Gamma_M$ is the quasiparticle peak width. The broadening of the
off-diagonal spectra, $\delta_M$, is reduced compared to $\tilde\Gamma_M$ due to
$d$-wave symmetry.
Using the approximative formulas from the last sections
at $\epsilon = -\tilde E_M$, we obtain
(with $\alpha = g^2I_0/\pi $)
\begin{eqnarray}
\tilde Z_M=1+\frac{\lambda_M^{(N)}}{\tilde Z_N}+
\lambda_M^{(c)}+
\frac{\alpha}{\tilde Z_M\Omega_{res}\tilde E_M}
\frac{\tilde E_M}{\Omega_{res}+2\tilde E_M} \\
\tilde \Delta_M=\frac{\Delta_M}{\tilde Z_M}+
\frac{\alpha}{\tilde Z_M^2\Omega_{res} \tilde E_M}
\frac{\Omega_{res}+\tilde E_M}{\Omega_{res}+2\tilde E_M}\tilde \Delta_M \\
\tilde \xi_M=\frac{\xi_M}{\tilde Z_M}-
\frac{\alpha}{\tilde Z_M^2\Omega_{res}\tilde E_M}
\frac{\Omega_{res}+\tilde E_M}{\Omega_{res}+2\tilde E_M}\tilde \xi_M
\end{eqnarray}
where $\lambda_M^{(c)}$ denotes renormalizations due to the spin fluctuation
continuum and the incoherent part of the spectral function,
and $\lambda_M^{(N)}$ the contribution coming from the nodal regions (these
contributions are renormalized with the nodal renormalization factor 
$\tilde Z_N$, which is smaller than $\tilde Z_M$).
The last two equations merely express the measurable quantities
$\tilde \Delta_M$ and $\tilde \xi_M$ as functions of the bare quantities
$\Delta_M$ and $\xi_M$. The first equation can
be solved, giving for small $\Omega_{res}$ 
and not too small $g^2I_0$
a quasiparticle weight
$\propto \sqrt{\Omega_{res}(\Omega_{res}+2\tilde E_M)}$.
Note that we derived this set of equations for the case where $\tilde \Gamma_M$
is neglected, which describes the slightly underdoped region.
When $\Omega_{res}$ becomes comparable to
$\tilde \Gamma_M$, these equations have to be modified.

It should be remarked, though, that using these equations in the absence 
of vertex corrections usually give poorer results than those presented in 
this paper using bare Green's functions.\cite{Vilk97}

\subsection{Bilayer splitting}

For bilayer compounds, the dispersion can be split into
bonding ($b$) and antibonding ($a$) bands. Accordingly, the self energy for
each band is defined as $\Sigma_{k,\epsilon }^{(b)}$ and
$\Sigma_{k,\epsilon }^{(a)}$.
Similarly, the spin susceptibility is now a matrix in the bonding-antibonding
indices, having elements diagonal ($\chi_{aa}$, $\chi_{bb}$)
and off-diagonal $(\chi_{ba}$, $\chi_{ab}$) in the bonding-antibonding 
representation. The components of the spin susceptibility transforming
even and odd with respect to the plane indices are given by
$\chi_e=\chi_{aa}+\chi_{bb}$ and $\chi_o=\chi_{ab}+\chi_{ba}$.
For identical planes, we have $\chi_{aa}=\chi_{bb}$ and 
$\chi_{ab}=\chi_{ba}$.
The measured susceptibility is then given by
\begin{equation}
\chi=\chi_e \cos^2 \frac{q_zd}{2} + \chi_o \sin^2 \frac{q_zd}{2}
\end{equation}
where $d$ is the separation of the layers within a bilayer.
If we write the self energy for a single layer as 
$\chi \ast \hat G$
(the hat denotes the 2x2 particle hole space), which
is a functional of the spin susceptibility
$\chi $ and the Gor'kov-Green's function $\hat G$,
then we have formally for the two-layer system
\begin{eqnarray}
\label{bilayer}
\hat\Sigma^{(b)} &=& \chi_{e} \ast \hat G^{(b)}
+ \chi_{o} \ast \hat G^{(a)}
\nonumber \\
\hat\Sigma^{(a)} &=& \chi_{e} \ast \hat G^{(a)}
+ \chi_{o} \ast \hat G^{(b)}
\end{eqnarray}
For the resonance part, which only has a $\chi_{o}$ component,
this means that fermionic excitations of
the antibonding band determine the self energy for the bonding band
and vice versa. The calculations presented in this paper hold for the case of
bilayer systems if we assume identical dispersions for
bonding and antibonding bands. Even small bilayer splittings
of the order of 10 meV or less do not matter, as they do not
qualitatively alter the spectral form of the self energy.
For larger bilayer splittings, the self energy is larger for
the bonding band, because it is determined by the van Hove singularity
near the chemical potential in the antibonding band. Thus, stronger
renormalizations are expected in the bonding band for this case,
which tends to decrease the bonding-antibonding splitting.
This effect of reducing the bilayer splitting should be strongest
in underdoped compounds, where the effect of the resonance mode
is strongest. In overdoped compounds, the bilayer splitting should
be less affected by spin fluctuations.
Our prediction is that if a bilayer splitting is
observed, then the peak-dip-hump structure should be 
stronger for the bonding band with the higher binding energy peak.
This is consistent with the data of Ref.~\onlinecite{Feng01a}.
The {\it onset} of strong fermionic damping should be independent
of the band index, as it is given by scattering to the nodes,
and thus occurs at the fixed energy $\Omega_{res}$.

In this paper, we have elected not to explicitly include bilayer 
splitting effects in our calculations.  The primary reason is that although 
all ARPES groups now detect the presence of bilayer splitting for 
heavily overdoped samples, the various groups disagree on its presence for 
optimal and underdoped samples \cite{Kaminski02}.   Recently, we have 
performed calculations including bilayer splitting and are able to 
reproduce a number of unusual spectral anomalies seen in heavily 
overdoped ARPES spectra \cite{Eschrig02}.  These calculations further 
confirm the picture advocated in this paper, in that the spectral 
anomalies imply a mode which has odd symmetry with respect to the
layer index of the bilayer, a unique 
property of the magnetic resonance observed by neutrons.  For further 
details, the reader is referred to Ref.~\onlinecite{Eschrig02}.

\subsection{Doping dependence}
In this section, we deal with the doping dependence of the spectral lineshape
near the $M$ point of the Brillouin zone. As there are many parameters which 
change
with doping in different ways, it could turn out to be a meaningless task
to adjust all of those parameters and at the same time make a sensible 
prediction.
But, fortunately,  all changes with doping 
lead to spectral changes which go in the same direction. This `fortuitous'
accident allows us to make some general predictions from the theory we use.
To see this, we turn again to Figs. \ref{figspecom} and \ref{figspecg}.
From there we see that the quasiparticle weight decreases with
decreasing $\Omega_{res}/\Delta_M$, and with increasing coupling constant
$g^2w_Q$.
The quasiparticle scattering rate increases with decreasing 
$\Omega_{res}/\Delta_M$. And the hump energy disperses to higher binding
energies for increasing coupling constant and increasing $\xi_M$.
Thus, in our model, going from overdoping to underdoping amounts
to a decreasing quasiparticle weight, an increasing 
quasiparticle scattering rate, and an increasing hump binding energy.

\begin{figure}
\centerline{
\epsfxsize=0.45\textwidth{\epsfbox{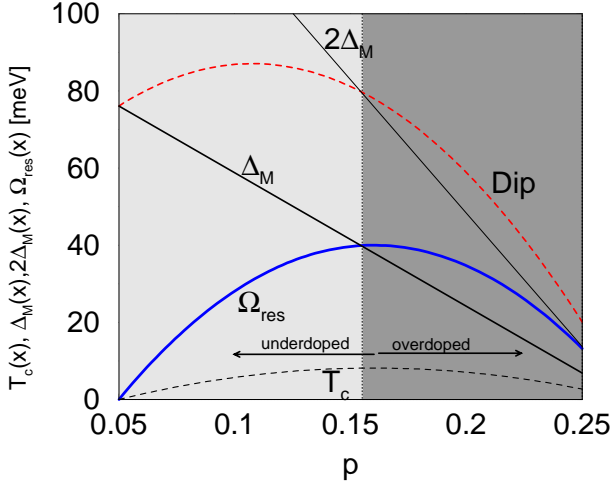}}
}
\caption{
\label{phasediagram}
In the dark gray shaded region, corresponding to overdoping,
quasiparticles peaks are well defined. In the light gray shaded region,
corresponding to underdoping, the peak weight is
strongly reduced, and an incoherent part due to scattering 
from the spin fluctuation mode is dominant. 
The resonance
energy, shown as a thick line, is bounded from above by twice the maximal
gap energy, $\Omega_{res}<2\Delta_{M}$, and approaches it on the 
overdoped side. 
The position of optimal doping, at maximal $T_c$ and $\Omega_{res}$,
roughly coincides with the point where $\Delta_M$ as a function of doping
crosses $\Omega_{res}$.
}
\end{figure}

The important parameter, 
as we see from this study,
is the ratio $\Omega_{res}/\Delta_{M}$,
the ratio of the mode energy to the maximal superconducting $d$-wave gap.
We distinct two regions: the first, where $\Omega_{res}/\Delta_{M}>1$,
and the second, where $\Omega_{res}/\Delta_{M}<1$. 
The situation is schematically shown in the phase diagram in 
Fig.~\ref{phasediagram}.
The curves shown are calculated using the formulas
(we relate $T_c$ to the hole doping level in the Cu-O$_2$ planes
in the usual manner\cite{Presland91})
\begin{eqnarray}
T_c &=& 95\mbox{ K}\left(1 - 82.6 (p-0.16)^2\right)\\
\Delta_{M} &=& 38\mbox{ meV}\left(1 - 9.1 (p-0.16)\right)\\
\Omega_{res}&=& 40 \mbox{ meV}\left(1 - 82.6 (p-0.16)^2\right)
\end{eqnarray}
All these quantities approach zero on the overdoped side at $p=0.27$.
Optimal doping corresponds to $p=0.16$.
Note that $\Omega_{res}=4.9 T_c$ in agreement with 
Ref.~\onlinecite{Zasadzinski01}.
The $\Delta_M$ variation was based on ARPES data.\cite{Campuzano99,Mesot99}

As can be seen, the separation between overdoped and
underdoped regions roughly coincides with the regions where
$\Omega_{res}> \Delta_M$ and $\Omega_{res}<\Delta_M$, respectively.
The dip onset is given by $\Omega_{res}+\Delta_A$. As
$\Delta_A$ is about the same as $\Delta_M$, we have shown in 
Fig.~\ref{phasediagram} the line for $\Omega_{res}+\Delta_{M}$ as a dashed line,
which determines the position of the dip fairly accurately. 
The continuum in the spin fluctuation spectrum 
only affects electrons above $2\Delta_h$, which is near or above
the dip energy. 
One important observation is that the point of optimal doping for a Cu-O$_2$ 
plane roughly corresponds to the point where $\Omega_{res}/\Delta_{M}=1$. Thus, 
region I of Fig.~\ref{Zfac} is relevant to overdoped materials, and region II 
to underdoped materials.
Another experimental observation is that this ratio never exceeds the value 
two.  This is expected for an excitonic collective mode below a continuum 
edge.\cite{Zasadzinski01}

For a quantitative theory of the doping dependence the self consistency
issue becomes important. 
The coherent quasiparticle weight and the quasiparticle linewidth
are given by,
\begin{eqnarray}
&&z_M \approx \frac{1}{\tilde Z_M }
\left(\frac{1}{2}+\frac{|\tilde \xi_M|}{2\tilde E_M} \right) \\
&&\tilde\Gamma_{M} \approx \frac{g^2w_{MN}}{\pi \tilde v_N \tilde v_{\Delta}}
\frac{\tilde E_M - \Omega_{res}}{ \tilde Z_M } 
\Theta (\tilde E_M - \Omega_{res})
\end{eqnarray}
where $\tilde Z_M \equiv \tilde Z_M (-\tilde E_M)$ is the only quantity not 
available from
experiment. We can eliminate it, to obtain the relation
\begin{equation}
\tilde\Gamma_{M} \approx 2z_M\tilde E_M
\frac{g^2w_{MN}}{\pi \tilde v_N \tilde v_{\Delta}}
\frac{\tilde E_M-\Omega_{res}}{\tilde E_M+|\tilde \xi_M|} 
\Theta (\tilde E_M-\Omega_{res})
\end{equation}

Note that experimentally both $z_M\tilde E_M$ and (possibly) 
$\tilde v_{\Delta}$ scale
with $k_B T_c$.\cite{Ding00,Mesot99} So, the quasiparticle width is
dominated by the difference $\tilde E_M - \Omega_{res}$. 
Quasiparticles
are sharp at the overdoped side where $\tilde E_M<\Omega_{res}$, and
an onset of quasiparticle scattering as a function of underdoping
takes place when $\tilde E_M = \Omega_{res}$. This point is slightly
beyond optimal doping.

\section{Coupling to the zone boundary half breathing optical phonon}

The sharp structure in the dispersion needs an explanation in terms of
an almost dispersionless feature which couples to the electrons.
Numerous phonons modes are seen in inelastic neutron scattering in
high $T_c$ cuprates. Most of them do not show indications of strong
coupling to electrons. 
Two special types of phonons have attracted attention: the Cu-O buckling mode,
which is attractive in the $d$-wave 
channel,\cite{Song95,Nazarenko96,Bulut96,Dahm96,Sakai97,Nunner99} and the 
Cu-O breathing mode,
which is repulsive in the $d$-wave 
channel.\cite{Nazarenko96,Bulut96,Dahm96,Nunner99}
Typically, the absolute values of the pairing interactions
in the $B_{1g}$ (`$d$-wave') channel
for both types of vibrations are smaller than 0.1 eV, in the 
$A_{1g}$ (`$s$-wave') channel
about 0.5-1 eV; for spin fluctuations, the corresponding numbers
are in the $d$-wave channel 0.5-1 eV and in the $s$-wave channel
1-2 eV.\cite{Nunner99}
The total electron-phonon coupling constant in the $s$-wave channel
amounts to $\lambda_s \approx 
0.4-0.6$,\cite{Friedl90,Litvinchuk91,Andersen96,Jepsen98,Nunner99}
and in the $d$-wave channel to 
$\lambda_d \approx 0.3$.\cite{Andersen96,Jepsen98}
Thus, phonons are not likely to be responsible for 
the high transition temperature.

It was argued recently that strong coupling of electrons to
the zone boundary half breathing phonon may be responsible for the
anomalies in the dispersion. It is known for some time that this
phonon shows a dispersion which is strongly renormalized midway between the 
zone boundary and the zone center when entering the superconducting state.
These findings show that the zone boundary half breathing phonon is affected
by superconductivity. 
It was suggested to be responsible for the renormalizations of the dispersion
observed in ARPES.\cite{Lanzara01}
This zone boundary half breathing longitudinal optical phonon
is a Cu-O bond stretching mode with an energy between 50 and 100 meV.
Its dispersion is very strong in the middle of the branch, and it was
suggested that a discontinuity develops there in the metallic 
state.\cite{McQueeney99} 
The first measurements concentrated on lanthanum cuprates, but recently
YBa$_2$Cu$_3$O$_{7-\delta}$ was also studied.\cite{Petrov00,McQueeney01}
The displacements involve oscillations of the oxygen atoms
in phase between the two planes in the bilayer.
The results for optimally doped YBa$_2$Cu$_3$O$_{7-\delta}$ are the following:
The dispersion of the zone edge mode in
the superconducting state shows a `break' at $(0,\pi/2)$ (and equivalent
points), with an almost dispersionless branch (at $\sim 55$ meV)
between $\vec q=(0,\pi/2)$ and $\vec q=(0,\pi)$, and
a dispersive branch (68 meV to 72 meV) between
$\vec q=(0,0)$ and $\vec q=(0,\pi/2)$.\cite{McQueeney99,Petrov00,McQueeney01}
Experimental investigation showed that the 
dispersionless branch extends over a region $\pi/2 < q_x < \pi ,
-0.1 \pi < q_y < 0.1 \pi$ (and analogously 
for $q_x $ and $q_y$ interchanged).\cite{McQueeney99}
The dispersionless branch was only observed for bond stretching along
the $a$ direction (perpendicular to the chains).
The dispersions of the longitudinal bond-stretching phonon branches were
found to show no apparent temperature dependence.\cite{McQueeney01}
The phonon intensity was found to show significant temperature dependence
below $T_c$.\cite{Petrov00} 
Phonon weight is transferred from a position
halfway to the zone boundary (in a range between 55 meV and 70 meV) to
the zone center (70-75 meV) and the zone boundary (50-55 meV).
This transfer sets in at $T_c$ and increases with decreasing temperature.

The coupling strength, $g_b(\vec{q})$,
goes to zero for small momentum transfer $\vec{q}$.
Furthermore, in the model of Ref.~\onlinecite{Shen01},
the coupling vanishes near the $\vec{q}=(\pi,\pi)$ point,
thus having minimum impact on the electrons
near the $M$ point of their Brillouin zone.
This is in stark contrast to the resonance mode model, and can certainly
not explain the effects at the $M$ points. It is, however, possible that
they contribute to the renormalization of the nodal dispersion.
The maximal coupling strength was theoretically estimated to
$g_b\approx 0.04 $eV,\cite{Nunner99}, but in some models is enhanced by vertex
corrections.\cite{Shen01}

\begin{figure}
\centerline{
\epsfxsize=0.22\textwidth{\epsfbox{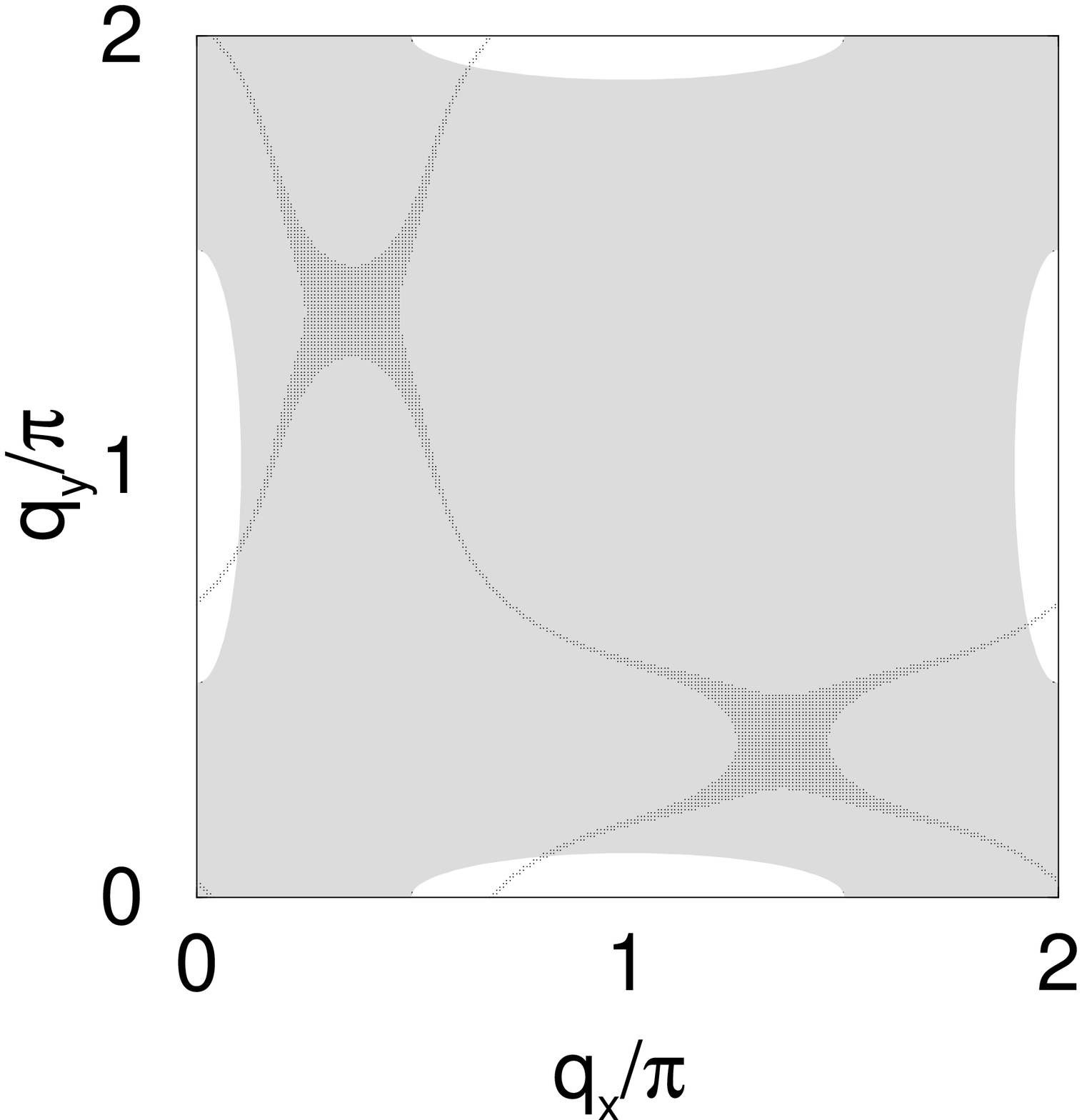}}
\epsfxsize=0.22\textwidth{\epsfbox{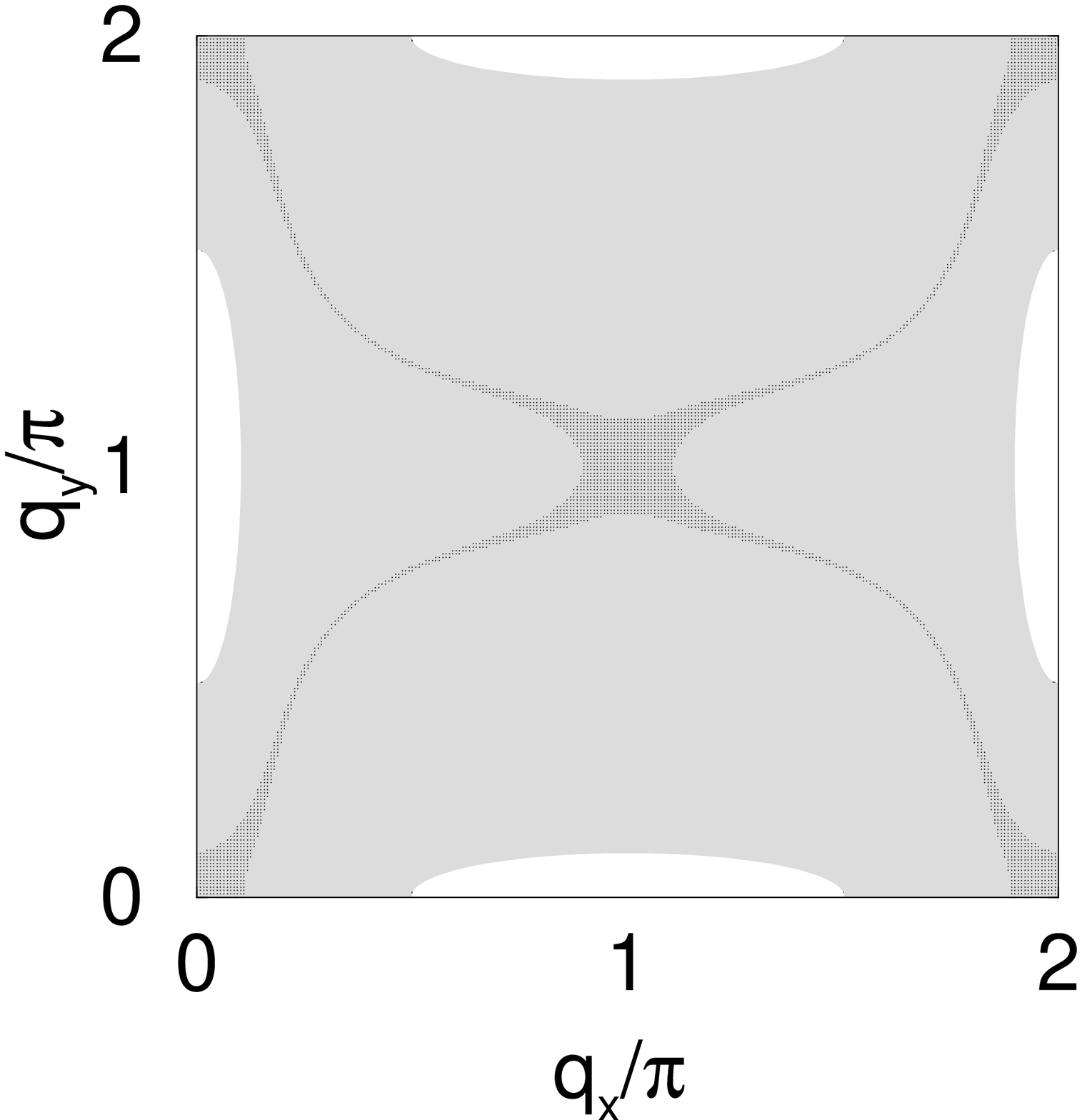}}
}
\caption{
\label{schematicphon}
As in Figs.~\ref{schematic1} d) and \ref{schematic}, but the white areas now 
denote the
experimentally determined regions where the dispersionless
half breathing zone boundary phonon mode is present.
For nodal wave vectors (left) the main self energy contribution comes from
node-node scattering processes at small energies (near $-\Omega_{phon}$).
At the $M$ point (right) the self energy effects are negligible due to
geometric restrictions. Only higher anharmonic terms (with two phonon
processes which add up to a $(\pi,\pi)$ wavevector) could contribute at
fermionic wavevectors near the $M$ point.
}
\end{figure}

In Fig.~\ref{schematicphon}, we point out an important
difference to the magnetic mode. The magnetic mode
is peaked around $(\pi,\pi)$, whereas the 
spectral density of
half breathing phonon is 
peaked around the points $(0,\pi)$ and $(\pi,0)$. Because for the
imaginary part of the self energy only excitations near the Fermi surface
are important, there are geometric restrictions for the possible
scattering events. In the case of the magnetic mode, scattering was
dominated by processes connecting the $M$ points of the 
Brillouin zone, and these scattering processes are enhanced by the
presence of a van Hove singularity close to the chemical potential.
In the case of the half breathing phonon, the $M$ point electrons
are very ineffectively scattered by these phonons due to the Pauli
exclusion principle. The important points here are near the nodal regions
for the electrons.
Thus, the strongest effects are expected near (but not necessarily at) 
the nodes, 
not near the $M$-points, in contrast to what experiment shows.

It is possible that both processes play a role and dominate
in different regions of the Brillouin zone. Phonons would then
play some role for nodal electrons.

\section{Conclusions}
We suggest that the van Hove singularity
at the $M$ point of the Brillouin zone plays an important role in
determining the self energy effects observed in ARPES and tunneling
experiments.
The picture can be understood as follows:
the quasiparticle dispersion is fairly flat near the $M$ points of the
Brillouin zone, with a large effective mass in the $M-\Gamma$ direction, 
and is close to the chemical potential.
Because the continuum part of the spin fluctuation spectrum is
gapped up to energies $\sim 70$ meV, the
scattering at low energies is dominated by scattering
processes accompanied by emission of a spin fluctuation mode (which
lies below the gapped continuum at $\sim 40$ meV and has a sharp energy).
This coupling leads to cusps in the energy dependence of the self energy
in the range of 70-80 meV due to the effect of the van Hove singularities
at the $M$ and $A$ points. Because of the finite width
of the spin fluctuation mode in momentum, there are traces of 
these cusps for electrons at all positions near the Fermi surface.
The {\it position in energy} of these cusps are determined by electrons
near the $M$ and $A$ points only, which explains the isotropy around
the Fermi surface of
the energy scale of 70-80 meV where kink features in the dispersion 
are observed.
The {\it intensity} of this self energy effect is determined by the intensities
of the spin fluctuation mode at such momenta $\vec{q}$ which connect
the electron with momentum $\vec{k} $ 
to electrons near the $M$ point region. Thus, 
$\vec q=(\vec k-\vec k_M)$ modulo $(\vec{G})$, 
where $\vec{G} $ is a reciprocal lattice vector.
This intensity is large for $\vec{k} \approx \vec{k}_M$, but smaller for
$\vec{k} \approx \vec{k}_N$.
This explains the strong anisotropy of the magnitude of the 
effect around the Fermi surface.

\section{Acknowledgements}
We would like to thank Adam Kaminski and Juan Carlos Campuzano for discussions
concerning their photoemission data, and John Zasadzinski concerning
tunneling data. We acknowledge communications with Steve Kivelson
and Hae-Young Kee, and helpful discussions with
Andrey Chubukov.
This work was supported by the U.~S.~Dept.~of Energy, 
Office of Science, under Contract No.~W-31-109-ENG-38.

{\it Note added in proof.}  After submitting this paper, we became 
aware of an experimental paper which claims no momentum anisotropy in 
the linewidth for overdoped compounds \cite{Bogdanov02}.
This result is actually consistent with our picture, in that for
overdoped compounds, the spectral peak lies inside the
scattering rate gap, which can be appreciated from Fig.~21.

\appendix
\section{Self energies}
\label{app1}
In this appendix, we derive self energy expressions which allow us to
evaluate the real part of the self energy analytically in several special
cases, and have the numerical advantage of having eliminated all principal
value integrals. The procedure is a generalization of a method developed
by F. Marsiglio {\it et al.}\cite{Marsiglio88}
The self energy is given by,
\begin{equation}
\label{self}
\Sigma^R_{\epsilon,k}=
\frac{i}{2} g^2 \sum_{q,\omega } \left( G^R_{\epsilon-\omega,k-q}D^K_{\omega,q}
+G^K_{\epsilon-\omega,k-q} D^R_{\omega,q} \right)
\end{equation}
where $D=-\chi$.
In equilibrium, the Keldysh components are given by the simple
expressions
\begin{eqnarray}
D^K_{\omega,q}=\left(D^R_{\omega,q}-D^A_{\omega,q}\right)\coth 
\frac{\omega}{2T} 
=-iB_{\omega,q}(1+2b_{\omega})\quad \\
G^K_{\epsilon,k}=\left(G^R_{\epsilon,k}-G^A_{\epsilon,k}\right) 
\tanh\frac{\epsilon}{2T}
=-iA_{\epsilon,k}(1-2f_{\epsilon}) \quad
\end{eqnarray}
where $B_{\omega,q}$ and $A_{\epsilon,k}$ are the bosonic and fermionic 
spectral functions,
and $b_{\omega }$, $f_{\epsilon }$ their corresponding distribution functions,
respectively.
Note, that the Keldysh components are purely imaginary.

In evaluating these integrals,
the only numerical complication comes from the real parts of $\Sigma $
due to principal value integrals. 
We can eliminate those by using the following trick:
Note that in equilibrium the identities
\begin{eqnarray}
\label{id1}
&&\sum_{\omega} D^R_{\omega} G^K_{\epsilon-\omega} =
-i\sum_{\omega} \tanh \frac{\epsilon - \omega}{2T}
B_{\omega} G^R_{\epsilon-\omega} \nonumber \\
&&\quad +\sum_{\omega} \left( D^A_{\omega} G^R_{\epsilon-\omega} -
D^R_{\omega} G^A_{\epsilon-\omega} \right) \tanh \frac{\epsilon-\omega}{2T} \\
\label{id2}
&&\sum_{\omega} D^K_{\omega} G^R_{\epsilon-\omega} =
-i\sum_{\omega} \coth \frac{\omega }{2T} 
D^R_{\omega} A_{\epsilon-\omega} \nonumber \\
&&\quad +\sum_{\omega} \left( D^R_{\omega} G^A_{\epsilon-\omega} -
D^A_{\omega} G^R_{\epsilon-\omega} \right) \coth \frac{\omega}{2T}
\end{eqnarray}
hold which are easy to check. 
The convenient feature is that 
the second lines in Eqs.~\ref{id1} and \ref{id2}
can be converted into Matsubara
sums by noting that $D^A_{\omega }G^R_{\epsilon-\omega} $ is an analytic
function in the lower $\omega $ half plane, and analogously 
$D^R_{\omega }G^A_{\epsilon-\omega} $ analytic in the upper half plane.
Thus,
\begin{eqnarray}
&&\frac{i}{2} \sum_{\omega} \left( D^A_{\omega} G^R_{\epsilon-\omega} - 
D^R_{\omega} G^A_{\epsilon-\omega} \right) \tanh \frac{\epsilon-\omega}{2T} 
= \nonumber \\
&&\qquad -T \sum_{\epsilon_n} D^M(\epsilon -i\epsilon_n) G^M(i\epsilon_n) \\
&&\frac{i}{2} \sum_{\omega} \left( D^R_{\omega} G^A_{\epsilon-\omega} - 
D^A_{\omega} G^R_{\epsilon-\omega} \right) \coth \frac{\omega}{2T} 
= \nonumber \\
&&\qquad -T \sum_{\omega_m} D^M(i\omega_m) G^M(\epsilon - i\omega_m)
\end{eqnarray}
where $D^M(\epsilon -i\epsilon_n) $ and $G^M(\epsilon - i\omega_m)$ are
smooth functions (except at $\omega_m=0$, which is treated separately,
see below). So, the self energy Eq.~\ref{self} has the two
alternative equivalent forms (the first form was found in 
Ref.~\onlinecite{Marsiglio88}),
\begin{eqnarray}
\Sigma^R_{\epsilon,k}&=&g^2 \Bigg[
\sum_{\omega,q} 
B_{\omega,q} 
\rho^T_{\omega,\epsilon-\omega}
G^R_{\epsilon-\omega,k-q} 
\nonumber \\
&&- T \sum_{\epsilon_n,q} G^M_{k-q}(i\epsilon_n)D^M_q(\epsilon-i\epsilon_n)
\Bigg] \\
\Sigma^R_{\epsilon,k}&=&g^2 \Bigg[
\sum_{\omega,q} 
\left(
D^R_{\omega,q} \rho^T_{\omega,\epsilon-\omega} 
- \mbox{Re} D^R_{0,q} \cdot \frac{T}{\omega } 
\right) 
A_{\epsilon-\omega,k-q} 
\nonumber \\
&&- T \sum_{\omega_m\ne 0,q} G^M_{k-q}(\epsilon-i\omega_m)D^M_q(i\omega_m)
\Bigg]
\end{eqnarray}
where the population factor $\rho^T_{\omega,\epsilon-\omega}$ is given by,
\begin{eqnarray}
\rho^T_{\omega,\epsilon-\omega}=\frac{1}{2} \left(\coth \frac{\omega}{2T}+
\tanh \frac{\epsilon-\omega}{2T} \right)
\end{eqnarray}
Note that the terms containing Matsubara sums are pure real quantities.

Let us examine the simple case
\begin{eqnarray}
A_{\epsilon,k} = 2\pi \delta (\epsilon -\xi_k)
\end{eqnarray}
which gives, using the second expression
\begin{eqnarray}
\Sigma^R_{\epsilon,k}&=&g^2 
\sum_q 
\Bigg( 
D^R_{\epsilon-\xi_{k-q},q} 
\rho^T_{\epsilon-\xi_{k-q},\xi_{k-q}}
- \mbox{Re} D^R_{0,q} \frac{T}{\epsilon-\xi_{k-q}} 
\nonumber \\
&&- T\sum_{\omega_m \ne 0} \frac{D^M_q(i\omega_m)}{\epsilon-i\omega_m-
\xi_{k-q}}
\Bigg)
\end{eqnarray}
Finally, for the case that the bosonic mode has the
simple form
\begin{equation}	
B_{\omega,q}=2 w_q\left[ \delta (\omega-\Omega )-\delta (\omega+\Omega ) 
\right]
\end{equation}	
the first expression leads to
\begin{eqnarray}
\Sigma^R_{\epsilon,k}=\frac{g^2}{\pi} 
\sum_q w_q \Bigg(
\rho^T_{\Omega,\epsilon-\Omega}
G^R_{\epsilon-\Omega, k-q} 
-\rho^T_{-\Omega,\epsilon+\Omega}
G^R_{\epsilon+\Omega, k-q} \nonumber \\
-T \sum_{\epsilon_n} G^M_{k-q} (i\epsilon_n) \left(
\frac{1}{\epsilon-i\epsilon_n-\Omega} - \frac{1}{\epsilon-i\epsilon_n+\Omega}
\right) \Bigg) \nonumber \\
\end{eqnarray}
The last sum can be performed for the case of a Green's function of the
form
\begin{eqnarray}
G^R_{\epsilon,k}&=&\frac{1}{\epsilon-\xi_k+i\Gamma_k} \\
G^M_k(i\epsilon_n) &=& \frac{1}{i\epsilon_n-\xi_k+i\Gamma_k \mbox{sign} 
(\epsilon_n)}
\end{eqnarray}
leading to
\begin{eqnarray}
&&-T \sum_{\epsilon_n} G^M_{k-q} (i\epsilon_n) \left(
\frac{1}{\epsilon-i\epsilon_n-\Omega} - \frac{1}{\epsilon-i\epsilon_n+\Omega}
\right) \nonumber\\
&&= \pi \mbox{Re} \Bigg[
\frac{i}{\epsilon-\Omega-\xi_{k-q}+i\Gamma_{k-q}} \nonumber \\ 
&&\times \left\{
\Psi \left(\frac{1}{2} + i\frac{\epsilon-\Omega}{2\pi T} \right)
-\Psi \left(\frac{1}{2} + \frac{\Gamma_{k-q}+i\xi_{k-q}}{2\pi T} \right)
\right\} \nonumber \\
&&\qquad 
-\frac{i}{\epsilon+\Omega-\xi_{k-q}+i\Gamma_{k-q}} \nonumber \\ 
&&\times \left\{
\Psi \left(\frac{1}{2} + i\frac{\epsilon+\Omega}{2\pi T} \right)
-\Psi \left(\frac{1}{2} + \frac{\Gamma_{k-q}+i\xi_{k-q}}{2\pi T} \right)
\right\} \Bigg] \nonumber \\
\end{eqnarray}

\end{document}